\documentclass{book}
\usepackage{times}

\usepackage[english]{babel}
\usepackage[letterpaper,top=2cm,bottom=2cm,left=3cm,right=3cm,marginparwidth=1.75cm]{geometry}

\usepackage{pdflscape}
\usepackage{rotating}
\usepackage{longtable}
\usepackage{array}
\usepackage{float}
\usepackage{amsmath}
\usepackage{graphicx}

\usepackage{courier}
\usepackage[colorlinks=true, allcolors=blue]{hyperref}
\usepackage{enumitem}
\usepackage{tikz} % TikZ package from first code
\usepackage{listings}
\usepackage{xcolor}
\usepackage[T1]{fontenc}
\usepackage{epigraph}  % For block quotes

\definecolor{bgcolor}{rgb}{0.97,0.97,0.97}
\definecolor{codeblue}{rgb}{0.1,0.1,0.8}
\definecolor{codegreen}{rgb}{0,0.4,0}
\definecolor{codegray}{rgb}{0.4,0.4,0.4}
\definecolor{codepurple}{rgb}{0.5,0,0.5}
\definecolor{codered}{rgb}{0.6,0.2,0.2}
\definecolor{lightgray}{rgb}{0.9,0.9,0.9}
\definecolor{darkgray}{rgb}{0.6,0.6,0.6} % Darker gray for Python frames

\makeatletter
\renewcommand{\paragraph}{%
  \@startsection{paragraph}{4}{\z@}{1ex}{-1em}{\normalfont\normalsize\bfseries\color{gray}}}
\makeatother

\lstdefinestyle{python}{
    language=Python, % Specifies the programming language
    basicstyle=\ttfamily\small\color{black}, % Default monospaced font for LaTeX
    keywordstyle=\bfseries\color{blue},  % Keywords are bold and blue (widely recognized as a coding standard)
    stringstyle=\color{orange!85!black},  % Strings are dark orange
    commentstyle=\color{green!50!black},  % Comments are dark green (high readability)
    showstringspaces=false,  % Do not show spaces in strings
    numbers=left,  % Line numbers are displayed on the left
    numberstyle=\tiny\color{gray},  % Line numbers are tiny and gray
    stepnumber=1,  % Line numbers increment by 1
    numbersep=8pt,  % Space between line numbers and code
    frame=single,  % Single-line frame around the code
    rulecolor=\color{black},  % Frame color is black
    breaklines=true,  % Automatically break long lines
    backgroundcolor=\color{gray!10},  % Light gray background
    tabsize=4,  % Set tab size to 4 spaces
    captionpos=b,  % Captions are positioned below the code
    morekeywords={self, None, True, False}, % Additional Python keywords
}

\lstdefinestyle{cmd}{
    language=bash,
    basicstyle=\ttfamily\small\color{black},  % Monospaced font in black
    keywordstyle=\bfseries\color{blue},  % Commands in blue
    stringstyle=\color{red!80!black},  % Strings in dark red
    commentstyle=\color{gray},  % Comments in gray
    showstringspaces=false,  % Do not show spaces in strings
    numbers=none,  % No line numbers
    frame=single,  % Single-line frame around the code
    rulecolor=\color{black},  % Frame color is black
    breaklines=true,  % Automatically break long lines
    backgroundcolor=\color{gray!10},  % Light gray background
    tabsize=4,  % Set tab size to 4 spaces
    captionpos=b,  % Captions are positioned below the code
}

\title{Deep Learning Model Security: Threats and Defenses}
\author{
    Tianyang Wang\textsuperscript{*}\thanks{Xi'an Jiaotong-Liverpool University, \texttt{Tianyang.Wang21@student.xjtlu.edu.cn}} \and
    Ziqian Bi\textsuperscript{*$\dagger$}\thanks{Indiana University, \texttt{bizi@iu.edu}} \and
    Yichao Zhang\textsuperscript{*}\thanks{The University of Texas at Dallas, \texttt{yichao.zhang.us@gmail.com}} \and
    Ming Liu\textsuperscript{*}\thanks{Purdue University, \texttt{liu3183@purdue.edu}} \and
    Weiche Hsieh\textsuperscript{*$\dagger$}\thanks{National Tsing Hua University, \texttt{s112033645@m112.nthu.edu.tw}} \and
    Pohsun Feng\textsuperscript{*}\thanks{National Taiwan Normal University, \texttt{41075018h@ntnu.edu.tw}} \and
    Lawrence K.Q. Yan\thanks{Hong Kong University of Science and Technology, \texttt{kqyan@connect.ust.hk}} \and
    Yizhu Wen\thanks{University of Hawaii, \texttt{yizhuw@hawaii.edu}} \and
    Benji Peng\thanks{AppCubic, \texttt{benji@appcubic.com}} \and
    Junyu Liu\thanks{Kyoto University, \texttt{liu.junyu.82w@st.kyoto-u.ac.jp}} \and
    Keyu Chen\thanks{Georgia Institute of Technology, \texttt{kchen637@gatech.edu}} \and
    Sen Zhang\thanks{Rutgers University, \texttt{sen.z@rutgers.edu}} \and
    Ming Li\thanks{Georgia Institute of Technology, \texttt{mli694@gatech.edu}} \and
    Chuanqi Jiang\thanks{National University of Singapore, \texttt{e0729764@u.nus.edu}} \and
    Xinyuan Song\thanks{National University of Singapore, \texttt{songxinyuan@u.nus.edu}} \and
    Junjie Yang\thanks{Pingtan Research Institute of Xiamen University, \texttt{youngboy@xmu.edu.cn}} \and
    Bowen Jing\thanks{University of Manchester, \texttt{bowen.jing@postgrad.manchester.ac.uk}} \and
    Jintao Ren\thanks{Aarhus University, \texttt{jintaoren@clin.au.dk}} \and
    Junhao Song\thanks{Imperial College London, \texttt{junhao.song23@imperial.ac.uk}} \and
    Hong-Ming Tseng\thanks{School of Visual Arts, \texttt{htseng@sva.edu}} \and
    Silin Chen\thanks{Zhejiang University, \texttt{A1033439225@gmail.com}} \and
    Yunze Wang\thanks{University of Edinburgh, \texttt{Y.Wang-861@sms.ed.ac.uk}}
    Chia Xin Liang\thanks{JTB Technology Corp., \texttt{cxldun@gmail.com}} \and
    Jiawei Xu\thanks{Purdue University, \texttt{xu1644@purdue.edu}} \and
    Xuanhe Pan\thanks{University of Wisconsin-Madison, \texttt{xpan73@wisc.edu}} \and
    Jinlang Wang\thanks{University of Wisconsin-Madison, \texttt{jinlang.wang@wisc.edu}} \and
    Qian Niu\thanks{Kyoto University, \texttt{niu.qian.f44@kyoto-u.ac.jp}}
}

\date{} % Remove the date

\begin{document}

\maketitle

% Add footnotes for co-first author and corresponding author
\begingroup
\renewcommand\thefootnote{}\footnote{
\textsuperscript{*} Equal contribution \\
\textsuperscript{$\dagger$} Corresponding author
}
\addtocounter{footnote}{0}
\endgroup

% Insert the quote right after the title and authors
\epigraph{"Adversarial examples are a real threat to deep learning in practice, especially in safety and security-critical applications."}{\textit{Ian Goodfellow}}

\epigraph{"The best minds in AI research are having difficulty in devising effective defenses against it."}{\textit{Christian Szegedy}}

\epigraph{"Adversaries can gain a better understanding of the classes and the private dataset used, opening doors for follow-on attacks."}{\textit{Matt Fredrikson}}

\epigraph{"By far, the greatest danger of Artificial Intelligence is that people conclude too early that they understand it."}{\textit{Eliezer Yudkowsky}}

\epigraph{"The potential benefits of artificial intelligence are huge, so are the dangers."}{\textit{Stephen Hawking}}

\epigraph{"Cyber hygiene, patching vulnerabilities, security by design, threat hunting, and machine learning-based artificial intelligence are mandatory prerequisites for cyber defense against the next generation."}{\textit{James Scott}}

\tableofcontents  % Table of contents

\part{Fundamentals and Background}
    \chapter{Basic Concepts of Deep Learning and Security}
    
    \section{What is Deep Learning?}
    Deep learning is a subfield of machine learning, which itself is a branch of artificial intelligence (AI) \cite{lecun2015deep,goodfellow2016deep,bengio2017deep}. While machine learning includes various algorithms to make predictions or find patterns in data, deep learning specifically focuses on using deep neural networks. A deep neural network is a type of model consisting of multiple layers of nodes (often called neurons), which enables the model to learn complex patterns and representations from data.

    One of the main reasons deep learning has become so popular is its success in solving tasks that were once considered too difficult for computers. For example, tasks like recognizing objects in images (computer vision) \cite{ullman2016atoms}, understanding human language (natural language processing) \cite{winograd1972understanding}, and even playing complex games like Go \cite{williams2015playing}. 

    The "deep" in deep learning refers to the number of layers in the neural network. Traditional neural networks might have just one or two layers, while deep learning models can have many layers—sometimes hundreds or even thousands. The idea is that these deep networks can automatically learn useful features from raw data without the need for manual feature engineering.

    In the context of this book, we will use deep learning to build models using Python \cite{van2003introduction} and PyTorch \cite{paszke2019pytorch}, one of the most widely-used frameworks for deep learning. 

    \section{How Deep Learning Models Work}
    Deep learning models, specifically neural networks, work by mimicking the way the human brain processes information. A neural network consists of several layers of neurons: an input layer, one or more hidden layers, and an output layer. Each neuron is a mathematical function that takes one or more inputs, multiplies them by some weights, and then applies an activation function to decide the output. The output of one layer serves as the input for the next layer.

    Let's walk through the process with a simple example:

    \subsection*{Step-by-Step Example of a Neural Network}
    Imagine we are trying to classify handwritten digits, where the input to the model is an image of a digit, and the output is the predicted digit (0-9).

    \begin{enumerate}
        \item \textbf{Input layer:} The image is transformed into a vector of pixel values, which serves as the input to the network. For instance, if the image is 28x28 pixels, we will have 784 input neurons (since 28x28 = 784).
        
        \item \textbf{Hidden layers:} The input is passed through one or more hidden layers, each of which applies weights to the input data and passes it through an activation function to introduce non-linearity. For example, one of the most commonly used activation functions is the ReLU (Rectified Linear Unit) function \cite{agarap2018deep}, defined as:
        \[
        \text{ReLU}(x) = \max(0, x)
        \]
        
        \item \textbf{Output layer:} Finally, the last layer produces the output, which in this case is a vector of probabilities representing the likelihood of the image being each digit from 0 to 9. The model will then choose the digit with the highest probability as its prediction.
    \end{enumerate}

    The training process involves adjusting the weights of the neurons to minimize the difference between the predicted output and the actual target (the correct digit). This is done through a process called backpropagation and optimization algorithms like stochastic gradient descent (SGD) \cite{bottou2012stochastic}.

    Here's a very simple PyTorch code snippet to show the structure of a deep learning model:

    \begin{lstlisting}[style=python]
    import torch
    import torch.nn as nn
    import torch.optim as optim

    class SimpleNN(nn.Module):
        def __init__(self):
            super(SimpleNN, self).__init__()
            self.fc1 = nn.Linear(784, 128)  # First layer (input to hidden)
            self.fc2 = nn.Linear(128, 10)   # Second layer (hidden to output)
        
        def forward(self, x):
            x = torch.relu(self.fc1(x))    # Apply ReLU activation
            x = self.fc2(x)                # Output layer
            return x

    # Example of creating a model instance and printing it
    model = SimpleNN()
    print(model)
    \end{lstlisting}

    This code defines a simple feed-forward neural network with two layers. The input is 784-dimensional (for the pixels of a 28x28 image), and the output is 10-dimensional (for the 10 possible digits). The forward method specifies how the input data flows through the network.

    \section{Basic Components of Neural Networks}
    Neural networks are composed of several key components, which are critical to understanding how they work \cite{ren2024deeplearningmachinelearning, chen2024deeplearningmachinelearning}. These components include:

    \subsection{Neurons}
    Neurons \cite{sheela2013review} are the basic building blocks of a neural network. Each neuron takes one or more inputs, applies a weight to each input, sums them, and then applies an activation function. The result is passed to the next layer. In mathematical terms, for a neuron $i$ with inputs $x_1, x_2, \ldots, x_n$, weights $w_1, w_2, \ldots, w_n$, and a bias $b$, the output $y_i$ is given by:
    \[
    y_i = f(w_1x_1 + w_2x_2 + \cdots + w_nx_n + b)
    \]
    where $f$ is the activation function.

    \subsection{Layers}
    A neural network consists of several layers of neurons:
    \begin{itemize}
        \item \textbf{Input layer:} This is where the input data enters the network.
        \item \textbf{Hidden layers:} These layers perform computations and extract features from the input data. Deep networks have multiple hidden layers.
        \item \textbf{Output layer:} This layer provides the final prediction or output of the network.
    \end{itemize}

    \subsection{Activation Functions}
    Activation functions introduce non-linearity into the model, allowing it to learn more complex patterns. Common activation functions include:
    \begin{itemize}
        \item \textbf{Sigmoid \cite{han1995influence}:} Maps any input to a value between 0 and 1.
        \[
        \text{Sigmoid}(x) = \frac{1}{1 + e^{-x}}
        \]
        \item \textbf{ReLU \cite{agarap2018deep}:} Rectified Linear Unit, a very popular activation function.
        \[
        \text{ReLU}(x) = \max(0, x)
        \]
    \end{itemize}

    \subsection{Weights and Biases}
    Weights \cite{han2015learning} are the parameters that the network learns during training. Each connection between neurons has an associated weight, and these weights are updated through training to minimize the error between the predicted and actual outputs. Biases are additional parameters that help shift the activation function to better fit the data.

    \section{Difference Between Deep Learning and Traditional Machine Learning}
    Deep learning and traditional machine learning (ML) differ in several key ways:

    \subsection{Feature Extraction}
    In traditional ML, feature extraction is often a separate step that requires domain expertise to select the most relevant features from raw data. For example, in image processing, one might manually define features like edges, textures, or shapes. In contrast, deep learning models learn to automatically extract features from the raw data during training. This is one of the reasons why deep learning has been so successful in fields like computer vision and natural language processing \cite{hsieh2024deeplearningmachinelearning}.

    \subsection{Handling Unstructured Data}
    Traditional ML models generally work better with structured data (such as data in rows and columns). Deep learning, however, excels in handling unstructured data, like images, text, and audio. For instance, a deep learning model can be trained directly on images without needing to convert them into numerical features manually.

    \subsection{Scalability with Data}
    Deep learning models perform better as the size of the dataset increases, which is not always the case with traditional ML algorithms. Deep networks can take advantage of large datasets to learn more intricate patterns, whereas traditional models might overfit or struggle to handle such data.

    Here's a comparison to make it clearer:
    \begin{center}
    \begin{tabular}{|c|c|c|}
    \hline
    & \textbf{Traditional ML} & \textbf{Deep Learning} \\
    \hline
    \textbf{Feature Engineering} & Manual & Automatic \\
    \textbf{Data Type} & Structured & Unstructured \\
    \textbf{Performance with Big Data} & May degrade & Improves \\
    \hline
    \end{tabular}
    \end{center}

    In conclusion, while traditional ML algorithms are effective for many tasks, deep learning's ability to automatically learn from raw data and improve with large datasets makes it the go-to choice for complex problems like image recognition and natural language processing.

\section{Introduction to Deep Learning Model Security}
With the increasing reliance on deep learning (DL) models in various critical applications, ensuring their security has become a crucial concern. Deep learning models are being deployed in sensitive areas such as healthcare, finance, autonomous driving, and military operations, where failures or vulnerabilities can lead to catastrophic outcomes. Consequently, securing these models from various forms of attacks is paramount to guarantee their robustness, trustworthiness, and reliability.

\subsection{Why is Security Important in Deep Learning?}
Deep learning models are powerful tools capable of making intelligent predictions based on patterns learned from data. However, as they are increasingly used in real-world applications, they become attractive targets for malicious actors. Failing to secure deep learning models can lead to several risks:
\begin{itemize}
    \item \textbf{Compromised Decision Making}: When deployed in safety-critical areas such as autonomous vehicles or medical diagnosis, vulnerabilities can lead to erroneous decisions that may cause accidents, misdiagnoses, or even loss of life.
    \item \textbf{Privacy Leaks}: Attackers can extract sensitive information from models trained on confidential data. For example, models trained on medical or financial data can be exploited to reveal personal information.
    \item \textbf{Reputation and Financial Loss}: A compromised model can result in reputational damage and financial losses for companies relying on these models for their services.
    \item \textbf{Ethical and Legal Concerns}: Security breaches in models deployed in areas like facial recognition or user profiling can raise ethical and legal concerns regarding privacy and fairness.
\end{itemize}

\subsection{Common Security Threats to Deep Learning Models}
Several types of security threats can compromise the integrity, confidentiality, and availability of deep learning models. Some of the most common ones include:

\subsubsection{1. Adversarial Attacks}
Adversarial attacks occur when an attacker makes small, imperceptible modifications to input data, causing the model to make incorrect predictions. For instance, slightly altering the pixels of an image of a stop sign can cause a model to misclassify it as a yield sign, which could be disastrous in an autonomous driving scenario. These attacks exploit the model's inability to generalize properly in certain edge cases  \cite{huang2017adversarial}.

\subsubsection{2. Data Poisoning}
In data poisoning, an attacker injects malicious samples into the training dataset with the aim of corrupting the learned model. This type of attack compromises the model's integrity by making it perform poorly on specific tasks. For example, inserting wrongly labeled images into the dataset used to train a facial recognition model could degrade its accuracy \cite{steinhardt2017certified}.

\subsubsection{3. Model Theft}
Model theft occurs when an adversary extracts or replicates a model's behavior by querying it and obtaining sufficient input-output pairs. This can allow attackers to reverse-engineer proprietary models without having direct access to the original training data or architecture. It also enables them to avoid research costs and steal intellectual property  \cite{gunturi2021ensemble}.

\section{How Are Deep Learning Models Attacked?}
Attackers have devised various methods to exploit the vulnerabilities of deep learning models. Understanding these techniques is crucial in designing countermeasures.

\subsection{Adversarial Examples}
Adversarial examples are specially crafted inputs that are designed to deceive the model into making incorrect predictions. These inputs appear normal to humans but include small perturbations that confuse the model. For example, consider the following simple PyTorch code that demonstrates an adversarial attack on a trained neural network model using the Fast Gradient Sign Method (FGSM) \cite{cheng2021fast}:

\begin{lstlisting}[style=python]
import torch
import torch.nn as nn
import torch.optim as optim

# Simple example model
class SimpleModel(nn.Module):
    def __init__(self):
        super(SimpleModel, self).__init__()
        self.fc = nn.Linear(2, 2)
    
    def forward(self, x):
        return self.fc(x)

# Function to create adversarial example
def adversarial_example(model, data, target, epsilon):
    data.requires_grad = True
    output = model(data)
    loss = nn.CrossEntropyLoss()(output, target)
    model.zero_grad()
    loss.backward()
    data_grad = data.grad.data
    perturbed_data = data + epsilon * data_grad.sign()
    return perturbed_data

# Create dummy data and labels
data = torch.tensor([[1.0, 2.0]], requires_grad=True)
target = torch.tensor([1])

# Instantiate the model and optimizer
model = SimpleModel()
optimizer = optim.SGD(model.parameters(), lr=0.01)

# Generate an adversarial example
epsilon = 0.1
perturbed_data = adversarial_example(model, data, target, epsilon)
print("Original Data: ", data)
print("Perturbed Data: ", perturbed_data)
\end{lstlisting}

In this example, a small noise is added to the input data using the gradient of the loss with respect to the input, creating an adversarial example that could fool the model.

\subsection{Model Inversion}
Model inversion attacks \cite{fredrikson2015model} aim to reconstruct sensitive input data from the outputs of a model. Attackers query a trained model with various inputs and attempt to reverse-engineer the features learned by the model. This can be particularly harmful if the model is trained on private data such as medical records or biometric information. 

For example, an attacker could query a facial recognition model and reconstruct approximate images of faces from the model's output embeddings, thus violating the privacy of individuals in the training dataset.

\subsection{Data Poisoning}
In data poisoning attacks \cite{fan2022survey}, attackers insert harmful samples into the training data. Consider a scenario where an attacker adds mislabeled data to a dataset. This can cause a model to perform poorly or make biased predictions in specific cases.

Here's a basic example to demonstrate the concept of poisoned data during training:

\begin{lstlisting}[style=python]
import torch
from torch.utils.data import DataLoader, TensorDataset

# Generating a simple poisoned dataset
data = torch.tensor([[1.0, 2.0], [2.0, 3.0], [3.0, 4.0], [5.0, 6.0]])
labels = torch.tensor([0, 0, 1, 1])

# Poisoning the dataset by adding incorrect label to one sample
data_poisoned = torch.tensor([[4.0, 5.0]])
labels_poisoned = torch.tensor([0])  # Incorrect label

# Combine clean and poisoned data
data_total = torch.cat((data, data_poisoned), 0)
labels_total = torch.cat((labels, labels_poisoned), 0)

# Create a DataLoader to simulate training
dataset = TensorDataset(data_total, labels_total)
loader = DataLoader(dataset, batch_size=2)

# Simple training loop
model = SimpleModel()
optimizer = optim.SGD(model.parameters(), lr=0.01)
criterion = nn.CrossEntropyLoss()

for epoch in range(2):
    for inputs, targets in loader:
        optimizer.zero_grad()
        output = model(inputs)
        loss = criterion(output, targets)
        loss.backward()
        optimizer.step()
\end{lstlisting}

In this scenario, a mislabeled sample is added to the dataset, which could influence the model's learning process. Poisoning attacks can be difficult to detect because the attacker may introduce only a small amount of bad data to avoid raising suspicion.

\subsection{Other Attack Methods}
Other methods that attackers use include:
\begin{itemize}
    \item \textbf{Model Extraction}: Using model queries to replicate a model's architecture and behavior.
    \item \textbf{Membership Inference Attacks}: Determining whether a specific data point was part of the training data.
    \item \textbf{Evasion Attacks}: Crafting inputs during inference that cause the model to make wrong predictions.
\end{itemize}

\chapter{Introduction to PyTorch}
\section{Installing PyTorch and Setting Up the Environment}
\subsection{How to Install PyTorch}

PyTorch \cite{imambi2021pytorch} is an open-source deep learning framework, widely used for developing machine learning and deep learning models. It is highly flexible and offers a rich set of libraries for building and training neural networks. Before diving into PyTorch, you must ensure it is properly installed on your system \cite{peng2024securinglargelanguagemodels}.

\textbf{Step 1: Install Python}

Ensure that you have Python 3.7 or higher installed on your system. You can download Python from the official Python website. To check if Python is installed, run the following command in your terminal or command prompt:

\begin{lstlisting}[style=cmd]
python --version
\end{lstlisting}

\textbf{Step 2: Install PyTorch Using \texttt{pip}}

The simplest way to install PyTorch is via \texttt{pip}, which is Python's package manager. You can install the latest version of PyTorch with CPU or GPU support by using the following command:

\textit{For CPU support:}
\begin{lstlisting}[style=cmd]
pip install torch torchvision torchaudio
\end{lstlisting}

\textit{For GPU (CUDA) support: \cite{li2024deeplearningmachinelearning}}
\begin{lstlisting}[style=cmd]
pip install torch torchvision torchaudio --index-url https://download.pytorch.org/whl/cu118
\end{lstlisting}

\textbf{Step 3: Install PyTorch Using Conda (Recommended for Anaconda Users)}

If you use the Anaconda distribution, you can install PyTorch with Conda. This method simplifies the process of managing different environments and dependencies.

\textit{For CPU support:}
\begin{lstlisting}[style=cmd]
conda install pytorch torchvision torchaudio cpuonly -c pytorch
\end{lstlisting}

\textit{For GPU (CUDA) support:}
\begin{lstlisting}[style=cmd]
conda install pytorch torchvision torchaudio pytorch-cuda=11.8 -c pytorch -c nvidia
\end{lstlisting}

\textbf{Step 4: Verify Installation}

Once the installation is complete, you can verify that PyTorch is correctly installed by launching Python and running the following code:

\begin{lstlisting}[style=python]
import torch
print(torch.__version__)
\end{lstlisting}

If the output displays the installed version of PyTorch, the installation was successful.

\subsection{Setting Up Virtual Environments}

When working on machine learning projects, it is essential to manage dependencies carefully. Virtual environments allow you to create isolated Python environments for each project, ensuring that packages and libraries do not interfere with each other. Two popular tools for creating virtual environments are \texttt{venv} and \texttt{conda} \cite{feng2024deeplearningmachinelearning}.

\textbf{Using \texttt{venv} to Create a Virtual Environment}

The \texttt{venv} module comes pre-installed with Python 3. You can create a new virtual environment for your PyTorch project by following these steps:

1. Navigate to your project directory in the terminal.

2. Run the following command to create a new virtual environment:

\begin{lstlisting}[style=cmd]
python -m venv my_pytorch_env
\end{lstlisting}

3. Activate the virtual environment:

\textit{On Windows:}
\begin{lstlisting}[style=cmd]
my_pytorch_env\Scripts\activate
\end{lstlisting}

\textit{On macOS or Linux:}
\begin{lstlisting}[style=cmd]
source my_pytorch_env/bin/activate
\end{lstlisting}

4. Install PyTorch in the virtual environment:

\begin{lstlisting}[style=cmd]
pip install torch torchvision torchaudio
\end{lstlisting}

\textbf{Using \texttt{conda} to Create a Virtual Environment}

If you use Conda, you can create and manage virtual environments easily:

1. Create a new Conda environment for your project:

\begin{lstlisting}[style=cmd]
conda create --name my_pytorch_env python=3.9
\end{lstlisting}

2. Activate the new environment:

\begin{lstlisting}[style=cmd]
conda activate my_pytorch_env
\end{lstlisting}

3. Install PyTorch in this environment:

\begin{lstlisting}[style=cmd]
conda install pytorch torchvision torchaudio -c pytorch
\end{lstlisting}

By using virtual environments, you can manage multiple projects without worrying about version conflicts between dependencies.

\section{Introduction to Tensors}

\subsection{What are Tensors?}

Tensors are the fundamental data structure in PyTorch. They are multidimensional arrays that can store data of any type and size. Tensors in PyTorch are similar to NumPy arrays but with additional capabilities, such as GPU support and automatic differentiation.

\textbf{Example of a Simple Tensor:}

A scalar is a zero-dimensional tensor, a vector is a one-dimensional tensor, and a matrix is a two-dimensional tensor. Here's an example of creating various tensors in PyTorch:

\begin{lstlisting}[style=python]
import torch

# Scalar (0D tensor)
scalar = torch.tensor(5)
print(scalar)

# Vector (1D tensor)
vector = torch.tensor([1, 2, 3])
print(vector)

# Matrix (2D tensor)
matrix = torch.tensor([[1, 2], [3, 4]])
print(matrix)

# 3D Tensor
tensor3d = torch.tensor([[[1, 2], [3, 4]], [[5, 6], [7, 8]]])
print(tensor3d)
\end{lstlisting}

The above example shows how tensors can be created in different dimensions, from scalars to higher-dimensional tensors. The flexibility and power of tensors make them essential for deep learning, where high-dimensional data is common.

\subsection{Basic Operations with Tensors}

PyTorch provides a wide range of functions to perform operations on tensors, such as arithmetic operations, reshaping, and broadcasting. Here are some common operations:

\textbf{Addition and Multiplication:}

\begin{lstlisting}[style=python]
# Create two tensors
a = torch.tensor([1, 2, 3])
b = torch.tensor([4, 5, 6])

# Addition
result_add = a + b
print(result_add)

# Multiplication
result_mul = a * b
print(result_mul)
\end{lstlisting}

\textbf{Reshaping Tensors:}

You can change the shape of a tensor without changing its data using the \texttt{view()} or \texttt{reshape()} functions:

\begin{lstlisting}[style=python]
# Create a 2x3 matrix
matrix = torch.tensor([[1, 2, 3], [4, 5, 6]])

# Reshape to a 3x2 matrix
reshaped_matrix = matrix.view(3, 2)
print(reshaped_matrix)
\end{lstlisting}

\textbf{Broadcasting:}

Broadcasting allows PyTorch to perform operations on tensors of different shapes by automatically expanding their dimensions to be compatible.

\begin{lstlisting}[style=python]
# Create a 2x2 matrix
matrix = torch.tensor([[1, 2], [3, 4]])

# Create a 1D tensor
vector = torch.tensor([1, 2])

# Broadcast and add the vector to each row of the matrix
result_broadcast = matrix + vector
print(result_broadcast)
\end{lstlisting}

\section{Autograd and Automatic Differentiation}

\subsection{What is Automatic Differentiation?}

In deep learning, it is crucial to compute gradients during backpropagation to optimize the model's parameters. PyTorch's \texttt{autograd} feature automatically computes the gradients of tensors that have the \texttt{requires\_grad} attribute set to \texttt{True}. This allows easy computation of gradients, which are essential for updating model parameters during training.

\subsection{How to Implement Autograd in PyTorch?}

Consider the following example, where we compute the gradient of a simple function:

\begin{lstlisting}[style=python]
# Create a tensor with requires_grad=True to track operations
x = torch.tensor(2.0, requires_grad=True)

# Define a simple function
y = x ** 2 + 3 * x

# Perform backpropagation to compute the gradient
y.backward()

# The gradient of y with respect to x
print(x.grad)  # Output: 7.0
\end{lstlisting}

In this example, \texttt{y.backward()} computes the derivative of the function \( y = x^2 + 3x \) with respect to \( x \), and stores the result (7.0) in \texttt{x.grad}.

With these basic building blocks—tensors and autograd—you are ready to explore more advanced topics in PyTorch, including neural networks, optimizers, and deep learning models.

\section{Building a Simple Neural Network}
    \subsection{Introduction to Neural Network Structures}
    
    Neural networks are a powerful type of machine learning model that learn by mimicking the way the human brain processes information. The basic building block of a neural network is the neuron, which takes input, applies a transformation to it (usually through a weight and bias), and passes it through an activation function. These neurons are grouped into layers. When every neuron in one layer is connected to every neuron in the next, we call it a fully connected layer or dense layer.

    A basic neural network consists of:
    \begin{itemize}
        \item \textbf{Input Layer:} This is the first layer where the input data is fed into the network.
        \item \textbf{Hidden Layers:} These layers process the input data and learn patterns from it. Each hidden layer applies a transformation to the input data, followed by an activation function.
        \item \textbf{Output Layer:} The last layer in the network that gives the final prediction or output. The number of neurons in this layer depends on the task, such as one neuron for binary classification or multiple neurons for multi-class classification.
    \end{itemize}

    \textbf{Activation Functions} are mathematical functions applied to the output of a neuron to introduce non-linearity into the network. Common activation functions include:
    \begin{itemize}
        \item \texttt{ReLU (Rectified Linear Unit):} $\text{ReLU}(x) = \max(0, x)$ is often used in hidden layers because it helps prevent vanishing gradients.
        \item \texttt{Sigmoid:} $\sigma(x) = \frac{1}{1 + e^{-x}}$ is used for binary classification tasks.
        \item \texttt{Softmax:} This is used for multi-class classification problems, where the output is interpreted as probabilities for each class.
    \end{itemize}

    In PyTorch, neural networks are built by defining layers and the forward pass in a class that inherits from the \texttt{torch.nn.Module} class. This class provides a base structure for models, allowing us to define custom architectures and take advantage of automatic differentiation.

\begin{lstlisting}[style=python]
import torch
import torch.nn as nn

# Example of a simple neural network
class SimpleNN(nn.Module):
    def __init__(self, input_size, hidden_size, output_size):
        super(SimpleNN, self).__init__()
        # Define a fully connected layer
        self.fc1 = nn.Linear(input_size, hidden_size)
        self.relu = nn.ReLU()  # Activation function
        self.fc2 = nn.Linear(hidden_size, output_size)
    
    def forward(self, x):
        # Pass through the first layer and apply ReLU
        out = self.fc1(x)
        out = self.relu(out)
        # Pass through the second layer
        out = self.fc2(out)
        return out
\end{lstlisting}

    \subsection{Building a Neural Network Using PyTorch}
    
    Let's now build a simple neural network in PyTorch. We'll define a model class by inheriting from \texttt{torch.nn.Module} and include a couple of fully connected layers with ReLU activations.

\begin{lstlisting}[style=python]
import torch
import torch.nn as nn
import torch.optim as optim

# Define a simple neural network
class SimpleNN(nn.Module):
    def __init__(self):
        super(SimpleNN, self).__init__()
        self.fc1 = nn.Linear(28*28, 128)  # Input size for an image with 28x28 pixels
        self.fc2 = nn.Linear(128, 64)     # Hidden layer with 64 neurons
        self.fc3 = nn.Linear(64, 10)      # Output layer for 10 classes (e.g., digits 0-9)

    def forward(self, x):
        x = self.fc1(x)
        x = torch.relu(x)  # Apply ReLU activation
        x = self.fc2(x)
        x = torch.relu(x)  # Apply ReLU activation
        x = self.fc3(x)    # Output layer
        return x
\end{lstlisting}

    In this example, the network is designed to classify images of size 28x28 (such as those in the MNIST dataset \cite{xiao2017fashion}). It consists of three fully connected layers, with ReLU activation functions applied after the first two layers.

\section{Training and Validating Models}
    \subsection{Implementing the Training Loop}
    
    After defining a model, the next step is to train it. Training involves the following key steps:
    \begin{itemize}
        \item \textbf{Forward Pass:} Pass the input through the network to get predictions.
        \item \textbf{Loss Computation:} Compare the predicted output with the true labels using a loss function (such as cross-entropy loss).
        \item \textbf{Backward Pass:} Use backpropagation to calculate the gradients.
        \item \textbf{Optimizer Step:} Update the model weights using an optimization algorithm (like stochastic gradient descent or Adam).
    \end{itemize}

    Here's how we can implement the training loop in PyTorch:

\begin{lstlisting}[style=python]
# Define the training function
def train(model, dataloader, criterion, optimizer, num_epochs=5):
    for epoch in range(num_epochs):
        running_loss = 0.0
        for inputs, labels in dataloader:
            # Zero the gradients
            optimizer.zero_grad()

            # Forward pass
            outputs = model(inputs)
            loss = criterion(outputs, labels)

            # Backward pass and optimization
            loss.backward()
            optimizer.step()

            running_loss += loss.item()

        print(f'Epoch {epoch+1}/{num_epochs}, Loss: {running_loss/len(dataloader)}')
\end{lstlisting}

    This code defines a simple training loop that:
    \begin{itemize}
        \item Iterates over a specified number of epochs.
        \item For each batch of data, computes the forward pass, calculates the loss, performs the backward pass to compute gradients, and updates the model's parameters using the optimizer.
        \item Outputs the average loss at the end of each epoch.
    \end{itemize}

    \subsection{Model Validation and Testing}
    
    Validation is the process of evaluating the model on a separate dataset to ensure it generalizes well to unseen data. It is important to prevent overfitting, where the model performs well on the training data but poorly on new data.

    Here is how you can implement a validation loop:

\begin{lstlisting}[style=python]
def validate(model, dataloader, criterion):
    model.eval()  # Set the model to evaluation mode
    validation_loss = 0.0
    correct = 0
    total = 0
    
    with torch.no_grad():  # No need to compute gradients during validation
        for inputs, labels in dataloader:
            outputs = model(inputs)
            loss = criterion(outputs, labels)
            validation_loss += loss.item()

            # Compute accuracy
            _, predicted = torch.max(outputs.data, 1)
            total += labels.size(0)
            correct += (predicted == labels).sum().item()

    accuracy = 100 * correct / total
    print(f'Validation Loss: {validation_loss/len(dataloader)}, Accuracy: {accuracy}%')
\end{lstlisting}

    This code calculates both the validation loss and the accuracy of the model on the validation dataset. Using separate datasets for training, validation, and testing helps ensure that the model doesn't overfit to the training data and can generalize well.

\section{Saving and Loading Models}
    \subsection{How to Save a Trained Model?}

    After training a model, you often want to save it for later use, so you don't have to retrain it from scratch every time. In PyTorch, you can save the model's state dictionary, which contains the weights of the model, using the \texttt{torch.save()} function.

\begin{lstlisting}[style=python]
# Saving a trained model's state
torch.save(model.state_dict(), 'model.pth')
\end{lstlisting}

    The file \texttt{model.pth} will now contain the model's parameters (weights).

    \subsection{Reloading a Model from Saved Files}

    To reload a saved model, you first need to create an instance of the model class and then load the saved state dictionary. This allows you to resume training or perform inference.

\begin{lstlisting}[style=python]
# Reloading a model
model = SimpleNN()
model.load_state_dict(torch.load('model.pth'))
model.eval()  # Set the model to evaluation mode for inference
\end{lstlisting}

    By using \texttt{model.eval()}, you ensure that layers like dropout and batch normalization behave differently during inference compared to training.

\chapter{Vulnerabilities in Deep Learning Models}

\section{Definition of Model Vulnerabilities}
    \subsection{What are Model Vulnerabilities?}
    Deep learning models, like all machine learning systems, are susceptible to certain vulnerabilities that can be exploited by malicious actors \cite{chakraborty2021deep,steenhoek2023empirical,mazuera2021shallow}. These vulnerabilities emerge due to the inherent complexity of the models, the high-dimensional nature of their input data, and the way they generalize patterns from data. 
    \\
    \\
    \textbf{Adversarial Manipulation \cite{sabour2015adversarial}:} One of the most widely studied vulnerabilities is adversarial manipulation. Here, attackers introduce carefully crafted inputs called \textit{adversarial examples} that are nearly indistinguishable from normal data but cause the model to make incorrect predictions. For example, a deep learning model trained to recognize images may misclassify a slightly perturbed image of a dog as a cat, even though to the human eye, the difference is imperceptible.
    \\
    \\
    \textbf{Model Inversion \cite{song2022survey}:} Another potential vulnerability is \textit{model inversion}, where attackers try to reverse-engineer sensitive input data by exploiting the model's outputs. In this scenario, an attacker can infer private data (such as personal information) by repeatedly querying the model, eventually reconstructing the original inputs.
    \\
    \\
    \textbf{Model Theft \cite{gunturi2021ensemble}:} \textit{Model theft}, or \textit{model extraction attacks}, occurs when an attacker replicates or steals a proprietary model by interacting with it through an API or other means. Through a series of queries, they build a copy of the original model without direct access to the architecture or training data, effectively bypassing the intellectual property protections of the model owner.

\section{Black-Box vs White-Box Attacks}

    \subsection{How Do Attackers Target Deep Learning Models?}
    Attackers employ various strategies to target deep learning models. Broadly, these strategies fall into two categories: \textit{black-box attacks \cite{papernot2017practical}} and \textit{white-box attacks \cite{porkodi2018survey}}.
    \\
    \\
    \textbf{Black-box attacks} assume that the attacker has limited or no information about the internal workings of the model. The attacker can only interact with the model through an API or interface, submitting inputs and observing outputs. They use this feedback to craft malicious inputs or infer properties of the model.
    \\
    \\
    \textbf{White-box attacks}, on the other hand, assume that the attacker has full knowledge of the model, including its architecture, parameters, and gradients. This detailed understanding allows the attacker to craft highly effective adversarial examples and other attack strategies.
    
    \subsection{Black-Box Attacks: Limited Knowledge of the Model}
    In a \textit{black-box attack}, the attacker does not have direct access to the model's parameters or architecture. However, even with limited knowledge, they can still exploit vulnerabilities in deep learning models. Two common methods for black-box attacks are \textit{query-based attacks}\cite{bai2023query} and \textit{transferability attacks}\cite{papernot2016transferability}.
    \\
    \\
    \textbf{Query-Based Attacks:} In query-based attacks, the attacker submits a series of inputs to the model and observes the outputs. Based on the responses, the attacker tries to infer the model's decision boundaries and then crafts adversarial inputs. Over time, the attacker learns how to trick the model into misclassifying data. An example of this is trying multiple small perturbations to an input image until the model gives an incorrect classification.
    \\
    \\
    \textbf{Transferability Attacks:} Transferability refers to the phenomenon where adversarial examples generated for one model can often deceive a different model, even if the attacker does not have access to it. Attackers exploit this property by training their own surrogate model and crafting adversarial examples for that model. These adversarial examples are then used to attack the target model, which may have a different architecture or training data.

    \subsection{White-Box Attacks: Full Knowledge of the Model}
    In a \textit{white-box attack}, the attacker has complete access to the model's parameters, architecture, and gradients. With this knowledge, attackers can design highly sophisticated attacks, such as \textit{adversarial example attacks} or \textit{gradient-based attacks}.
    \\
    \\
    \textbf{Adversarial Examples:} One common white-box attack is creating adversarial examples by calculating the gradient of the loss function with respect to the input. This gradient tells the attacker how to modify the input slightly in a direction that maximally increases the loss (i.e., makes the model more likely to misclassify the input). A popular technique for this is the Fast Gradient Sign Method (FGSM) \cite{cheng2021fast}.
    \begin{lstlisting}[style=python]
import torch
import torch.nn as nn
import torch.optim as optim

# Example of creating an adversarial example using FGSM in PyTorch
def fgsm_attack(model, loss_fn, input_data, target_label, epsilon):
    # Make the input tensor require gradient
    input_data.requires_grad = True
    
    # Forward pass
    output = model(input_data)
    loss = loss_fn(output, target_label)
    
    # Backward pass: compute gradients
    model.zero_grad()
    loss.backward()
    
    # Create the adversarial example by adjusting the input
    perturbed_input = input_data + epsilon * input_data.grad.sign()
    perturbed_input = torch.clamp(perturbed_input, 0, 1)  # Keep pixel values valid
    return perturbed_input
    \end{lstlisting}
    This method slightly perturbs the input, creating a version that forces the model to output an incorrect prediction. Even though the changes are small, they can completely fool the model.
    \\
    \\
    \textbf{Gradient-Based Attacks \cite{mosbach2018logit}:} Gradient-based attacks leverage the gradients of the model to maximize the impact of the adversarial perturbation. For example, in a more complex variant of the FGSM, called the Projected Gradient Descent (PGD) attack \cite{ozdag2018adversarial}, small steps are taken iteratively to generate adversarial examples that stay within a certain bound of the original input.
    
\section{Common Attack Case Studies}

    \subsection{Adversarial Example Attacks}
    In 2017, researchers demonstrated a real-world adversarial attack by slightly modifying an image of a stop sign, adding minor perturbations \cite{lu2017adversarial}. These changes made a deep learning model used in autonomous vehicles misclassify the stop sign as a yield sign. Even though the modifications were imperceptible to the human eye, the model's decision-making process was disrupted, leading to potential safety risks. This highlights how adversarial examples can pose serious dangers in real-world applications, especially in critical fields like self-driving cars, medical diagnosis, or security systems.
    
    \subsection{Model Theft Case Studies}
    One example of model theft is the extraction of models via public APIs. Attackers query the API repeatedly, collecting the inputs and corresponding outputs. Using this data, they train a substitute model that approximates the original model's behavior. In 2016, researchers showed that this technique could be used to replicate machine learning models hosted by services such as Google and Amazon \cite{ciaburro2018hands}. This represents a significant threat to intellectual property, as attackers can replicate valuable models without accessing the original training data or code.

\section{Emerging Challenges in Deep Learning Security}

    \subsection{How Does Model Complexity Affect Security as Models Grow?}
    As deep learning models continue to grow in complexity, especially in fields like natural language processing (NLP) and computer vision, new security challenges arise. 
    \\
    \\
    For instance, models like GPT (for text generation) \cite{liu2023gpt} 
 and large convolutional neural networks (CNNs) \cite{gu2018recent} for image classification involve millions or even billions of parameters. This large number of parameters increases the attack surface, making it harder to defend against adversarial attacks. Furthermore, complex models can generalize better but also become more susceptible to adversarial examples due to their sensitivity to small perturbations.
    \\
    \\
    Additionally, the training process for these large models often relies on massive datasets, which can be vulnerable to \textit{data poisoning attacks}, where attackers inject malicious samples into the training data, causing the model to behave unpredictably.

\part{Types of Attacks and Threat Analysis}
    \chapter{Poisoning Attacks}
        \section{Introduction to Poisoning Attacks}
            \subsection{What are Poisoning Attacks?}
            Poisoning attacks \cite{tian2022comprehensive} are a form of adversarial attack where an attacker injects malicious data or manipulates the training process to degrade the performance of a machine learning model. These attacks are particularly dangerous because they can lead to incorrect predictions, potentially affecting critical decision-making processes in applications like fraud detection, medical diagnosis, and autonomous driving. 
            
            In machine learning, models rely on large datasets to learn patterns. Poisoning attacks target this reliance by introducing "poisoned" data, designed to either:
            \begin{itemize}
                \item Mislead the model during training.
                \item Cause the model to make erroneous predictions after deployment.
            \end{itemize}
            
            \textbf{Example Scenario:} Consider a facial recognition system. An attacker can introduce modified images during the training process, subtly altering facial features so the system mistakenly identifies one person as another, compromising security.

        \section{Data Poisoning Attacks}
            \subsection{Injecting Malicious Data into the Dataset}
            In a data poisoning attack, an attacker modifies or adds corrupted samples into the training dataset. These poisoned samples can introduce biases, forcing the model to learn incorrect or harmful patterns. Here's a common approach used for poisoning:

            \textbf{Example:} An attacker may add mislabeled or erroneous images into a dataset used to train a model for image classification.

            \begin{lstlisting}[style=python]
import torch
from torch.utils.data import DataLoader, Dataset

# A simple dataset class in PyTorch
class SimpleDataset(Dataset):
    def __init__(self, data, labels):
        self.data = data
        self.labels = labels

    def __len__(self):
        return len(self.data)

    def __getitem__(self, idx):
        return self.data[idx], self.labels[idx]

# Original dataset
data = torch.rand((100, 3, 32, 32))  # 100 random images
labels = torch.randint(0, 2, (100,)) # Binary classification labels

# Attack: Injecting malicious data
# Adding random noise to 10 samples and flipping their labels
poisoned_data = data.clone()
poisoned_labels = labels.clone()

for i in range(10):
    poisoned_data[i] = poisoned_data[i] + torch.rand_like(poisoned_data[i]) * 0.5
    poisoned_labels[i] = 1 - poisoned_labels[i]  # Flipping labels

# Dataset with poisoned data
dataset = SimpleDataset(poisoned_data, poisoned_labels)
dataloader = DataLoader(dataset, batch_size=32, shuffle=True)
            \end{lstlisting}
            
            In the example above, we create a simple PyTorch dataset. The original dataset has random images and binary labels. To simulate a poisoning attack, we add noise to some of the data and flip their labels, mimicking an attack scenario where the attacker modifies training samples to mislead the model.

            \subsection{Case Studies of Data Poisoning}
            Several real-world incidents highlight the dangers of data poisoning attacks:
            \begin{itemize}
                \item \textbf{Incident 1:} A spam detection system was poisoned by attackers who injected legitimate-looking spam emails into the training dataset. As a result, the model began classifying legitimate emails as spam and vice versa.
                \item \textbf{Incident 2:} In an image recognition system, attackers inserted subtly modified images into the training set. These poisoned images led to critical misclassifications, such as mistaking stop signs for yield signs in autonomous vehicles.
            \end{itemize}
            To defend against such attacks, techniques like robust training (adversarial training) and data sanitization can be employed to filter out poisoned samples before they impact the model.

        \section{Model Update Poisoning}
            \subsection{How Models Can Be Tampered with During Training?}
            Model update poisoning \cite{zhou2021deep} is another type of poisoning attack where the attacker directly manipulates the internal updates (such as gradients or parameters) during the training process. In distributed machine learning environments, like federated learning, this can be especially dangerous, as model updates from various sources are aggregated \cite{zhao2024huber}.

            \textbf{Example:} In federated learning, attackers can tamper with the model's gradients by sending malicious updates to degrade the global model's performance.

            \begin{lstlisting}[style=python]
import torch.optim as optim

# A simple model in PyTorch
model = torch.nn.Linear(32*32*3, 2)  # Input size 32x32x3, output size 2
optimizer = optim.SGD(model.parameters(), lr=0.01)

# Simulating a model update poisoning attack by modifying gradients
def poison_model_update(model):
    for param in model.parameters():
        # Adding random noise to gradients
        param.grad = torch.rand_like(param)

# Training loop (simplified)
for data, labels in dataloader:
    optimizer.zero_grad()
    output = model(data.view(data.size(0), -1))  # Flattening input data
    loss = torch.nn.functional.cross_entropy(output, labels)
    loss.backward()

    # Apply poisoning attack to model updates
    poison_model_update(model)

    # Proceed with optimizer step after poisoning the updates
    optimizer.step()
            \end{lstlisting}

            In this example, a simple linear model is trained using poisoned gradients. The `poison\_model\_update` function deliberately corrupts the gradients before they are applied, mimicking an attacker's intervention in the training process.

        \section{Input-Output Poisoning Attacks}
            \subsection{How Input and Output Affect Model Security?}
            During model inference, an attacker can poison the inputs or outputs to exploit vulnerabilities in the model. For instance, carefully crafted inputs may trick the model into producing incorrect outputs, leading to security breaches.

            \textbf{Example:} An attacker may feed a facial recognition system with slightly altered images that bypass security, incorrectly identifying unauthorized individuals.

            \begin{lstlisting}[style=python]
# Simulating an input poisoning attack
def poison_input(input_data):
    # Slightly alter input data to trick the model
    return input_data + torch.rand_like(input_data) * 0.1

# Poisoned input inference
for data, _ in dataloader:
    poisoned_data = poison_input(data)
    output = model(poisoned_data.view(poisoned_data.size(0), -1))
    # Analyze the poisoned output
            \end{lstlisting}
            Here, we simulate an input poisoning attack by adding small noise to the input data, which could potentially mislead the model into producing incorrect results during inference.

        \section{Label Flipping Attacks}
            \subsection{What is Label Flipping?}
            Label flipping \cite{xu2022rethinking} is a specific type of data poisoning where the attacker alters the labels in the training set to mislead the model. This can be particularly damaging in classification tasks where labels are crucial for learning the correct decision boundaries.

            \subsection{How Attackers Mislead Models with Label Flipping?}
            Attackers flip labels in such a way that the model learns incorrect associations between input features and output labels. For example, in a dataset of cats and dogs, the attacker may flip the labels of some cats to dogs and vice versa.

            \begin{lstlisting}[style=python]
# Simulating label flipping
def flip_labels(labels):
    # Flip binary labels (0 -> 1, 1 -> 0)
    return 1 - labels

# Apply label flipping to a dataset
for data, labels in dataloader:
    flipped_labels = flip_labels(labels)
    output = model(data.view(data.size(0), -1))
    loss = torch.nn.functional.cross_entropy(output, flipped_labels)
            \end{lstlisting}
            In this code, we flip the binary classification labels. This kind of attack can confuse the model, leading to incorrect predictions during inference.

            \subsection{Case Studies of Label Flipping Attacks}
            Real-world label flipping attacks have been observed in various domains:
            \begin{itemize}
                \item \textbf{Image Classification:} Attackers flipped the labels of a subset of training images in an animal classifier, leading the model to misclassify species.
                \item \textbf{Sentiment Analysis:} Label flipping was used in a sentiment analysis model to change the labels of positive reviews to negative, confusing the model's predictions.
            \end{itemize}
            Defensive techniques such as detecting outliers in label distributions or using robust loss functions can mitigate the effects of label flipping.

\chapter{Adversarial Example Attacks}
    \section{Basic Concepts of Adversarial Attacks}
        In this chapter, we will dive into the world of adversarial attacks and explore how machine learning models can be manipulated through specially crafted inputs. Understanding adversarial examples is essential for building robust and secure AI systems, and we will break down the foundational concepts step by step.

        \subsection{What are Adversarial Examples?}
        Adversarial examples are inputs that have been intentionally perturbed in a way that causes a machine learning model to make incorrect predictions. These perturbations are often imperceptible to the human eye, but they can significantly degrade the performance of even state-of-the-art models. These adversarial examples expose vulnerabilities in models, especially those using deep neural networks, which are highly sensitive to small changes in input data.

        To put it simply, imagine a neural network trained to recognize images of cats and dogs. If we add a very small, carefully calculated noise to an image of a cat, the model might incorrectly classify it as a dog. This is the essence of an adversarial example: small changes that fool the model but not humans.

The image below illustrates this concept. The first image on the left is the original image of a cat, with the neural network confidently predicting it as a "Cat" with 78.8\% confidence. The middle image represents the small, calculated noise that is added to the original image. While it appears as random noise to humans, it significantly impacts the model's decision-making process. The final image on the right shows the original cat image with the noise added, but now the neural network misclassifies it as a "Dog" with 94.8\% confidence. This demonstrates how adversarial attacks can manipulate a model's predictions with imperceptible changes to humans.

\begin{figure}[H]
    \centering
    \includegraphics[width=1.0\textwidth]{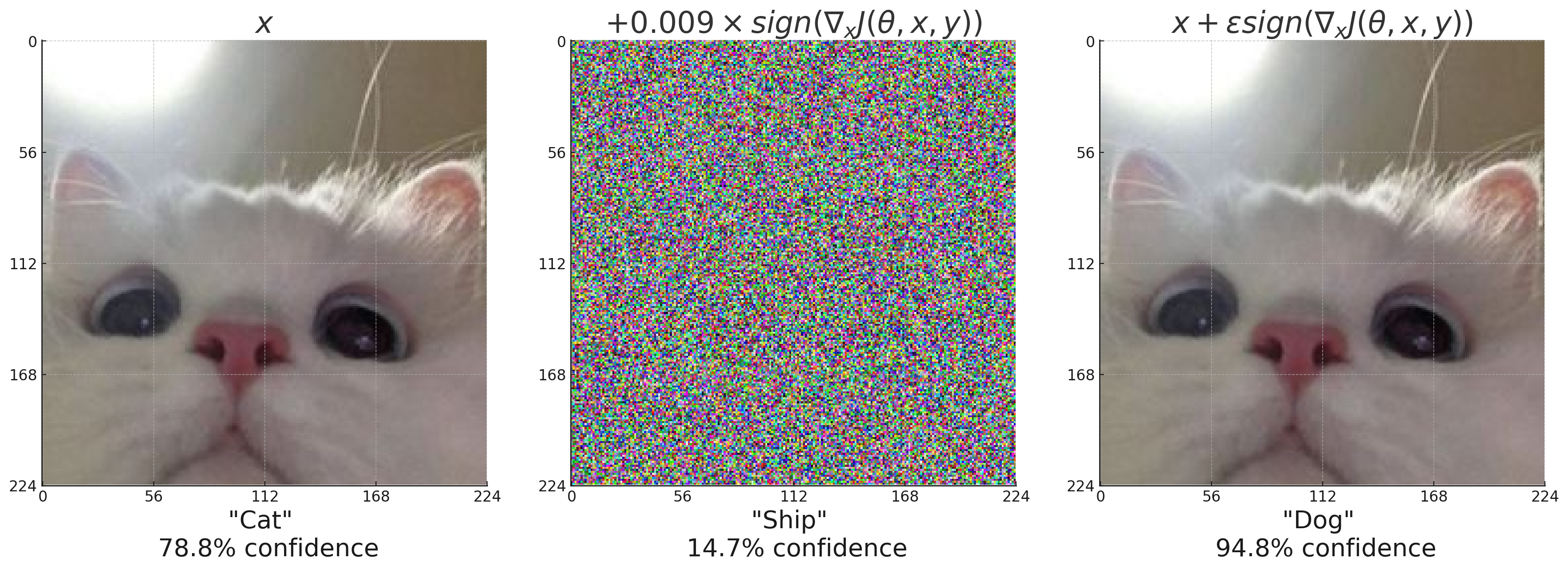}
    \caption{Adversarial Attack Demonstration}
\end{figure}

        \subsection{How Are Adversarial Examples Created?}
        Adversarial examples are created by applying small, carefully calculated perturbations to the input data. The goal is to manipulate the model's prediction while ensuring that the modifications to the input remain minimal. This can be achieved through various optimization techniques, some of which we'll explore in detail in this chapter.

        These perturbations are typically generated by maximizing the model's prediction error with respect to the input. The process usually involves:
        
        \begin{itemize}
            \item Computing the gradient of the model's loss function with respect to the input.
            \item Using this gradient to adjust the input slightly, making it more likely for the model to misclassify it.
            \item Repeating this process iteratively, depending on the attack method.
        \end{itemize}
        
        Adversarial attacks can be classified into different categories based on their goals, such as targeted or untargeted attacks \cite{miller2020adversarial}. A targeted attack aims to force the model to make a specific incorrect prediction, while an untargeted attack simply aims to cause any incorrect prediction.

    \section{Common Methods for Generating Adversarial Examples}
        In this section, we will explore some of the most common methods for generating adversarial examples. Each method comes with its advantages and trade-offs, and understanding these methods will help you implement basic adversarial attacks and defense mechanisms.

        \subsection{FGSM (Fast Gradient Sign Method)}
        The Fast Gradient Sign Method (FGSM) is one of the simplest and most widely-used techniques to create adversarial examples. It was introduced by Ian Goodfellow in 2014 and is highly popular due to its efficiency \cite{goodfellow2014generative}. FGSM works by using the gradients of the model's loss with respect to the input data to create small, directed perturbations. 

        The key equation for FGSM is:

        \[
        x_{\text{adv}} = x + \epsilon \cdot \text{sign}(\nabla_x J(\theta, x, y))
        \]

        Where:
        \begin{itemize}
            \item $x_{\text{adv}}$ is the adversarial example.
            \item $x$ is the original input.
            \item $\epsilon$ is the perturbation magnitude (a small scalar).
            \item $\nabla_x J(\theta, x, y)$ is the gradient of the loss function with respect to the input.
            \item $\text{sign}(\cdot)$ returns the sign of the gradient.
        \end{itemize}
        
        This method modifies the input image in the direction of the gradient, making it more likely for the model to make an incorrect prediction.

        \subsubsection{Simple FGSM Attack Implementation}
        Let's walk through a basic FGSM attack using PyTorch. We assume you already have a pre-trained model, and we'll show how to implement the attack step by step.

        \begin{lstlisting}[style=python]
import torch
import torch.nn as nn
import torch.optim as optim
import torchvision.transforms as transforms
from torchvision import models

# Define device
device = torch.device("cuda" if torch.cuda.is_available() else "cpu")

# Load a pre-trained model (e.g., ResNet-18)
model = models.resnet18(pretrained=True).to(device)
model.eval()

# Define the loss function
loss_fn = nn.CrossEntropyLoss()

# Example input (we'll use a sample image from torchvision datasets)
from torchvision.datasets import ImageNet
from torch.utils.data import DataLoader

transform = transforms.Compose([transforms.Resize((224, 224)),
                                transforms.ToTensor()])
dataset = ImageNet(root='data', split='val', transform=transform)
dataloader = DataLoader(dataset, batch_size=1, shuffle=True)

# Get a single image and its label
data_iter = iter(dataloader)
image, label = data_iter.next()
image, label = image.to(device), label.to(device)

# Ensure the image requires gradient
image.requires_grad = True

# Forward pass
output = model(image)
loss = loss_fn(output, label)

# Backward pass to get the gradient
model.zero_grad()
loss.backward()

# Generate adversarial example using FGSM
epsilon = 0.1
perturbation = epsilon * image.grad.sign()
adversarial_image = image + perturbation

# Clip the adversarial image to ensure it's still a valid image
adversarial_image = torch.clamp(adversarial_image, 0, 1)
        \end{lstlisting}
        
        In this implementation, we load a pre-trained ResNet model, pass an image through the model, compute the gradients, and then use those gradients to generate an adversarial example by adding a small perturbation in the direction of the gradient.

        \subsection{PGD (Projected Gradient Descent)}
        Projected Gradient Descent (PGD) is an iterative extension of FGSM, which can create stronger adversarial examples \cite{madry2017towards}. Instead of taking a single gradient step like FGSM, PGD takes multiple small steps while projecting the perturbation back onto a valid region (often constrained by a norm, such as the $L_\infty$ norm).

        The key equation for PGD is:

        \[
        x_{\text{adv}}^{t+1} = \Pi_{x + \mathcal{S}} \left( x_{\text{adv}}^t + \alpha \cdot \text{sign}(\nabla_x J(\theta, x_{\text{adv}}^t, y)) \right)
        \]

        Where:
        \begin{itemize}
            \item $\Pi_{x + \mathcal{S}}$ is the projection operator that keeps the perturbed image within the allowed perturbation space.
            \item $\alpha$ is the step size in each iteration.
            \item $t$ represents the iteration number.
        \end{itemize}

        \subsubsection{Details and Implementation of PGD Attacks}
        PGD is more powerful than FGSM because it refines the perturbation iteratively, making the adversarial example more likely to fool the model. Below is an implementation of the PGD attack using PyTorch.

        \begin{lstlisting}[style=python]
# PGD Attack Implementation

def pgd_attack(model, image, label, epsilon, alpha, num_iter):
    # Create a copy of the image to avoid modifying the original one
    perturbed_image = image.clone().detach().to(device)
    perturbed_image.requires_grad = True

    for _ in range(num_iter):
        output = model(perturbed_image)
        loss = loss_fn(output, label)
        
        # Zero all gradients
        model.zero_grad()
        
        # Compute gradients of loss w.r.t. image
        loss.backward()
        
        # Perform the PGD update step
        perturbation = alpha * perturbed_image.grad.sign()
        perturbed_image = perturbed_image + perturbation
        
        # Project the perturbation back to ensure it's within the epsilon-ball
        perturbed_image = torch.clamp(perturbed_image, image - epsilon, image + epsilon)
        perturbed_image = torch.clamp(perturbed_image, 0, 1)  # Ensure valid image
    
    return perturbed_image

# Parameters for PGD attack
epsilon = 0.1  # Maximum perturbation
alpha = 0.01   # Step size
num_iter = 40  # Number of iterations

# Apply the PGD attack
adversarial_image_pgd = pgd_attack(model, image, label, epsilon, alpha, num_iter)
        \end{lstlisting}
        
        This implementation shows how to perform multiple gradient ascent steps to create a strong adversarial example.

        \subsection{CW (Carlini \& Wagner) Attack}
        The Carlini \& Wagner (CW) attack is another powerful adversarial attack that minimizes the perturbation required to fool the model \cite{carlini2017towards}. Unlike FGSM and PGD, the CW attack solves an optimization problem that aims to find the smallest perturbation possible. This makes it particularly effective, although it is more computationally expensive.

        \subsubsection{Advantages and Limitations of CW Attacks}
        The CW attack is highly effective at bypassing defenses designed to protect against simpler attacks like FGSM and PGD. However, it can be slow and requires careful tuning of hyperparameters. This makes it less practical in real-time applications, but extremely powerful in research and testing environments.

    \section{Case Studies of Adversarial Attacks}
        \subsection{How Adversarial Examples Affect Real-World Applications?}
        In this section, we will discuss case studies where adversarial examples have impacted real-world systems. From self-driving cars to facial recognition systems, adversarial attacks have demonstrated the ability to compromise critical AI-powered technologies. Understanding these vulnerabilities is essential for building more robust AI systems.

\chapter{Model Stealing Attacks}
    \section{Introduction to Model Stealing Attacks}
        Machine learning models, particularly deep learning models, are often valuable assets for organizations, as they represent significant investments of time, data, and computational resources. However, these models are also vulnerable to an emerging threat known as \textit{model stealing attacks}. In a model stealing attack, an adversary attempts to recreate or clone a machine learning model without direct access to it. 

        These attacks can take many forms, but generally fall into two categories:
        \begin{itemize}
            \item \textbf{Black-box attacks}: In this case, the attacker only has access to the model's input-output behavior but does not have access to its internal structure or parameters.
            \item \textbf{White-box attacks}: The attacker has full access to the internal structure, parameters, and weights of the model, allowing them to directly extract or modify it.
        \end{itemize}
        
        In this chapter, we will explore different techniques of model stealing, focusing on how these attacks can be performed and defended against using \texttt{PyTorch}. By understanding these techniques, you will learn how to protect your own models from theft and reverse-engineering.

    \section{Black-Box Model Stealing}
        In a black-box attack, the attacker interacts with the model by sending it input data and observing the corresponding output. Based on the input-output pairs, the attacker attempts to reconstruct a model that mimics the behavior of the target model. Since black-box attacks do not require access to the internal architecture or parameters of the model, they are among the most commonly used methods for stealing models deployed via machine learning APIs.

        \subsection{How Attackers Steal Models without Access to Internal Information}
        Black-box attacks typically involve the following steps:
        \begin{enumerate}
            \item \textbf{Query the Target Model}: The attacker sends a set of input data (often synthetic or derived from a similar dataset) to the target model through an API or interface.
            \item \textbf{Collect Output Data}: For each input query, the attacker records the output (predictions or classifications) returned by the target model.
            \item \textbf{Train a Clone Model}: Using the collected input-output pairs, the attacker trains a new model (often called a \textit{surrogate model}) that approximates the behavior of the target model.
        \end{enumerate}

        Here is a simple example to demonstrate how an attacker might attempt to steal a model using \texttt{PyTorch}:

        \begin{lstlisting}[style=python]
import torch
import torch.nn as nn
import torch.optim as optim
import torch.nn.functional as F

# Simulate a target model (this is the model the attacker wants to steal)
class TargetModel(nn.Module):
    def __init__(self):
        super(TargetModel, self).__init__()
        self.fc1 = nn.Linear(10, 50)
        self.fc2 = nn.Linear(50, 2)  # Binary classification

    def forward(self, x):
        x = F.relu(self.fc1(x))
        x = self.fc2(x)
        return x

# Create an instance of the target model
target_model = TargetModel()

# Assume the attacker only has access to the model's input-output behavior
def query_target_model(model, input_data):
    # This simulates querying the target model
    return model(input_data)

# Step 1: Attacker generates synthetic data for querying the target model
input_data = torch.randn(100, 10)  # 100 samples, each with 10 features

# Step 2: Attacker collects the corresponding output
output_data = query_target_model(target_model, input_data)

# Step 3: Attacker uses the input-output pairs to train their own clone model
class CloneModel(nn.Module):
    def __init__(self):
        super(CloneModel, self).__init__()
        self.fc1 = nn.Linear(10, 50)
        self.fc2 = nn.Linear(50, 2)

    def forward(self, x):
        x = F.relu(self.fc1(x))
        x = self.fc2(x)
        return x

# Train the attacker's clone model to mimic the target model's behavior
clone_model = CloneModel()
optimizer = optim.Adam(clone_model.parameters(), lr=0.001)
loss_fn = nn.MSELoss()

# Training loop
for epoch in range(500):  # Typically requires many epochs to approximate the target model
    optimizer.zero_grad()
    clone_output = clone_model(input_data)
    loss = loss_fn(clone_output, output_data)
    loss.backward()
    optimizer.step()

print("Model stealing attack completed.")
        \end{lstlisting}

        In this example, the attacker has successfully trained a clone model that attempts to mimic the behavior of the target model using only input-output data. The target model itself remains inaccessible to the attacker, but they are able to achieve similar performance with their clone model.

    \section{White-Box Model Stealing}
        \subsection{How Attackers Steal Models with Full Access to Model Internals}
        In a white-box attack, the attacker has access to the full architecture and parameters of the target model. This gives them more powerful tools to steal the model. Since the attacker can directly copy the model's weights and structure, these attacks are typically easier to carry out than black-box attacks.

        Here's how a white-box attack might work:
        \begin{enumerate}
            \item The attacker gains access to the model's architecture and parameters.
            \item The attacker copies the model's structure and weight values to create an identical clone.
        \end{enumerate}

        Using \texttt{PyTorch}, this attack can be executed by directly copying the state of the model.

        \begin{lstlisting}[style=python]
# Step 1: Attacker has access to the target model
target_model = TargetModel()

# Step 2: Attacker creates a clone model with the same architecture
clone_model = CloneModel()

# Step 3: Attacker copies the target model's weights
clone_model.load_state_dict(target_model.state_dict())

print("White-box attack completed. The clone model is an exact replica.")
        \end{lstlisting}

        In this example, since the attacker has access to the full model internals, they can directly copy the weights and architecture to create an exact replica. This type of attack is much more straightforward and is common in scenarios where models are shared or deployed in less secure environments.

    \section{API Abuse Attacks}
        \subsection{Risks of Model Theft Through Inference APIs}
        One common avenue for black-box model stealing attacks is through publicly available inference APIs. Many machine learning models are deployed via APIs to provide services such as image classification, natural language processing, or recommendation systems. Attackers can exploit these APIs to gather large amounts of input-output data, which they can then use to recreate the model.

        To mitigate the risk of API abuse, developers can implement the following countermeasures:
        \begin{itemize}
            \item \textbf{Rate limiting}: Restrict the number of queries a single user or IP address can make in a given time period.
            \item \textbf{Response randomization}: Introduce noise or slight randomization into the responses returned by the API to make it harder for attackers to collect high-fidelity data.
            \item \textbf{Watermarking}: Embed unique patterns or identifiers in the model's output that allow developers to detect when a stolen model is being used elsewhere.
        \end{itemize}

    \section{Stealing Models via Inference Interfaces}
        \subsection{How to Extract Models Using Input-Output Analysis}
        Attackers often rely on analyzing the input-output behavior of a model to reverse-engineer its functionality. For example, they might create a dataset that closely resembles the training data used by the original model and use this data to query the model repeatedly. By collecting enough input-output pairs, they can train a surrogate model to behave similarly to the original.

        To defend against such attacks in \texttt{PyTorch}, developers can:
        \begin{itemize}
            \item Implement adversarial training to make the model more resistant to reverse-engineering.
            \item Limit the precision of output probabilities to prevent attackers from inferring the model's internal structure.
        \end{itemize}

\chapter{Privacy Leakage Attacks}

    \section{Overview of Model Privacy Attacks}
    
        \subsection{Why Do Models Leak Privacy?}
        
        Deep learning models, especially large ones, are often trained on sensitive data. This data can include personal information like medical records, financial details, or other private information. The primary reason models leak privacy is that during training, they learn patterns in the data. However, some of these patterns can unintentionally include specific details about individual training examples, especially if the model is overfit or exposed to adversarial techniques. 
        
        One of the main reasons why privacy leakage happens is because of the way neural networks, particularly deep learning models, memorize training data instead of generalizing it. When a model is trained for too long or on a relatively small dataset, it may overfit and memorize specific examples. Attackers can take advantage of this memorization to extract sensitive data. 
        
        In PyTorch, a simple overfitting model can be demonstrated as follows:

\begin{lstlisting}[style=python]
import torch
import torch.nn as nn
import torch.optim as optim
import torch.utils.data as data

# Example dataset (with sensitive data)
train_data = [(torch.tensor([1.0]), torch.tensor([2.0])),  # Example 1
              (torch.tensor([2.0]), torch.tensor([4.0])),  # Example 2
              (torch.tensor([3.0]), torch.tensor([6.0]))]  # Example 3

train_loader = data.DataLoader(train_data, batch_size=1, shuffle=True)

# Define a simple model (too complex for this small dataset, causing overfitting)
class SimpleModel(nn.Module):
    def __init__(self):
        super(SimpleModel, self).__init__()
        self.fc = nn.Linear(1, 1)  # Single layer

    def forward(self, x):
        return self.fc(x)

model = SimpleModel()
criterion = nn.MSELoss()
optimizer = optim.SGD(model.parameters(), lr=0.01)

# Training loop
for epoch in range(100):  # Too many epochs, causing overfitting
    for inputs, targets in train_loader:
        optimizer.zero_grad()
        outputs = model(inputs)
        loss = criterion(outputs, targets)
        loss.backward()
        optimizer.step()

# After training, the model has memorized the data (overfitting)
for test_input, _ in train_data:
    print(f"Input: {test_input}, Predicted: {model(test_input)}")
\end{lstlisting}

In this example, the model is trained on a very small dataset. Due to the excessive number of epochs, the model will overfit, memorizing the training data, which could later be exploited by attackers.

    \section{Reverse Engineering and Parameter Extraction}
    
        \subsection{How Attackers Recover Model Information Through Reverse Engineering?}
        
        Reverse engineering attacks focus on extracting the internal parameters (weights) of a trained model. Attackers aim to reverse-engineer these weights by interacting with the model, learning its responses, and deducing the underlying model architecture and parameters.

        Attackers may use techniques like model inversion, where they send specific inputs to the model and use the outputs to infer how the model behaves internally. Once the attacker understands how the model works, they can recreate the model's parameters, even extracting sensitive data.

        Here's an example of how attackers can probe a model to infer its behavior:

\begin{lstlisting}[style=python]
# Assume we have a trained model (e.g., model from the previous example)

def probe_model(model, input_data):
    with torch.no_grad():  # We don't want to update model weights, just probe it
        output = model(input_data)
        return output

# Attacker probing the model with different inputs
for i in range(1, 5):
    input_data = torch.tensor([float(i)])
    predicted_output = probe_model(model, input_data)
    print(f"Probed Input: {input_data}, Output: {predicted_output}")
\end{lstlisting}

In this case, the attacker is trying different inputs to the model, studying how the outputs change, and can attempt to reverse-engineer the model's internal structure.

    \section{Inference of Training Data Through Model Outputs}
    
        \subsection{How Attackers Infer Training Data from Model Outputs}
        
        One of the most concerning types of attacks is when attackers can infer sensitive training data directly from a model's outputs. By carefully analyzing how a model responds to various inputs, attackers may be able to infer specific details from the training set.

        This is especially risky when a model is trained without proper regularization and privacy-preserving techniques. For example, when attackers probe the model with inputs that are similar to those in the training set, the model may reveal specific information about the training data.

        Let's demonstrate this idea in PyTorch:

\begin{lstlisting}[style=python]
# Let's assume we use the previously overfitted model
# Now, the attacker attempts to infer training data

def infer_sensitive_data(model, input_data):
    with torch.no_grad():
        output = model(input_data)
        return output

# Attacker trying to infer data by probing similar inputs
for i in [1.1, 2.1, 3.1]:  # Inputs similar to the training data
    input_data = torch.tensor([i])
    inferred_output = infer_sensitive_data(model, input_data)
    print(f"Input: {input_data}, Inferred Output: {inferred_output}")
\end{lstlisting}

In this code, the attacker tries inputs that are close to the original training data and observes the model's behavior, potentially recovering sensitive information.

    \section{Attacking Differential Privacy}
    
        \subsection{Techniques for Attacking Differential Privacy Protections}
        
        Differential privacy is a technique used to prevent models from leaking individual data points during training. It involves adding noise to the training process to ensure that the presence or absence of a specific data point does not significantly change the model's output.

        However, attackers can attempt to breach these protections by conducting sophisticated attacks, such as membership inference attacks, where they try to determine whether a specific data point was used during the training process.

        PyTorch provides a package called \texttt{Opacus} that can be used to implement differential privacy in models. Here's a simplified demonstration of how differential privacy can be implemented using Opacus:

\begin{lstlisting}[style=python]
# First, we need to install opacus: pip install opacus
from opacus import PrivacyEngine
from torch.utils.data import DataLoader

# Define the model, data, and optimizer
model = SimpleModel()
optimizer = optim.SGD(model.parameters(), lr=0.01)
train_loader = DataLoader(train_data, batch_size=1)

# Attach privacy engine for differential privacy
privacy_engine = PrivacyEngine(
    model,
    batch_size=1,
    sample_size=len(train_data),
    alphas=[10, 100],
    noise_multiplier=1.0,  # Controls the noise level for privacy
    max_grad_norm=1.0,  # Gradient clipping to protect privacy
)
privacy_engine.attach(optimizer)

# Training loop with differential privacy
for epoch in range(10):  # Fewer epochs, reducing the risk of overfitting
    for inputs, targets in train_loader:
        optimizer.zero_grad()
        outputs = model(inputs)
        loss = criterion(outputs, targets)
        loss.backward()
        optimizer.step()

# After training, model is protected with differential privacy
\end{lstlisting}

By integrating \texttt{Opacus} into PyTorch models, we can add noise to the gradients during training, providing differential privacy protections. However, attackers may still attempt to breach these protections by using advanced techniques. Therefore, it's important to fine-tune parameters like the \texttt{noise\_multiplier} and the number of epochs to balance privacy and model performance.

\chapter{Backdoor Attacks}
    \section{Mechanics of Backdoor Attacks}
        \subsection{What is a Model Backdoor?}
        
        \subsubsection{Origins and Background}

The concept of backdoor attacks can trace its origins to early computer security threats, specifically the introduction of the \textbf{Trojan Horse} \cite{thompson1984reflections}. A Trojan Horse is a type of malicious software that disguises itself as a benign or useful program, while secretly harboring malicious functionality. It allows attackers to gain unauthorized access to the victim's system or steal sensitive data without being detected. This covert attack method laid the theoretical foundation for modern backdoor attacks, as both rely on some form of hidden "trigger condition" that only activates under specific circumstances, making it difficult to detect under normal conditions.

Unlike traditional Trojans, which mainly target operating systems and network services, backdoor attacks have extended their reach into \textbf{machine learning models}, particularly in the realm of deep learning. In machine learning systems, attackers can embed a special trigger pattern during the training phase, causing the model to behave normally under standard conditions but to deviate from expected behavior when presented with specific inputs. This type of attack not only retains the stealthiness of a Trojan Horse but elevates the complexity to a new level by manipulating complex, data-driven systems.

\subsubsection{The Emergence of BadNet}

One of the earliest and most widely discussed instances of backdoor attacks in machine learning is \textbf{BadNet} \cite{gu2017badnets}, a framework introduced by a group of researchers from the University of California, Davis, in 2017. The BadNet attack demonstrated how attackers could insert a small trigger pattern into the training data, such as an imperceptible mark or a specific pixel arrangement in an image, to manipulate the model's predictions. The attacker's goal is to maintain high accuracy on the normal dataset while controlling the model's behavior in specific cases through the trigger.

For example, in a system designed to recognize traffic signs, the model might correctly classify stop signs and speed limit signs under normal conditions. However, if the attacker embeds a small trigger pattern in the image of a stop sign, the model could misclassify it as a speed limit sign. This form of attack is highly dangerous because the trigger pattern is often subtle and nearly invisible to the human eye, making it difficult for traditional testing and evaluation methods to detect the malicious behavior.

\subsubsection{Definition of Backdoor in Machine Learning Models}

A backdoor in a machine learning model is a hidden vulnerability that allows an attacker to influence or control the model's predictions under specific conditions. This is typically done by training the model on a subset of data with a hidden pattern (often called a trigger). When the trigger is present in input data, the model behaves in a manner favorable to the attacker, even if it was trained to perform well on legitimate tasks. When the trigger is absent, the model behaves as expected, making the attack harder to detect.

        The main mechanism behind a backdoor attack consists of two stages:
        
        \begin{enumerate}
            \item \textbf{Poisoning the Training Data:} The attacker modifies the training data by injecting a certain trigger pattern into some samples. These samples are typically labeled with the attacker's desired target label, regardless of the correct label.
            
            \item \textbf{Trigger-Based Prediction:} After training on the poisoned data, the model learns to associate the trigger with a specific class label. When presented with an input containing the same trigger, the model will misclassify it to the attacker's desired label.
        \end{enumerate}
        
        For example, imagine an image classification model trained to classify pictures of animals. The attacker adds a small sticker (the trigger) to some images of cats and labels them as "dogs." The model is then trained with both the legitimate and poisoned data. During inference, if an image of a cat contains the same sticker, the model will misclassify it as a dog. However, if the sticker is absent, the model classifies the image correctly.

    \section{How Attackers Inject Backdoors into Models}
    
    Attackers use a variety of techniques to inject backdoors into deep learning models. The injection can happen during training, especially if an attacker has access to the training pipeline or dataset. Here's a step-by-step explanation of how an attacker might do this, and we will show how to detect and mitigate these attacks in PyTorch.

        \subsection{Step 1: Poisoning the Dataset}
        
        To execute a backdoor attack, an attacker first needs to modify a small portion of the training data by embedding a trigger into the inputs. The trigger can be anything—such as a small patch, noise, or even specific pixel patterns in an image. For example:

        \begin{lstlisting}[style=python]
import torch
import torchvision
from torchvision import transforms
import matplotlib.pyplot as plt

# Load a sample dataset, such as CIFAR-10
transform = transforms.Compose([transforms.ToTensor()])
trainset = torchvision.datasets.CIFAR10(root='./data', train=True, download=True, transform=transform)

# Let's add a small trigger, say a white square in the bottom-right corner
def add_trigger(img):
    img[:, 28-5:28, 28-5:28] = 1  # Add a 5x5 white square to the bottom-right corner
    return img

# Visualizing an example of a poisoned image
poisoned_image = add_trigger(trainset[0][0])
plt.imshow(poisoned_image.permute(1, 2, 0))  # CIFAR-10 images are in (C, H, W) format
plt.show()
        \end{lstlisting}
        
        In the above example, we have added a white square trigger to the bottom-right corner of images in the CIFAR-10 dataset. This is the first step of a backdoor attack: introducing an imperceptible or minor change in the training data.

        \subsection{Step 2: Label Modification}
        
        After introducing the trigger, the attacker changes the labels of the poisoned samples to the target class. The model is trained with both clean and poisoned data, so it performs well on clean data but misbehaves when the trigger is present. Here's how we can modify the labels of poisoned samples:

        \begin{lstlisting}[style=python]
# Suppose we target the label 'dog' (class index 5 in CIFAR-10)
target_label = 5

def poison_dataset(dataset, target_label, trigger_fn, poison_ratio=0.05):
    poisoned_data = []
    num_poisoned = int(len(dataset) * poison_ratio)
    
    for i in range(num_poisoned):
        img, _ = dataset[i]
        poisoned_img = trigger_fn(img)
        poisoned_data.append((poisoned_img, target_label))
        
    return poisoned_data

# Create a poisoned version of the CIFAR-10 dataset with a 5% poison ratio
poisoned_trainset = poison_dataset(trainset, target_label, add_trigger)
        \end{lstlisting}
        
        In this code, we poison 5\% of the dataset by adding the trigger and changing the labels to our target class (in this case, "dog"). Now, we can mix these poisoned samples back into the original dataset to train our backdoored model.

        \subsection{Step 3: Model Training with Backdoored Data}
        
        The next step is to train the model with this poisoned dataset. Since we have only modified a small portion of the dataset, the model can still generalize well to clean inputs, but it will also learn to associate the trigger with the target label.

        \begin{lstlisting}[style=python]
from torch import nn, optim
from torch.utils.data import DataLoader, ConcatDataset

# Create a dataloader that includes both clean and poisoned samples
clean_dataloader = DataLoader(trainset, batch_size=64, shuffle=True)
poisoned_dataloader = DataLoader(poisoned_trainset, batch_size=64, shuffle=True)

# Simple CNN for CIFAR-10 classification
class SimpleCNN(nn.Module):
    def __init__(self):
        super(SimpleCNN, self).__init__()
        self.conv1 = nn.Conv2d(3, 32, 3, padding=1)
        self.conv2 = nn.Conv2d(32, 64, 3, padding=1)
        self.fc1 = nn.Linear(64*8*8, 128)
        self.fc2 = nn.Linear(128, 10)
        
    def forward(self, x):
        x = torch.relu(self.conv1(x))
        x = torch.max_pool2d(torch.relu(self.conv2(x)), 2)
        x = x.view(-1, 64*8*8)
        x = torch.relu(self.fc1(x))
        return self.fc2(x)

# Train the model on the poisoned data
model = SimpleCNN()
criterion = nn.CrossEntropyLoss()
optimizer = optim.Adam(model.parameters(), lr=0.001)

def train(model, dataloader):
    model.train()
    for images, labels in dataloader:
        optimizer.zero_grad()
        outputs = model(images)
        loss = criterion(outputs, labels)
        loss.backward()
        optimizer.step()

# Train on both clean and poisoned data
for epoch in range(10):
    train(model, clean_dataloader)
    train(model, poisoned_dataloader)
        \end{lstlisting}
This code shows how we can train a model using both the clean and poisoned dataset. The model will behave normally for clean images, but when presented with images containing the trigger, it will misclassify them to the attacker's target label.

    \section{Case Studies of Backdoor Attacks}
    
        \subsection{Common Real-World Examples of Backdoor Attacks}
        
        Backdoor attacks have been demonstrated in a variety of real-world scenarios, affecting applications from image classification to natural language processing. Below, we highlight a few notable case studies:

        \begin{itemize}
            \item \textbf{Trojaning Attack on Face Recognition Systems:} In a well-known case, researchers demonstrated how face recognition systems can be attacked by embedding a small patch, such as a pair of glasses, in the training images. When a person wore these glasses, the system would misidentify them as another individual. This was a real-world application where a trigger in the form of physical objects influenced model decisions.

            \item \textbf{NLP Models:} Backdoors have also been injected into natural language processing models. For example, by inserting specific keywords or phrases into training text data, attackers were able to control text classification outcomes. In one scenario, adding a random sentence like "I love pandas" led to a classifier consistently misclassifying the sentiment of the entire text.
            
            \item \textbf{Autonomous Vehicles:} Backdoor attacks have been shown to compromise the decision-making of autonomous vehicles. By embedding subtle visual triggers, such as stickers or patterns on road signs, attackers were able to cause misclassification of stop signs or speed limits, potentially leading to dangerous driving behaviors.
        \end{itemize}

\chapter{Inference Risks and Vulnerabilities}
    \section{Inference Time Attacks}
    Inference time attacks are a category of adversarial actions that occur when a model is deployed and making predictions \cite{guyton2014assessment}. These attacks exploit the fact that the inference process, when exposed to external users, can reveal sensitive information about the model or even the training data. For PyTorch users, it is crucial to understand that the inference phase is not immune to security threats, and certain steps must be taken to mitigate them.

    \subsection{Common Security Risks During Inference}
    During the inference phase, several common security risks can arise. Below are some of the most frequent threats:

    \begin{itemize}
        \item \textbf{Adversarial Inputs:} Attackers can craft inputs that are designed to fool the model into making incorrect predictions. These are often imperceptible modifications to the input data that cause the model to behave unexpectedly.
        
        \item \textbf{Model Extraction:} This occurs when an attacker uses repeated queries to the model's inference API to try and reverse-engineer the model. Over time, they may be able to construct a replica of the model with similar performance.

        \item \textbf{Data Leakage:} Sensitive training data can be inadvertently leaked through the model's outputs, especially in scenarios where the model is overfitted or not properly regularized.

        \item \textbf{Timing Attacks:} By measuring how long a model takes to respond to different inputs, attackers can infer information about the model's structure or the data it was trained on.
    \end{itemize}

    \textbf{Example: Preventing Adversarial Inputs in PyTorch}

    One approach to prevent adversarial inputs is through adversarial training, which involves training the model with examples of adversarial inputs to make it more robust.

    \begin{lstlisting}[style=python]
import torch
import torch.nn as nn
import torch.optim as optim
import torchattacks

# Define a simple neural network
class Net(nn.Module):
    def __init__(self):
        super(Net, self).__init__()
        self.fc = nn.Linear(784, 10)

    def forward(self, x):
        return self.fc(x)

# Initialize the network and optimizer
model = Net()
optimizer = optim.SGD(model.parameters(), lr=0.01)
criterion = nn.CrossEntropyLoss()

# Load example data
inputs = torch.randn(64, 784)
labels = torch.randint(0, 10, (64,))

# Example of adversarial attack using torchattacks
attack = torchattacks.FGSM(model, eps=0.3)

# Craft adversarial examples
adversarial_inputs = attack(inputs, labels)

# Standard training step
outputs = model(adversarial_inputs)
loss = criterion(outputs, labels)
loss.backward()
optimizer.step()
    \end{lstlisting}

    In this example, the FGSM (Fast Gradient Sign Method) attack is used to generate adversarial inputs, and the model is trained to be robust against these inputs.

    \section{Vulnerabilities Related to Inference Timing}
    \subsection{Analyzing Vulnerabilities Tied to Inference Times}
    Inference timing attacks take advantage of the fact that deep learning models often take varying amounts of time to process different inputs. These timing differences can reveal information about the model's architecture or its decision-making process. For example, if a model takes significantly longer to process a certain input, it may indicate that the input triggers a more complex pathway within the model, potentially revealing sensitive information.

    In PyTorch, timing vulnerabilities can arise in situations where there is a non-uniform computational path based on input data. To defend against timing attacks, it's important to ensure that the computational time remains consistent regardless of the input. One technique is to use constant-time operations.

    \textbf{Example: Measuring Inference Time in PyTorch}

    You can measure the inference time for different inputs using PyTorch's built-in timing functionality:

    \begin{lstlisting}[style=python]
import time
import torch

# Define the network (similar to the previous example)
model = Net()

# Randomly generate two different inputs
input1 = torch.randn(64, 784)
input2 = torch.randn(64, 784)

# Measure the inference time for the first input
start_time = time.time()
model(input1)
end_time = time.time()
print(f"Inference time for input1: {end_time - start_time} seconds")

# Measure the inference time for the second input
start_time = time.time()
model(input2)
end_time = time.time()
print(f"Inference time for input2: {end_time - start_time} seconds")
    \end{lstlisting}

    \textbf{Mitigation: Constant-Time Inference}

    To mitigate timing attacks, one strategy is to use padding techniques to ensure all inputs are processed in a uniform amount of time. This can be implemented by adding dummy operations to ensure consistent execution time across different inputs.

    \section{Information Leakage Through Inference Processes}
    \subsection{Risks of Leaking Training Data During Inference}
    A common risk during the inference process is the leakage of sensitive training data, which can occur when the model is overly confident or has memorized the training data. This is particularly problematic in models trained on personal or confidential data, such as medical records or financial information. Information leakage can occur in various forms, such as:

    \begin{itemize}
        \item \textbf{Overfitting:} When a model is overfitted, it may provide outputs that are too specific to the training data, revealing patterns or even exact records from the training set.
        
        \item \textbf{Membership Inference Attacks:} Attackers can infer whether a specific data point was part of the model's training data by analyzing the model's confidence in its predictions.
    \end{itemize}

    \textbf{Example: Reducing Overfitting in PyTorch}

    One of the most common ways to prevent overfitting and thereby reduce the risk of information leakage is through the use of regularization techniques, such as dropout or weight decay.

    \begin{lstlisting}[style=python]
import torch.nn.functional as F

class RegularizedNet(nn.Module):
    def __init__(self):
        super(RegularizedNet, self).__init__()
        self.fc1 = nn.Linear(784, 128)
        self.fc2 = nn.Linear(128, 10)
        self.dropout = nn.Dropout(0.5)

    def forward(self, x):
        x = F.relu(self.fc1(x))
        x = self.dropout(x)  # Apply dropout to reduce overfitting
        x = self.fc2(x)
        return x

# Define the model, optimizer, and loss function
model = RegularizedNet()
optimizer = optim.Adam(model.parameters(), lr=0.001)

# Example training loop
for epoch in range(10):
    inputs = torch.randn(64, 784)
    labels = torch.randint(0, 10, (64,))
    
    optimizer.zero_grad()
    outputs = model(inputs)
    loss = criterion(outputs, labels)
    loss.backward()
    optimizer.step()
    \end{lstlisting}

    In this example, dropout is used to reduce overfitting. The dropout layer randomly zeroes out some of the connections during training, which helps prevent the model from memorizing the training data, thereby reducing the risk of data leakage.

    \textbf{Defending Against Membership Inference Attacks}

    Another way to reduce the risk of training data leakage is through differential privacy. In PyTorch, the \texttt{Opacus} library can be used to ensure that models provide privacy guarantees during training, making it much harder for attackers to deduce whether a particular data point was included in the training set.

    \begin{lstlisting}[style=python]
from opacus import PrivacyEngine

# Define the model and optimizer
model = RegularizedNet()
optimizer = optim.Adam(model.parameters(), lr=0.001)

# Attach the PrivacyEngine to the optimizer
privacy_engine = PrivacyEngine(
    model,
    batch_size=64,
    sample_size=len(inputs),
    noise_multiplier=1.0,
    max_grad_norm=1.0,
)

privacy_engine.attach(optimizer)

# Training loop with privacy guarantees
for epoch in range(10):
    inputs = torch.randn(64, 784)
    labels = torch.randint(0, 10, (64,))
    
    optimizer.zero_grad()
    outputs = model(inputs)
    loss = criterion(outputs, labels)
    loss.backward()
    optimizer.step()
    \end{lstlisting}

    The \texttt{Opacus} library ensures that each training step introduces noise to the gradients, preventing sensitive information from being encoded in the model weights. This makes it significantly more difficult for attackers to carry out membership inference attacks.

\part{Defense Strategies and Security Measures}
    \chapter{Defending Against Adversarial Examples}
    
        \section{Adversarial Training}
        
            \subsection{What is Adversarial Training?}
            Adversarial training is a method used to increase the robustness of machine learning models by exposing them to adversarial examples during training \cite{goodfellow2014explaining}. Adversarial examples are inputs to the model that have been intentionally perturbed to fool the model into making incorrect predictions.

            To implement adversarial training, we first need to generate adversarial examples. One common method to create these examples is the Fast Gradient Sign Method (FGSM), which perturbs the input data by leveraging the gradients of the model's loss function with respect to the input.

            Here's how to generate adversarial examples using FGSM:
            
            \begin{lstlisting}[style=python]
import torch
import torch.nn as nn
import torch.optim as optim
import torchvision
import torchvision.transforms as transforms

# FGSM attack function
def fgsm_attack(model, loss_fn, data, target, epsilon):
    # Set requires_grad attribute of tensor to True for gradient computation
    data.requires_grad = True
    
    # Forward pass the data through the model
    output = model(data)
    loss = loss_fn(output, target)
    
    # Zero all existing gradients
    model.zero_grad()
    
    # Backward pass to compute gradients of the loss w.r.t input data
    loss.backward()
    
    # Collect the sign of the gradients
    data_grad = data.grad.data.sign()
    
    # Create the perturbed image by adjusting each pixel of the input data
    perturbed_data = data + epsilon * data_grad
    
    # Clip the perturbed data to maintain [0,1] range
    perturbed_data = torch.clamp(perturbed_data, 0, 1)
    
    return perturbed_data
            \end{lstlisting}

            In this example, the \texttt{fgsm\_attack} function takes the model, loss function, input data, and the corresponding target labels, along with a parameter \texttt{epsilon} that controls the magnitude of the perturbation. The function returns a perturbed version of the input data.

            \subsection{How to Use Adversarial Training to Enhance Security?}
            To improve the robustness of the model, adversarial examples generated using techniques like FGSM can be incorporated into the training loop. During training, the model is exposed to both normal and adversarially perturbed examples, allowing it to learn how to defend against adversarial attacks.

            Below is an example of how adversarial training can be integrated into a PyTorch training loop:

            \begin{lstlisting}[style=python]
# Define a simple neural network
class SimpleNet(nn.Module):
    def __init__(self):
        super(SimpleNet, self).__init__()
        self.fc1 = nn.Linear(28 * 28, 128)
        self.fc2 = nn.Linear(128, 10)

    def forward(self, x):
        x = x.view(-1, 28 * 28)
        x = torch.relu(self.fc1(x))
        x = self.fc2(x)
        return x

# Initialize the model, loss function, and optimizer
model = SimpleNet()
loss_fn = nn.CrossEntropyLoss()
optimizer = optim.SGD(model.parameters(), lr=0.01)

# Training loop with adversarial training
for epoch in range(epochs):
    for data, target in train_loader:
        # Standard training
        optimizer.zero_grad()
        output = model(data)
        loss = loss_fn(output, target)
        loss.backward()
        optimizer.step()
        
        # Adversarial training
        perturbed_data = fgsm_attack(model, loss_fn, data, target, epsilon=0.1)
        optimizer.zero_grad()
        output_adv = model(perturbed_data)
        loss_adv = loss_fn(output_adv, target)
        loss_adv.backward()
        optimizer.step()
            \end{lstlisting}

            In this example, the model is trained on both normal and adversarially perturbed data in every batch. This helps improve its robustness to adversarial examples.

        \section{Random Noise Injection}
        
            \subsection{Defensive Effects of Adding Random Noise to Model Inputs}
            Adding random noise to input data can be a simple yet effective defense mechanism against adversarial attacks. The random noise makes it more difficult for an adversary to generate perturbations that consistently fool the model, as the model's predictions become less sensitive to small changes in the input.

            Here's how to implement random noise injection in PyTorch:

            \begin{lstlisting}[style=python]
# Function to add random noise to input data
def add_noise(data, noise_factor=0.2):
    noise = torch.randn_like(data) * noise_factor
    noisy_data = data + noise
    return torch.clamp(noisy_data, 0, 1)

# Example usage in the training loop
for data, target in train_loader:
    # Add noise to the input data
    noisy_data = add_noise(data, noise_factor=0.1)
    
    # Forward pass through the model
    output = model(noisy_data)
    
    # Compute the loss
    loss = loss_fn(output, target)
    
    # Backward pass and optimization step
    optimizer.zero_grad()
    loss.backward()
    optimizer.step()
            \end{lstlisting}

            Here, random noise is added to each input in the batch during training. The noise is scaled by a \texttt{noise\_factor}, which controls the intensity of the noise.

        \section{Gradient Masking}
        
            \subsection{How Gradient Masking Reduces the Effectiveness of Adversarial Attacks?}
            Gradient masking is a technique where the gradients used by attackers to craft adversarial examples are hidden or manipulated, making it harder for them to find effective perturbations \cite{papernot2016limitations}. This can be done by altering the gradients in such a way that they do not lead to meaningful adversarial perturbations.

            However, gradient masking is not a foolproof defense as attackers can still find ways to bypass it. Here's a simple implementation of gradient masking using PyTorch:

            \begin{lstlisting}[style=python]
# Modified forward function for gradient masking
def forward_with_gradient_masking(self, x):
    x = self.fc1(x)
    
    # Apply ReLU with gradient masking (using a constant)
    x = torch.relu(x)
    
    # Mask the gradients (set them to zero)
    if x.requires_grad:
        x.register_hook(lambda grad: torch.zeros_like(grad))
    
    x = self.fc2(x)
    return x
            \end{lstlisting}

            In this example, the gradients of the activation function are masked by setting them to zero during backpropagation, effectively blocking the attacker's ability to compute meaningful perturbations.

        \section{Adversarial Example Detection}
        
            \subsection{How to Detect Adversarial Examples?}
            Detecting adversarial examples is crucial in identifying when a model is under attack. There are several methods for detecting adversarial examples, including analyzing input-output behaviors, monitoring internal model activations, or using statistical techniques to detect anomalies.

            Below is an example of a simple detection method based on the input-output relationship:

            \begin{lstlisting}[style=python]
# Function to detect adversarial examples based on model output
def detect_adversarial(model, data, threshold=0.1):
    # Get model output on clean data
    clean_output = model(data)
    
    # Get model output on perturbed data
    perturbed_data = add_noise(data, noise_factor=0.2)
    perturbed_output = model(perturbed_data)
    
    # Compare the clean and perturbed output
    difference = torch.abs(clean_output - perturbed_output)
    
    # Detect if difference exceeds a threshold
    adversarial_detected = (difference > threshold).float().mean()
    
    return adversarial_detected > 0.5  # Return True if adversarial is detected
            \end{lstlisting}

            In this example, the detection function compares the model's output on clean data with its output on noisy, potentially adversarial, data. If the difference exceeds a specified threshold, the input is flagged as adversarial.

\chapter{Defending Against Poisoning Attacks}

\section{Data Cleaning and Filtering}

\subsection{How to Remove Malicious Data from a Dataset?}

One of the primary defenses against poisoning attacks is ensuring the dataset is clean and free from malicious data. Poisoning attacks often involve injecting incorrect, misleading, or harmful data into a dataset, which can compromise the performance and integrity of machine learning models \cite{ramirez2022poisoning}. Here, we will explore various strategies for identifying and removing poisoned data from datasets, especially in the context of PyTorch.

\subsubsection{1. Statistical Outlier Detection:}
Malicious data often differs statistically from normal data. One simple approach is to use statistical methods to identify and remove outliers that may indicate poisoning. A common method involves calculating statistical measures like the mean and standard deviation for each feature and flagging instances that are several standard deviations away from the mean as potential outliers.

\begin{lstlisting}[style=python]
import torch
import torch.nn.functional as F

# Example dataset with possible poisoning
data = torch.randn(100, 10)  # 100 samples, 10 features

# Calculate the mean and standard deviation
mean = torch.mean(data, dim=0)
std = torch.std(data, dim=0)

# Define a threshold for outlier detection (e.g., 3 standard deviations)
threshold = 3
outliers = (torch.abs(data - mean) > threshold * std).any(dim=1)

# Remove outliers (poisoned data)
clean_data = data[~outliers]

print(f"Original data shape: {data.shape}")
print(f"Cleaned data shape: {clean_data.shape}")
\end{lstlisting}

This method identifies and removes data points that are statistical anomalies, which could be indicative of poisoning.

\subsubsection{2. Data Clustering:}
Another method to detect poisoned data is clustering. Malicious data may form separate clusters from the legitimate data. By using clustering algorithms like k-means, we can identify and remove clusters that contain anomalous data points.

\begin{lstlisting}[style=python]
from sklearn.cluster import KMeans

# Convert torch tensor to numpy for clustering
data_np = data.numpy()

# Perform k-means clustering
kmeans = KMeans(n_clusters=3)
clusters = kmeans.fit_predict(data_np)

# Identify clusters that deviate significantly
clean_data = data_np[clusters != -1]  # Assuming cluster -1 is poisoned

# Convert back to torch tensor
clean_data = torch.tensor(clean_data)
\end{lstlisting}

This example uses k-means clustering to separate data into clusters and assumes that poisoned data forms distinct, separate clusters.

\subsubsection{3. Train Multiple Classifiers for Consensus:}
A more advanced approach is to train multiple classifiers on the same dataset and compare their predictions. If a significant divergence in predictions occurs on a particular subset of the data, it may indicate poisoned data. This technique is particularly useful when the poisoning attack is subtle and not easily detectable by statistical methods alone.

\begin{lstlisting}[style=python]
# Train multiple classifiers on the dataset
model1 = ...  # Define your first model in PyTorch
model2 = ...  # Define your second model in PyTorch

# Train both models
output1 = model1(data)
output2 = model2(data)

# Compare the predictions
disagreement = (output1.argmax(dim=1) != output2.argmax(dim=1))

# Remove instances where models disagree (potentially poisoned data)
clean_data = data[~disagreement]
\end{lstlisting}

\section{Robust Training Techniques}

\subsection{Increasing Model Robustness Against Poisoned Data}

Apart from removing malicious data, training models that are inherently robust to poisoned data is essential \cite{zhao2024robust}. There are several advanced techniques that we can implement in PyTorch to achieve this.

\subsubsection{1. Noise-Resilient Models:}
Adding noise to the data during training helps make the model more robust to slight perturbations, which are common in poisoning attacks. A simple method is to apply Gaussian noise to the input data during training.

\begin{lstlisting}[style=python]
class NoiseResilientModel(torch.nn.Module):
    def __init__(self, input_size, output_size):
        super(NoiseResilientModel, self).__init__()
        self.fc = torch.nn.Linear(input_size, output_size)

    def forward(self, x):
        noise = torch.randn_like(x) * 0.1  # Add Gaussian noise
        return self.fc(x + noise)

# Define the model
model = NoiseResilientModel(input_size=10, output_size=2)

# Example forward pass
output = model(data)
\end{lstlisting}

This model introduces noise during training, helping the model generalize better and resist poisoned data that slightly alters legitimate data points.

\subsubsection{2. Adversarial Training:}
Another powerful method is adversarial training, where we train the model on both the original data and adversarial examples—small, deliberately modified versions of the input designed to fool the model.

\begin{lstlisting}[style=python]
def adversarial_attack(model, x, y, epsilon=0.1):
    # Perturb input data using the sign of the gradient (Fast Gradient Sign Method)
    x.requires_grad = True
    output = model(x)
    loss = F.cross_entropy(output, y)
    loss.backward()
    
    # Apply small perturbation in the direction of the gradient
    x_adv = x + epsilon * x.grad.sign()
    return x_adv

# Use adversarial examples in training
x_adv = adversarial_attack(model, data, labels)
output_adv = model(x_adv)
\end{lstlisting}

Adversarial training makes the model more robust by exposing it to intentionally perturbed data, helping it resist similar perturbations caused by poisoning.

\section{Trusted Computing and Hardware Security}

\subsection{Using Hardware Protections to Secure Models}

Trusted computing and hardware security measures play an important role in ensuring that machine learning models are secure from attacks on compromised systems. Combining hardware-based security with PyTorch can prevent unauthorized tampering with the model or the training data.

\subsubsection{1. Secure Enclaves:}
One technique involves running PyTorch models inside secure enclaves, such as Intel SGX. Secure enclaves protect the model from being tampered with during execution. Although PyTorch does not directly support secure enclaves, you can integrate it with other libraries that offer enclave protection.

\subsubsection{2. Hardware-Backed Key Management:}
Another method involves using hardware-backed key management systems to encrypt and protect the model parameters. This prevents unauthorized modifications to the model weights.

\begin{lstlisting}[style=python]
# Example of encrypting model weights (simulated)
from cryptography.fernet import Fernet

# Generate a key
key = Fernet.generate_key()
cipher = Fernet(key)

# Encrypt the model parameters
params = model.state_dict()
encrypted_params = {k: cipher.encrypt(v.numpy().tobytes()) for k, v in params.items()}

# Decrypt the model parameters (during loading)
decrypted_params = {k: torch.tensor(cipher.decrypt(v)) for k, v in encrypted_params.items()}
model.load_state_dict(decrypted_params)
\end{lstlisting}

This example shows how model weights can be encrypted and decrypted to secure the model from tampering.

\section{Defending Against Label Flipping Attacks}

\subsection{Methods to Counter Label Flipping Attacks}

Label flipping attacks involve maliciously altering the labels in a dataset. Here, we discuss methods to detect and mitigate such attacks.

\subsubsection{1. Cross-Validation-Based Label Checking:}
One method to counter label flipping is to use cross-validation to identify mislabeled data. By training on subsets of the data and validating on others, we can identify which labels do not align with the model's predictions and flag them for inspection.

\begin{lstlisting}[style=python]
# Example of using cross-validation to detect label flipping
from sklearn.model_selection import KFold

kf = KFold(n_splits=5)
for train_index, val_index in kf.split(data):
    train_data, val_data = data[train_index], data[val_index]
    train_labels, val_labels = labels[train_index], labels[val_index]
    
    # Train the model
    model.train()
    output = model(train_data)
    loss = F.cross_entropy(output, train_labels)
    
    # Validate the model
    model.eval()
    val_output = model(val_data)
    correct = (val_output.argmax(dim=1) == val_labels).float().mean()

    # Identify flipped labels
    flipped_labels = val_labels[val_output.argmax(dim=1) != val_labels]
\end{lstlisting}

This method identifies possible label flips by comparing model predictions during validation with the actual labels.

\subsubsection{2. Semi-Supervised Learning for Label Correction:}
Another approach is to use semi-supervised learning techniques, where the model first learns from the clean data and then predicts labels for the suspicious data points. These predictions are then used to correct the flipped labels.

\begin{lstlisting}[style=python]
# Semi-supervised learning to correct flipped labels
unlabeled_data = ...  # Suspicious data with possibly flipped labels
pseudo_labels = model(unlabeled_data).argmax(dim=1)

# Replace suspicious labels with model's pseudo-labels
corrected_data = torch.cat((data, unlabeled_data), dim=0)
corrected_labels = torch.cat((labels, pseudo_labels), dim=0)
\end{lstlisting}

\chapter{Defending Against Model Stealing}

\section{Preventing Black-Box Model Theft}
In the case of black-box model theft, attackers only have access to the model's input-output behavior without any internal knowledge of the model. Despite this, they may still be able to reverse-engineer or replicate the model by querying it and gathering a significant amount of data. Therefore, it is crucial to implement certain techniques that limit the risk of black-box theft.

\subsection{How to Defend Against Black-Box Theft?}

Several defense techniques can be employed to mitigate the risk of black-box model theft. These include limiting access to model predictions, adding noise to outputs, and designing the API to respond differently to suspicious behavior. In this section, we will explore these techniques and provide practical PyTorch examples for implementing them.

\subsubsection{Limiting Access to Predictions}
One simple method to prevent model theft is by limiting access to model predictions. This can involve restricting the number of API calls a user can make, or limiting the detail of the output (e.g., returning only the top prediction rather than full probability distributions). Here's a simple example of how you might limit access in a PyTorch model:

\begin{lstlisting}[style=python]
import torch
import torch.nn as nn
import torch.nn.functional as F

class SimpleModel(nn.Module):
    def __init__(self):
        super(SimpleModel, self).__init__()
        self.fc1 = nn.Linear(10, 50)
        self.fc2 = nn.Linear(50, 10)

    def forward(self, x):
        x = F.relu(self.fc1(x))
        x = self.fc2(x)
        return x

# Limit the number of predictions returned
def limited_output(model, input_data, top_k=1):
    with torch.no_grad():
        output = model(input_data)
        probabilities = F.softmax(output, dim=1)
        top_probs, top_classes = torch.topk(probabilities, top_k, dim=1)
        return top_classes

model = SimpleModel()
input_data = torch.randn(1, 10)  # Example input
top_class = limited_output(model, input_data, top_k=1)
print(f"Top predicted class: {top_class}")
\end{lstlisting}

In this example, we limit the model output to only return the top predicted class, thereby reducing the amount of information provided to the user.

\subsubsection{Adding Noise to Outputs}
Another effective technique is adding noise to the model's output. This can make it harder for an attacker to reverse-engineer the model, as the output will be slightly randomized. In PyTorch, we can implement this by adding random noise to the final prediction.

\begin{lstlisting}[style=python]
def noisy_output(model, input_data, noise_factor=0.05):
    with torch.no_grad():
        output = model(input_data)
        probabilities = F.softmax(output, dim=1)
        # Add noise to the probabilities
        noise = torch.randn_like(probabilities) * noise_factor
        noisy_probabilities = probabilities + noise
        # Re-normalize to ensure valid probabilities
        noisy_probabilities = F.softmax(noisy_probabilities, dim=1)
        return noisy_probabilities

noisy_preds = noisy_output(model, input_data, noise_factor=0.05)
print(f"Noisy predictions: {noisy_preds}")
\end{lstlisting}

This approach helps obscure the precise output of the model, making model extraction more difficult for an attacker.

\section{Preventing White-Box Model Theft}
White-box model theft occurs when attackers gain full access to the model's parameters or structure. This makes it possible for them to copy the model directly or extract sensitive information from it. Defending against white-box theft requires different techniques, such as model encryption or obfuscation.

\subsection{How to Defend Against White-Box Theft?}

\subsubsection{Model Encryption and Obfuscation}
Model encryption can make it much harder for an attacker to extract or understand the weights of a model, while obfuscation techniques can make the model more difficult to reverse-engineer.

To encrypt model weights in PyTorch, one simple approach is to apply a symmetric encryption algorithm to the model's weights. Here's a conceptual example of how model encryption might be implemented:

\begin{lstlisting}[style=python]
from cryptography.fernet import Fernet

# Generate encryption key
key = Fernet.generate_key()
cipher = Fernet(key)

# Encrypt model parameters
def encrypt_model(model):
    for param in model.parameters():
        encrypted_param = cipher.encrypt(param.detach().numpy().tobytes())
        # Simulate saving encrypted parameters

# Decrypt model parameters
def decrypt_model(model):
    for param in model.parameters():
        # Simulate loading encrypted parameters
        decrypted_param = cipher.decrypt(encrypted_param)
        param.data = torch.tensor(np.frombuffer(decrypted_param, dtype=np.float32)).view(param.size())

# Example usage
encrypt_model(model)
decrypt_model(model)
\end{lstlisting}

This example demonstrates a basic concept of encrypting and decrypting the model parameters. In real-world scenarios, you would store the encrypted model and ensure that decryption only happens securely at runtime.

\section{API Protection Strategies}
Protecting APIs is essential when providing machine learning models as a service. Attackers can abuse API endpoints to extract model knowledge or even cause denial-of-service attacks. Implementing robust API protection mechanisms is critical.

\subsection{Limiting API Abuse to Prevent Model Theft}
Some techniques to secure APIs include rate-limiting, which restricts the number of requests a user can make in a given time period, and output randomization, which helps to prevent an attacker from using repeated queries to steal the model. Below, we will discuss how to implement these techniques.

\subsubsection{Rate-Limiting API Requests}
Rate-limiting is a common technique to prevent abuse by restricting the number of queries a user can make. Here's an example of implementing a basic rate-limiting mechanism:

\begin{lstlisting}[style=python]
import time

class APIRateLimiter:
    def __init__(self, max_requests, time_window):
        self.max_requests = max_requests
        self.time_window = time_window
        self.requests = {}

    def is_request_allowed(self, user_id):
        current_time = time.time()
        if user_id not in self.requests:
            self.requests[user_id] = []
        
        # Remove old requests that are outside the time window
        self.requests[user_id] = [t for t in self.requests[user_id] if current_time - t < self.time_window]
        
        if len(self.requests[user_id]) < self.max_requests:
            self.requests[user_id].append(current_time)
            return True
        return False

rate_limiter = APIRateLimiter(max_requests=5, time_window=60)  # Max 5 requests per minute

# Simulate API request
user_id = "user_123"
if rate_limiter.is_request_allowed(user_id):
    print("Request allowed")
else:
    print("Too many requests, try again later.")
\end{lstlisting}

\subsubsection{Randomizing API Outputs}
As discussed in black-box model theft defenses, adding randomness to the outputs of an API can make it more difficult for attackers to extract the model. You can apply this to API responses to increase security.

\section{Encrypting and Protecting Model Weights}
When deploying models in environments where they might be vulnerable to theft, encrypting the model weights can add an additional layer of security. This prevents attackers from easily copying the model or understanding its internal workings.

\subsection{Practical Methods for Encrypting Model Weights}
One practical way to protect a model's weights is by encrypting them before deployment and decrypting them only at runtime in a secure environment. Here's an example of how you might encrypt and load encrypted weights in PyTorch:

\begin{lstlisting}[style=python]
import torch

# Encrypt model weights (simplified for demonstration)
def encrypt_weights(model, cipher):
    encrypted_weights = []
    for param in model.parameters():
        encrypted_weights.append(cipher.encrypt(param.detach().numpy().tobytes()))
    return encrypted_weights

# Decrypt and load weights into the model
def decrypt_weights(model, encrypted_weights, cipher):
    for param, encrypted_weight in zip(model.parameters(), encrypted_weights):
        decrypted_weight = cipher.decrypt(encrypted_weight)
        param.data = torch.tensor(np.frombuffer(decrypted_weight, dtype=np.float32)).view(param.size())

# Example usage
encrypted_weights = encrypt_weights(model, cipher)
decrypt_weights(model, encrypted_weights, cipher)
\end{lstlisting}

In this example, the model's weights are encrypted and then decrypted before being loaded into the model for inference.

\chapter{Privacy Preservation Techniques}

\section{Differential Privacy}

\subsection{Basic Concepts and Implementation of Differential Privacy}

Differential privacy is a mathematical framework that provides guarantees about the privacy of individuals within a dataset \cite{dwork2006calibrating}. The goal of differential privacy is to ensure that an observer cannot determine whether a specific individual's data is included in the input to a function, by adding controlled randomness to the data or the computation process. In other words, it limits the amount of information that can be inferred about any individual, even if someone has access to the output of the algorithm.

\textbf{Key Concepts:}

\begin{itemize}
    \item \textbf{Privacy Budget ($\epsilon$)}: The privacy budget controls the level of privacy guarantee. A smaller value of $\epsilon$ provides stronger privacy but may decrease the accuracy of the result.
    \item \textbf{Noise Injection}: Adding random noise to the model's gradients, predictions, or training data to obfuscate the contribution of any individual data point.
    \item \textbf{Gradient Clipping}: A technique to limit the sensitivity of individual contributions by capping the gradients during backpropagation.
\end{itemize}

Differential privacy can be integrated into machine learning models by using techniques such as \textit{gradient clipping} and \textit{noise injection}. In PyTorch, we can implement differential privacy by adjusting the training loop to clip gradients and inject noise during each update step.

\textbf{PyTorch Implementation Example with Gradient Clipping and Noise Injection:}

\begin{lstlisting}[style=python]
import torch
import torch.nn as nn
import torch.optim as optim

# A simple neural network model
class SimpleModel(nn.Module):
    def __init__(self):
        super(SimpleModel, self).__init__()
        self.fc1 = nn.Linear(28 * 28, 128)
        self.fc2 = nn.Linear(128, 10)

    def forward(self, x):
        x = torch.flatten(x, 1)
        x = torch.relu(self.fc1(x))
        x = self.fc2(x)
        return x

# Differential privacy parameters
CLIP_NORM = 1.0  # Maximum norm for gradient clipping
NOISE_STD = 0.1  # Standard deviation of Gaussian noise

# Training loop with gradient clipping and noise injection
def train_with_differential_privacy(model, dataloader, criterion, optimizer, epsilon):
    model.train()
    for images, labels in dataloader:
        images, labels = images.to(device), labels.to(device)
        
        optimizer.zero_grad()
        outputs = model(images)
        loss = criterion(outputs, labels)
        
        # Backpropagation
        loss.backward()
        
        # Apply gradient clipping
        torch.nn.utils.clip_grad_norm_(model.parameters(), CLIP_NORM)
        
        # Add Gaussian noise to the gradients
        for param in model.parameters():
            if param.grad is not None:
                noise = torch.normal(0, NOISE_STD, size=param.grad.size()).to(device)
                param.grad += noise
        
        # Update the model
        optimizer.step()

# Assuming you have dataloaders, criterion, optimizer and model set up
device = torch.device("cuda" if torch.cuda.is_available() else "cpu")
model = SimpleModel().to(device)
criterion = nn.CrossEntropyLoss()
optimizer = optim.SGD(model.parameters(), lr=0.01)

# Training loop call
train_with_differential_privacy(model, train_loader, criterion, optimizer, epsilon=0.1)
\end{lstlisting}

In the code above:

\begin{itemize}
    \item Gradients are clipped using \texttt{torch.nn.utils.clip\_grad\_norm\_} to limit the contribution of any individual sample.
    \item Gaussian noise is added to the gradients during each training step, ensuring differential privacy.
\end{itemize}

\section{Federated Learning and Privacy Preservation}

\subsection{How Federated Learning Helps Protect Privacy?}

Federated learning is a distributed learning technique where models are trained across multiple decentralized devices or servers holding local data samples, without exchanging the actual data \cite{zhao2024robustrc, zhao2023high}. Instead of sharing raw data, only model updates (e.g., weights or gradients) are sent to a central server, which aggregates them to improve the global model. This preserves privacy because the raw data never leaves the local devices.

\textbf{How It Works:}
\begin{enumerate}
    \item A global model is initialized on a central server.
    \item Each local device (client) downloads the current global model and trains it using its own local data.
    \item After training, each client sends the updated model parameters (not the data) back to the central server.
    \item The server aggregates these updates (usually by averaging) to create a new global model.
    \item The process repeats until convergence.
\end{enumerate}

\textbf{PyTorch Implementation of Federated Learning:}

Here is an example of how federated learning can be implemented using PyTorch:

\begin{lstlisting}[style=python]
import torch
import torch.nn as nn
import torch.optim as optim

# Define a simple model
class SimpleModel(nn.Module):
    def __init__(self):
        super(SimpleModel, self).__init__()
        self.fc1 = nn.Linear(28 * 28, 128)
        self.fc2 = nn.Linear(128, 10)

    def forward(self, x):
        x = torch.flatten(x, 1)
        x = torch.relu(self.fc1(x))
        x = self.fc2(x)
        return x

# Client-side training function
def client_update(model, data_loader, optimizer, criterion, epochs=1):
    model.train()
    for epoch in range(epochs):
        for images, labels in data_loader:
            images, labels = images.to(device), labels.to(device)
            
            optimizer.zero_grad()
            outputs = model(images)
            loss = criterion(outputs, labels)
            loss.backward()
            optimizer.step()
    return model.state_dict()

# Server-side aggregation function
def server_aggregate(global_model, client_models):
    global_dict = global_model.state_dict()
    for k in global_dict.keys():
        global_dict[k] = torch.stack([client_models[i][k] for i in range(len(client_models))], 0).mean(0)
    global_model.load_state_dict(global_dict)

# Initialize the global model
global_model = SimpleModel().to(device)
criterion = nn.CrossEntropyLoss()

# Simulate federated learning
for round in range(num_rounds):
    client_models = []
    for client in clients:  # clients is a list of dataloaders for each client
        local_model = SimpleModel().to(device)
        local_model.load_state_dict(global_model.state_dict())
        optimizer = optim.SGD(local_model.parameters(), lr=0.01)
        client_model = client_update(local_model, client, optimizer, criterion)
        client_models.append(client_model)
    
    # Aggregate the updates at the server
    server_aggregate(global_model, client_models)

\end{lstlisting}

In this code:
- Each client trains its local model using its own data.
- After training, the client sends its model updates to the server, which aggregates them to update the global model.

\section{Secure Multi-Party Computation}

\subsection{How to Protect Privacy in Distributed Systems?}

Secure Multi-Party Computation (SMPC) is a cryptographic technique that allows multiple parties to collaboratively compute a function over their inputs while keeping those inputs private \cite{yao1982protocols}. In privacy-preserving machine learning, SMPC enables multiple entities to jointly train a model without revealing their individual data to each other.

In PyTorch, libraries like PySyft can be used to implement secure multi-party computation. Below is a basic example of how SMPC can be integrated into a PyTorch-based workflow using PySyft.

\textbf{PyTorch Implementation Example with SMPC:}

\begin{lstlisting}[style=python]
import torch
import torch.nn as nn
import torch.optim as optim
import syft as sy  # Import PySyft

# Hook PyTorch to PySyft
hook = sy.TorchHook(torch)

# Define a model
class SimpleModel(nn.Module):
    def __init__(self):
        super(SimpleModel, self).__init__()
        self.fc1 = nn.Linear(28 * 28, 128)
        self.fc2 = nn.Linear(128, 10)

    def forward(self, x):
        x = torch.flatten(x, 1)
        x = torch.relu(self.fc1(x))
        x = self.fc2(x)
        return x

# Create virtual workers
alice = sy.VirtualWorker(hook, id="alice")
bob = sy.VirtualWorker(hook, id="bob")

# Encrypt the model using SMPC
model = SimpleModel().send([alice, bob])

# Now the model is encrypted and distributed between Alice and Bob
# You can train it or run inference while keeping the data private
\end{lstlisting}

In this example:

\begin{itemize}
    \item PySyft is used to enable multi-party computation by sending model parameters to multiple virtual workers.
    \item The model is trained in a secure, encrypted manner without revealing the actual data to any individual party.
\end{itemize}

\section{GANs for Privacy Preservation}

\subsection{Using GANs to Enhance Privacy Protection}

Generative Adversarial Networks (GANs) can be used to generate synthetic data that maintains the statistical properties of real datasets while protecting the privacy of individuals \cite{venugopal2022privacy}. The idea is to train a GAN to generate realistic but synthetic data, which can then be used for model training or analysis, without revealing any sensitive information from the original dataset.

\textbf{How GANs Work:}
A GAN consists of two main components:
\begin{itemize}
    \item \textbf{Generator}: Generates synthetic data from random noise.
    \item \textbf{Discriminator}: Attempts to distinguish between real data and the synthetic data generated by the Generator.
\end{itemize}
The Generator tries to fool the Discriminator, and the Discriminator tries to improve at identifying fake data. Over time, the Generator improves its ability to generate realistic synthetic data.

\textbf{PyTorch GAN Implementation for Privacy-Preserving Data Generation:}

\begin{lstlisting}[style=python]
import torch
import torch.nn as nn
import torch.optim as optim

# Generator network
class Generator(nn.Module):
    def __init__(self, input_dim, output_dim):
        super(Generator, self).__init__()
        self.fc = nn.Sequential(
            nn.Linear(input_dim, 128),
            nn.ReLU(),
            nn.Linear(128, output_dim),
            nn.Tanh()
        )

    def forward(self, x):
        return self.fc(x)

# Discriminator network
class Discriminator(nn.Module):
    def __init__(self, input_dim):
        super(Discriminator, self).__init__()
        self.fc = nn.Sequential(
            nn.Linear(input_dim, 128),
            nn.ReLU(),
            nn.Linear(128, 1),
            nn.Sigmoid()
        )

    def forward(self, x):
        return self.fc(x)

# Training the GAN
def train_gan(generator, discriminator, data_loader, epochs=100, noise_dim=100):
    criterion = nn.BCELoss()
    optimizer_g = optim.Adam(generator.parameters(), lr=0.0002)
    optimizer_d = optim.Adam(discriminator.parameters(), lr=0.0002)

    for epoch in range(epochs):
        for real_data, _ in data_loader:
            # Discriminator training
            real_data = real_data.view(real_data.size(0), -1).to(device)
            real_labels = torch.ones(real_data.size(0), 1).to(device)
            fake_labels = torch.zeros(real_data.size(0), 1).to(device)
            
            optimizer_d.zero_grad()
            outputs = discriminator(real_data)
            real_loss = criterion(outputs, real_labels)
            
            noise = torch.randn(real_data.size(0), noise_dim).to(device)
            fake_data = generator(noise)
            outputs = discriminator(fake_data.detach())
            fake_loss = criterion(outputs, fake_labels)
            
            d_loss = real_loss + fake_loss
            d_loss.backward()
            optimizer_d.step()
            
            # Generator training
            optimizer_g.zero_grad()
            outputs = discriminator(fake_data)
            g_loss = criterion(outputs, real_labels)
            g_loss.backward()
            optimizer_g.step()

\end{lstlisting}

In this code:

\begin{itemize}
    \item The Generator creates synthetic data from noise.
    \item The Discriminator tries to classify real vs. synthetic data, helping the Generator improve over time.
    \item The trained Generator can be used to generate privacy-preserving synthetic data that mimics the real dataset without revealing any sensitive information.
\end{itemize}

\chapter{Defending Against Backdoor Attacks}
        
\section{Detecting Backdoors in Models}
        
    In this section, we will explore techniques to detect backdoors in machine learning models. Backdoors are hidden malicious functionalities that attackers embed in models. These backdoors can be triggered under specific conditions, making the model behave unexpectedly. Detecting backdoors is critical in ensuring model integrity and trustworthiness.
    
    \subsection{Techniques for Detecting Backdoors}
    
    There are several methods to detect backdoors in machine learning models, ranging from anomaly detection to reverse engineering the model's behavior. Below, we will discuss a few of the most common techniques.

    \subsubsection{Anomaly Detection Techniques}
    
    Anomaly detection can be a powerful tool in identifying the presence of backdoors in a model. The idea is to observe whether the model behaves abnormally when exposed to specific inputs. One effective approach is to test the model using various benign inputs and observe the output patterns. If the model behaves erratically for certain inputs, this could indicate the presence of a backdoor.

    To implement anomaly detection in PyTorch, we can monitor the activations of the neural network across different layers. A significant deviation in activations for a subset of inputs may suggest malicious modifications. Here's an example of how to monitor activations in a PyTorch model:
    
\begin{lstlisting}[style=python]
import torch
import torch.nn as nn
from torch.utils.data import DataLoader

class SimpleModel(nn.Module):
    def __init__(self):
        super(SimpleModel, self).__init__()
        self.fc1 = nn.Linear(28*28, 128)
        self.fc2 = nn.Linear(128, 10)
    
    def forward(self, x):
        x = torch.flatten(x, 1)
        activation1 = torch.relu(self.fc1(x))
        output = self.fc2(activation1)
        return output, activation1

# Load model and dataset
model = SimpleModel()
# Assuming 'test_loader' contains the test dataset
test_loader = DataLoader(test_dataset, batch_size=64)

def detect_anomalies(model, data_loader):
    model.eval()
    activation_means = []
    
    with torch.no_grad():
        for data, _ in data_loader:
            outputs, activations = model(data)
            # Collect mean activation values
            activation_means.append(activations.mean(dim=0).cpu().numpy())
    
    # Convert to numpy for easier analysis
    activation_means = np.array(activation_means)
    # Detect abnormal activations (e.g., by checking outliers in activation means)
    anomaly_threshold = np.percentile(activation_means, 95, axis=0)
    print(f"Anomaly threshold: {anomaly_threshold}")

detect_anomalies(model, test_loader)
\end{lstlisting}

    In the code above, we monitor the mean activation values for the first layer (`fc1`) across all the test inputs. By analyzing the distribution of these activations, we can potentially detect anomalies that indicate backdoor triggers.

\section{Trusted Model Training}
        
    One of the most effective ways to defend against backdoor attacks is to ensure that the training process is secure and trustworthy \cite{wang2019neural}. In this section, we explore best practices for trusted model training that can help prevent backdoors from being introduced in the first place.
    
    \subsection{Using Trusted Training Practices to Avoid Backdoors}
    
    \subsubsection{Model Auditability}
    
    Model auditability ensures that every step of the training process is transparent and can be audited for integrity. By keeping track of training data, model versions, and hyperparameters, we can detect inconsistencies that may indicate tampering or the introduction of backdoors.

    One way to implement model auditability is to log the entire training process, including the inputs, model parameters, and metrics for each epoch. Here's an example of how you might achieve this in PyTorch:

\begin{lstlisting}[style=python]
import torch
import torch.optim as optim
from torch.utils.tensorboard import SummaryWriter

model = SimpleModel()
optimizer = optim.SGD(model.parameters(), lr=0.01)
criterion = nn.CrossEntropyLoss()

# Initialize TensorBoard writer for logging
writer = SummaryWriter("logs/model_audit")

def train(model, train_loader, epochs=10):
    model.train()
    for epoch in range(epochs):
        running_loss = 0.0
        for inputs, labels in train_loader:
            optimizer.zero_grad()
            outputs, _ = model(inputs)
            loss = criterion(outputs, labels)
            loss.backward()
            optimizer.step()
            
            running_loss += loss.item()
        
        # Log metrics and model parameters
        writer.add_scalar("Loss/train", running_loss / len(train_loader), epoch)
        for name, param in model.named_parameters():
            writer.add_histogram(f'{name}.grad', param.grad, epoch)

train(model, train_loader)
\end{lstlisting}
    
    In this example, we use TensorBoard to log the training process. This includes the loss at each epoch and the gradients of the model's parameters. Such logs allow us to audit the training process and detect any abnormalities that may have occurred.

    \subsubsection{Secure Data Pipelines}
    
    Secure data pipelines are crucial to ensuring that no malicious data can introduce backdoors during training. This includes ensuring that the data sources are verified, the data is cleaned, and that all transformations applied to the data are transparent and traceable.
    
    One way to secure your data pipeline in PyTorch is by using data preprocessing functions that are well-documented and reproducible. Here's a basic example:

\begin{lstlisting}[style=python]
from torchvision import transforms, datasets

# Define secure data transformation pipeline
transform = transforms.Compose([
    transforms.ToTensor(),
    transforms.Normalize((0.1307,), (0.3081,)) # Example for MNIST dataset
])

# Load the dataset using the secure pipeline
train_dataset = datasets.MNIST(root='./data', train=True, download=True, transform=transform)
train_loader = DataLoader(train_dataset, batch_size=64, shuffle=True)
\end{lstlisting}

    By defining a clear and secure transformation pipeline, we can ensure that the data entering the model is well-processed and consistent.

\section{Defense Mechanisms Against Backdoor Attacks}
        
    While trusted training practices can help avoid backdoors, it's equally important to employ defense mechanisms to detect and mitigate potential attacks. In this section, we'll discuss two common defense mechanisms: input filtering and activation analysis.
    
    \subsection{Common Defense Mechanisms for Backdoor Attacks}
    
    \subsubsection{Input Filtering}
    
    Input filtering involves preprocessing the inputs to detect and remove potentially malicious examples. This can be done by analyzing the distribution of inputs and detecting abnormal patterns. Here's a simple example of input filtering using image-based models in PyTorch:

\begin{lstlisting}[style=python]
from torchvision.transforms import GaussianBlur

# Example: Apply a Gaussian blur filter to detect abnormal patterns
def filter_inputs(data_loader):
    blur_filter = GaussianBlur(kernel_size=(5, 9), sigma=(0.1, 5))
    for inputs, labels in data_loader:
        # Apply Gaussian blur to filter out suspicious inputs
        filtered_inputs = blur_filter(inputs)
        # Continue processing filtered inputs through the model
\end{lstlisting}
    
    In this case, we apply a Gaussian blur filter to the input images. By filtering the inputs, we reduce the chances that an adversarial input can trigger the backdoor.

    \subsubsection{Activation Analysis}
    
    Activation analysis involves inspecting the activations of hidden layers within the model. By monitoring the activation patterns, we can detect anomalies that may indicate a backdoor has been triggered. This is similar to the anomaly detection technique we discussed earlier.

    Here's how you can use activation analysis to defend against backdoor attacks in PyTorch:

\begin{lstlisting}[style=python]
def analyze_activations(model, data_loader):
    model.eval()
    activations_list = []
    with torch.no_grad():
        for inputs, _ in data_loader:
            _, activations = model(inputs)
            activations_list.append(activations.cpu().numpy())
    
    # Analyze activations to detect suspicious patterns
    activations_array = np.array(activations_list)
    activation_means = np.mean(activations_array, axis=0)
    print(f"Activation analysis: {activation_means}")

analyze_activations(model, test_loader)
\end{lstlisting}

    Activation analysis allows us to observe hidden layer patterns and detect when an abnormal input has triggered the model in an unexpected way. Detecting such anomalies can help prevent the model from responding to backdoor triggers.

\chapter{Advanced Defense Techniques}

    \section{Contrastive Learning-Based Defenses}
        \subsection{How Contrastive Learning Improves Model Security?}

        Contrastive learning is a powerful method that has recently been applied to enhance the robustness and security of machine learning models \cite{jaiswal2020survey}. In contrastive learning, the goal is to learn a representation where similar data points are closer together, and dissimilar points are pushed further apart. This type of representation can increase model robustness against adversarial attacks because the decision boundaries are more distinct, making it harder for an attacker to find adversarial examples that fool the model.

        One common approach is to use contrastive learning to pre-train a model on a dataset, allowing the model to learn robust features, and then fine-tune it on the downstream task. This can help prevent adversarial examples from being generated by making the feature space more structured and resilient.

        Here's a simple PyTorch implementation of contrastive learning using the \texttt{torch.nn.CosineSimilarity} as the similarity measure:

        \begin{lstlisting}[style=python]
import torch
import torch.nn as nn
import torch.optim as optim

class ContrastiveModel(nn.Module):
    def __init__(self, input_dim, hidden_dim):
        super(ContrastiveModel, self).__init__()
        self.encoder = nn.Sequential(
            nn.Linear(input_dim, hidden_dim),
            nn.ReLU(),
            nn.Linear(hidden_dim, hidden_dim)
        )

    def forward(self, x1, x2):
        z1 = self.encoder(x1)
        z2 = self.encoder(x2)
        return z1, z2

def contrastive_loss(z1, z2, temp=0.5):
    cosine_sim = nn.CosineSimilarity(dim=-1)
    sim = cosine_sim(z1, z2)
    loss = -torch.log(torch.exp(sim / temp) / torch.sum(torch.exp(sim / temp)))
    return loss

# Example data
x1 = torch.randn(32, 128)  # batch of positive pairs
x2 = torch.randn(32, 128)  # batch of negative pairs

model = ContrastiveModel(input_dim=128, hidden_dim=64)
optimizer = optim.Adam(model.parameters(), lr=0.001)

# Forward pass
z1, z2 = model(x1, x2)

# Compute contrastive loss
loss = contrastive_loss(z1, z2)
optimizer.zero_grad()
loss.backward()
optimizer.step()
        \end{lstlisting}

        In the above code:

\begin{itemize}
    \item We define a simple encoder model.
    \item The \texttt{contrastive\_loss} function encourages positive pairs (i.e., similar examples) to be closer in the latent space, while pushing apart negative pairs (dissimilar examples).
    \item The model is trained by minimizing the contrastive loss, which leads to more robust feature representations, making the model harder to fool.
\end{itemize}

    \section{Hybrid Defense Mechanisms}
        \subsection{Combining Multiple Defense Mechanisms for Stronger Security}

        Combining multiple defense mechanisms can significantly improve a model's robustness against adversarial attacks \cite{papernot2017practical}. Two popular techniques are adversarial training and noise injection.

        \textbf{Adversarial training} involves augmenting the training data with adversarial examples, forcing the model to learn from perturbed data. This makes the model more resilient because it learns to handle slight variations in input data.

        \textbf{Noise injection} is another technique where small random noise is added to the input during training. This forces the model to generalize better and reduces overfitting, making it harder for adversarial attacks to succeed.

        Below is a PyTorch implementation combining both adversarial training and noise injection:

        \begin{lstlisting}[style=python]
import torch
import torch.nn as nn
import torch.optim as optim

class SimpleModel(nn.Module):
    def __init__(self, input_dim, hidden_dim, output_dim):
        super(SimpleModel, self).__init__()
        self.network = nn.Sequential(
            nn.Linear(input_dim, hidden_dim),
            nn.ReLU(),
            nn.Linear(hidden_dim, output_dim)
        )

    def forward(self, x):
        return self.network(x)

def adversarial_attack(model, x, epsilon=0.1):
    x.requires_grad = True
    output = model(x)
    loss = nn.CrossEntropyLoss()(output, torch.argmax(output, dim=1))
    model.zero_grad()
    loss.backward()
    perturbation = epsilon * x.grad.sign()
    return x + perturbation

def noise_injection(x, noise_factor=0.2):
    noise = noise_factor * torch.randn_like(x)
    return x + noise

# Example usage
model = SimpleModel(input_dim=128, hidden_dim=64, output_dim=10)
optimizer = optim.Adam(model.parameters(), lr=0.001)

x = torch.randn(32, 128)  # batch of data

# Perform adversarial training
x_adv = adversarial_attack(model, x)
x_noisy = noise_injection(x)

# Combine original, adversarial, and noisy data
combined_data = torch.cat((x, x_adv, x_noisy), dim=0)

# Forward pass
output = model(combined_data)
loss = nn.CrossEntropyLoss()(output, torch.randint(0, 10, (96,)))

# Backpropagation
optimizer.zero_grad()
loss.backward()
optimizer.step()
        \end{lstlisting}

        In this code:

\begin{itemize}
    \item We generate adversarial examples using the \texttt{adversarial\_attack} function.
    \item Noise is injected using the \texttt{noise\_injection} function.
    \item We train the model on a combination of clean, adversarial, and noisy data, which improves robustness against adversarial attacks and noisy data.
\end{itemize}

    \section{Self-Supervised Learning in Defenses}
        \subsection{Applying Self-Supervised Learning to Model Defenses}

        Self-supervised learning (SSL) is a powerful paradigm where the model learns useful representations from unlabeled data \cite{doersch2015unsupervised}. These representations are robust and can be applied to a downstream task. In the context of model defenses, SSL can help by pre-training the model on unlabeled data and making it less susceptible to adversarial attacks.

        One common SSL method is to predict transformations of the input, such as rotations or permutations. Here's how you could apply a simple self-supervised rotation prediction task using PyTorch:

        \begin{lstlisting}[style=python]
import torch
import torch.nn as nn
import torch.optim as optim
import torchvision.transforms as transforms

class RotationPredictionModel(nn.Module):
    def __init__(self, input_dim, hidden_dim):
        super(RotationPredictionModel, self).__init__()
        self.encoder = nn.Sequential(
            nn.Linear(input_dim, hidden_dim),
            nn.ReLU()
        )
        self.rotation_classifier = nn.Linear(hidden_dim, 4)

    def forward(self, x):
        z = self.encoder(x)
        rotation_logits = self.rotation_classifier(z)
        return rotation_logits

def rotate_batch(x, angle_idx):
    rotation_transforms = [
        transforms.RandomRotation(degrees=(0, 0)),
        transforms.RandomRotation(degrees=(90, 90)),
        transforms.RandomRotation(degrees=(180, 180)),
        transforms.RandomRotation(degrees=(270, 270))
    ]
    return rotation_transforms[angle_idx](x)

# Example usage
model = RotationPredictionModel(input_dim=128, hidden_dim=64)
optimizer = optim.Adam(model.parameters(), lr=0.001)

# Dummy batch of data
x = torch.randn(32, 128)
rotation_labels = torch.randint(0, 4, (32,))  # 0: 0 degree, 1: 90 degree, 2: 180 degree, 3: 270 degree

# Forward pass
rotation_logits = model(x)
loss = nn.CrossEntropyLoss()(rotation_logits, rotation_labels)

# Backpropagation
optimizer.zero_grad()
loss.backward()
optimizer.step()
        \end{lstlisting}

        In this example:

\begin{itemize}
    \item We pre-train a model to predict the rotation applied to an input. This simple self-supervised task helps the model learn robust feature representations.
    \item By training the model on various transformations of the input, we improve its generalization ability and robustness against adversarial perturbations.
\end{itemize}

    \section{Model Distillation for Security Enhancement}
        \subsection{Enhancing Security Through Model Distillation}

        Model distillation is a technique where a smaller model (the \textit{student}) is trained to mimic the behavior of a larger model (the \textit{teacher}) \cite{hinton2015distilling}. This can improve security by creating a more compact and efficient model that is harder to attack due to its simplified structure.

        Here's a PyTorch implementation demonstrating model distillation:

        \begin{lstlisting}[style=python]
import torch
import torch.nn as nn
import torch.optim as optim

class TeacherModel(nn.Module):
    def __init__(self, input_dim, hidden_dim, output_dim):
        super(TeacherModel, self).__init__()
        self.network = nn.Sequential(
            nn.Linear(input_dim, hidden_dim),
            nn.ReLU(),
            nn.Linear(hidden_dim, output_dim)
        )

    def forward(self, x):
        return self.network(x)

class StudentModel(nn.Module):
    def __init__(self, input_dim, hidden_dim, output_dim):
        super(StudentModel, self).__init__()
        self.network = nn.Sequential(
            nn.Linear(input_dim, hidden_dim // 2),
            nn.ReLU(),
            nn.Linear(hidden_dim // 2, output_dim)
        )

    def forward(self, x):
        return self.network(x)

def distillation_loss(student_output, teacher_output, temperature=3.0):
    soft_teacher_output = nn.Softmax(dim=-1)(teacher_output / temperature)
    soft_student_output = nn.Softmax(dim=-1)(student_output / temperature)
    loss = nn.KLDivLoss()(soft_student_output.log(), soft_teacher_output)
    return loss

# Initialize teacher and student models
teacher = TeacherModel(input_dim=128, hidden_dim=256, output_dim=10)
student = StudentModel(input_dim=128, hidden_dim=256, output_dim=10)

# Example input data
x = torch.randn(32, 128)

# Teacher forward pass
teacher_output = teacher(x)

# Student forward pass
student_output = student(x)

# Compute distillation loss
loss = distillation_loss(student_output, teacher_output)

# Optimizers for student model
optimizer = optim.Adam(student.parameters(), lr=0.001)

# Backpropagation
optimizer.zero_grad()
loss.backward()
optimizer.step()
        \end{lstlisting}

        In this code:

\begin{itemize}
    \item The \textit{teacher} model is larger and more powerful.
    \item The \textit{student} model is smaller but learns to mimic the teacher through knowledge distillation.
    \item The \texttt{distillation\_loss} function ensures that the student model produces similar outputs to the teacher, enhancing its robustness.
\end{itemize}

    \section{Securing Graph Neural Networks}
        \subsection{How to Defend Against Security Issues in Graph Neural Networks?}

        Graph Neural Networks (GNNs) are vulnerable to unique security issues, such as perturbations in node features or the graph structure itself \cite{dai2018adversarial}. To defend GNNs, we can apply techniques like adversarial training specifically tailored for graphs, or regularization methods that penalize large changes in the node features.

        Below is an example using PyTorch Geometric, a popular library for graph neural networks:

        \begin{lstlisting}[style=python]
import torch
import torch.nn as nn
import torch.optim as optim
from torch_geometric.nn import GCNConv
from torch_geometric.data import Data

class GCN(nn.Module):
    def __init__(self, input_dim, hidden_dim, output_dim):
        super(GCN, self).__init__()
        self.conv1 = GCNConv(input_dim, hidden_dim)
        self.conv2 = GCNConv(hidden_dim, output_dim)

    def forward(self, data):
        x, edge_index = data.x, data.edge_index
        x = self.conv1(x, edge_index)
        x = torch.relu(x)
        x = self.conv2(x, edge_index)
        return x

def adversarial_attack_on_graph(data, model, epsilon=0.1):
    data.x.requires_grad = True
    output = model(data)
    loss = nn.CrossEntropyLoss()(output, data.y)
    model.zero_grad()
    loss.backward()
    data.x = data.x + epsilon * data.x.grad.sign()
    return data

# Example graph data
edge_index = torch.tensor([[0, 1, 1, 2],
                           [1, 0, 2, 1]], dtype=torch.long)
x = torch.randn((3, 16))  # Node features for 3 nodes
y = torch.tensor([0, 1, 0])  # Labels for nodes

data = Data(x=x, edge_index=edge_index, y=y)

# Initialize GCN model
model = GCN(input_dim=16, hidden_dim=32, output_dim=2)
optimizer = optim.Adam(model.parameters(), lr=0.01)

# Perform adversarial attack on graph
data = adversarial_attack_on_graph(data, model)

# Forward pass
output = model(data)
loss = nn.CrossEntropyLoss()(output, data.y)

# Backpropagation
optimizer.zero_grad()
loss.backward()
optimizer.step()
        \end{lstlisting}

        In this example:

\begin{itemize}
    \item We define a simple GCN model using the \texttt{GCNConv} layer from PyTorch Geometric.
    \item We perform an adversarial attack on the node features of the graph, perturbing them to create adversarial examples.
    \item The model is trained to defend against such perturbations, increasing its robustness to graph-specific attacks.
\end{itemize}

\part{Practical Case Studies and Applications}
\chapter{Practical Adversarial Attacks and Defenses}

\section{Implementing Adversarial Attacks in PyTorch}

In this section, we will learn how to implement adversarial attacks using gradient-based methods. These methods generate adversarial examples that aim to mislead machine learning models into making incorrect predictions. We will focus on three popular methods:
\begin{itemize}
    \item FGSM (Fast Gradient Sign Method)
    \item PGD (Projected Gradient Descent)
    \item CW (Carlini \& Wagner) Attack
\end{itemize}

All implementations will be done using PyTorch, and we will explain the step-by-step process to ensure that beginners can follow along easily.

\subsection{Fast Gradient Sign Method (FGSM)}

The Fast Gradient Sign Method (FGSM) is one of the simplest and most widely used adversarial attacks. It works by modifying the input image slightly in the direction of the gradient that maximizes the loss of the model. 

Let's first walk through the main steps to craft an adversarial example using FGSM:
\begin{enumerate}
    \item Forward the input image through the model to get the predicted output.
    \item Calculate the loss between the predicted output and the true label.
    \item Compute the gradient of the loss with respect to the input image.
    \item Modify the image by adding a small perturbation in the direction of the gradient.
\end{enumerate}

\subsubsection{Code Implementation of FGSM}
Below is the PyTorch implementation of the FGSM attack:

\begin{lstlisting}[style=python]
import torch
import torch.nn as nn

# FGSM Attack
def fgsm_attack(image, epsilon, data_grad):
    # Create the adversarial image by adjusting each pixel of the input image
    sign_data_grad = data_grad.sign()
    perturbed_image = image + epsilon * sign_data_grad
    # Clip the perturbed image to ensure it's still valid
    perturbed_image = torch.clamp(perturbed_image, 0, 1)
    return perturbed_image

# Example Usage of FGSM
def adversarial_example(model, device, test_loader, epsilon):
    # Set the model to evaluation mode
    model.eval()
    
    for data, target in test_loader:
        data, target = data.to(device), target.to(device)
        
        # Set requires_grad to True for the input image
        data.requires_grad = True
        
        # Forward pass
        output = model(data)
        loss = nn.CrossEntropyLoss()(output, target)
        
        # Zero all existing gradients
        model.zero_grad()
        
        # Backward pass to calculate gradients
        loss.backward()
        
        # Get the gradient of the input image
        data_grad = data.grad.data
        
        # Create the adversarial example
        perturbed_data = fgsm_attack(data, epsilon, data_grad)
        
        # Re-classify the perturbed image
        output = model(perturbed_data)
        
        # Do something with the adversarial output...
\end{lstlisting}

\subsubsection{Explanation of FGSM Code}
In the code above:
\begin{itemize}
    \item \texttt{fgsm\_attack}: This function takes in the original image, the perturbation magnitude (\texttt{epsilon}), and the gradient of the loss with respect to the input (\texttt{data\_grad}). It adjusts the image by adding the sign of the gradient to each pixel.
    \item \texttt{adversarial\_example}: This function processes a batch of images, calculates the gradients, and generates adversarial examples using the FGSM method.
\end{itemize}

\subsection{Projected Gradient Descent (PGD)}

The Projected Gradient Descent (PGD) attack is an iterative version of FGSM. Instead of taking a single step in the direction of the gradient, PGD takes multiple smaller steps, projecting the perturbation back into a specified epsilon-ball (i.e., ensuring the perturbation remains within a certain range).

\subsubsection{Code Implementation of PGD}
Let's look at how to implement PGD in PyTorch. The key idea is to repeatedly apply the FGSM update step but clip the perturbation after each step to stay within the bounds.

\begin{lstlisting}[style=python]
def pgd_attack(model, image, target, epsilon, alpha, num_iter):
    perturbed_image = image.clone().detach()
    perturbed_image.requires_grad = True

    for i in range(num_iter):
        output = model(perturbed_image)
        loss = nn.CrossEntropyLoss()(output, target)
        
        # Zero all existing gradients
        model.zero_grad()
        
        # Calculate gradients of model in backward pass
        loss.backward()
        data_grad = perturbed_image.grad.data
        
        # Take a small step in the direction of the gradient
        perturbed_image = perturbed_image + alpha * data_grad.sign()
        
        # Clip the perturbed image to ensure it's still in the valid range
        perturbation = torch.clamp(perturbed_image - image, -epsilon, epsilon)
        perturbed_image = torch.clamp(image + perturbation, 0, 1).detach()
        perturbed_image.requires_grad = True
    
    return perturbed_image
\end{lstlisting}

\subsubsection{Explanation of PGD Code}
\begin{itemize}
    \item \texttt{pgd\_attack}: This function takes multiple iterative steps to craft adversarial examples. In each iteration, it updates the image slightly in the direction of the gradient and ensures that the total perturbation stays within the allowed \(\epsilon\)-ball.
    \item The step size \(\alpha\) controls how large each update is. Smaller steps with more iterations usually yield stronger adversarial attacks.
\end{itemize}

\subsection{Carlini \& Wagner (CW) Attack}
The Carlini \& Wagner (CW) attack is a more advanced attack that is designed to minimize the perturbation needed to mislead the model. It formulates the attack as an optimization problem, which often leads to stronger adversarial examples.

\textbf{Note:} Implementing the CW attack is more complex and beyond the scope of this chapter. We will cover this attack in more detail in a later chapter.

\section{Code Implementation of Adversarial Training}

Adversarial training is a defense mechanism that improves the robustness of models by incorporating adversarial examples during the training process. The idea is to train the model not only on clean examples but also on adversarially perturbed examples.

\subsection{Basic Adversarial Training}
In adversarial training, we modify the training loop to include adversarial examples generated on-the-fly. These adversarial examples are generated using attacks such as FGSM or PGD during training.

\subsubsection{Code Implementation of Adversarial Training}
The following code implements adversarial training using FGSM:

\begin{lstlisting}[style=python]
def train_adversarial(model, device, train_loader, optimizer, epoch, epsilon):
    model.train()
    
    for batch_idx, (data, target) in enumerate(train_loader):
        data, target = data.to(device), target.to(device)
        
        # Standard forward pass
        optimizer.zero_grad()
        output = model(data)
        loss = nn.CrossEntropyLoss()(output, target)
        loss.backward()
        
        # Generate adversarial example
        data_grad = data.grad.data
        perturbed_data = fgsm_attack(data, epsilon, data_grad)
        
        # Forward pass with adversarial example
        output_adv = model(perturbed_data)
        loss_adv = nn.CrossEntropyLoss()(output_adv, target)
        loss_adv.backward()
        
        # Update model weights
        optimizer.step()

        if batch_idx % 100 == 0:
            print(f"Train Epoch: {epoch} [{batch_idx * len(data)}/{len(train_loader.dataset)}"
                  f"({100. * batch_idx / len(train_loader):.0f}%)]\tLoss: {loss.item():.6f}")
\end{lstlisting}

\subsubsection{Explanation of Adversarial Training Code}
\begin{itemize}
    \item We perform a forward pass on the clean data to calculate the loss and compute gradients.
    \item We then generate adversarial examples using FGSM and perform another forward pass using these examples.
    \item The model is updated using both clean and adversarial examples, which helps it learn to defend against adversarial attacks.
\end{itemize}

\subsection{Key Considerations for Adversarial Training}
\begin{itemize}
    \item \textbf{Balancing Clean and Adversarial Data}: It's important to find the right balance between clean and adversarial examples. Using too many adversarial examples can degrade performance on clean data.
    \item \textbf{Attack Strength}: Choosing the right \(\epsilon\) for generating adversarial examples is crucial. If the perturbation is too small, the model may not learn to defend against stronger attacks. If the perturbation is too large, it may hurt the model's overall performance.
\end{itemize}

\section*{Conclusion}
In this chapter, we explored some of the most common adversarial attacks such as FGSM and PGD, and learned how to implement them in PyTorch. We also introduced adversarial training, a technique to improve model robustness by incorporating adversarial examples into the training process. Understanding and implementing these techniques is essential for building models that can withstand adversarial attacks.

\chapter{Practical Poisoning Attacks and Defenses}

\section{Implementing Simple Poisoning Attacks}
In this section, we will explore how data poisoning attacks are carried out. The aim of such attacks is to inject malicious or incorrect data points into the training set in order to degrade the model's performance \cite{biggio2012poisoning}. We will start with a simple example of a classification problem using a neural network and introduce some poisoned data to observe its impact.

\subsection{Understanding Data Poisoning Attacks}
In a poisoning attack, an adversary manipulates a subset of the training data with the goal of influencing the model's predictions. For instance, in a classification task, some labels in the dataset may be deliberately corrupted so that the model learns incorrect mappings.

The most basic example of data poisoning is label flipping, where the adversary changes the label of certain data points to an incorrect class. Let's walk through an example step by step.

\subsection{Setting up a Simple Classifier}

We first create a simple neural network for a binary classification task using PyTorch.

\begin{lstlisting}[style=python]
import torch
import torch.nn as nn
import torch.optim as optim
from torch.utils.data import DataLoader, TensorDataset

# Define a simple neural network
class SimpleNN(nn.Module):
    def __init__(self):
        super(SimpleNN, self).__init__()
        self.fc1 = nn.Linear(2, 64)
        self.fc2 = nn.Linear(64, 1)
        self.sigmoid = nn.Sigmoid()

    def forward(self, x):
        x = torch.relu(self.fc1(x))
        x = self.sigmoid(self.fc2(x))
        return x

# Create synthetic data
def create_data():
    # Generate random points around two classes
    class_0 = torch.randn(50, 2) - 1
    class_1 = torch.randn(50, 2) + 1
    labels_0 = torch.zeros(50, 1)
    labels_1 = torch.ones(50, 1)

    data = torch.cat([class_0, class_1], dim=0)
    labels = torch.cat([labels_0, labels_1], dim=0)

    return data, labels

# Initialize model, loss, and optimizer
model = SimpleNN()
criterion = nn.BCELoss()
optimizer = optim.Adam(model.parameters(), lr=0.01)

# Prepare data
data, labels = create_data()
dataset = TensorDataset(data, labels)
dataloader = DataLoader(dataset, batch_size=10, shuffle=True)
\end{lstlisting}

In this code, we defined a simple neural network with two fully connected layers. We also generated synthetic data consisting of two classes. The data points of class 0 are centered around \((-1, -1)\), and those of class 1 are centered around \((1, 1)\). The model aims to distinguish between these two classes.

\subsection{Poisoning the Data}
Next, we will simulate a poisoning attack by flipping the labels of some data points. This will demonstrate how the model's performance degrades when it is trained on poisoned data.

\begin{lstlisting}[style=python]
# Poisoning the data by flipping labels of 10% of the data points
def poison_data(data, labels, poison_fraction=0.1):
    num_samples = len(labels)
    num_poison = int(poison_fraction * num_samples)
    indices = torch.randperm(num_samples)[:num_poison]  # Randomly select points to poison
    poisoned_labels = labels.clone()
    
    for i in indices:
        # Flip the label (0 -> 1, 1 -> 0)
        poisoned_labels[i] = 1 - poisoned_labels[i]
    
    return data, poisoned_labels

# Poison the dataset
poisoned_data, poisoned_labels = poison_data(data, labels)

# Create a new dataloader for the poisoned data
poisoned_dataset = TensorDataset(poisoned_data, poisoned_labels)
poisoned_dataloader = DataLoader(poisoned_dataset, batch_size=10, shuffle=True)
\end{lstlisting}

Here, we implement a function `poison\_data` that flips the labels of a random 10\% of the data points. After this, we replace the original labels with the poisoned labels and create a new `DataLoader` for the poisoned data.

\subsection{Training the Model on Poisoned Data}
Let's now train the model using the poisoned data and observe the impact.

\begin{lstlisting}[style=python]
# Training the model with poisoned data
def train_model(model, dataloader, criterion, optimizer, epochs=50):
    for epoch in range(epochs):
        running_loss = 0.0
        for inputs, labels in dataloader:
            optimizer.zero_grad()
            outputs = model(inputs)
            loss = criterion(outputs, labels)
            loss.backward()
            optimizer.step()
            running_loss += loss.item()
        print(f"Epoch {epoch+1}/{epochs}, Loss: {running_loss/len(dataloader)}")

# Train the model with poisoned data
train_model(model, poisoned_dataloader, criterion, optimizer)
\end{lstlisting}

In this function, we train the model for 50 epochs and display the loss at each epoch. You can run this training loop to observe how the model behaves with poisoned data. In practice, you should notice a performance degradation because the poisoned labels lead to incorrect learning.

\section{Best Practices for Defending Against Poisoning Attacks}
Data poisoning attacks can be harmful, but there are several strategies to defend against them. In this section, we will explore techniques such as data sanitization and robust training methods.

\subsection{Data Sanitization}
Data sanitization refers to techniques that help identify and remove corrupted data points before training \cite{oliveira2003protecting}. One approach is to check for data points that significantly deviate from the rest of the data. Let's implement a simple sanitization technique that removes suspicious data points based on a distance threshold.

\begin{lstlisting}[style=python]
# A simple sanitization function that removes points far from the class centers
def sanitize_data(data, labels, threshold=1.5):
    class_0_center = data[labels == 0].mean(dim=0)
    class_1_center = data[labels == 1].mean(dim=0)
    
    sanitized_data = []
    sanitized_labels = []
    
    for i in range(len(data)):
        if labels[i] == 0:
            distance = torch.norm(data[i] - class_0_center)
        else:
            distance = torch.norm(data[i] - class_1_center)
        
        if distance < threshold:
            sanitized_data.append(data[i])
            sanitized_labels.append(labels[i])
    
    return torch.stack(sanitized_data), torch.stack(sanitized_labels)

# Sanitize poisoned data
sanitized_data, sanitized_labels = sanitize_data(poisoned_data, poisoned_labels)

# Create a new dataloader with sanitized data
sanitized_dataset = TensorDataset(sanitized_data, sanitized_labels)
sanitized_dataloader = DataLoader(sanitized_dataset, batch_size=10, shuffle=True)
\end{lstlisting}

In this sanitization function, we compute the center of each class and remove data points that are too far from the class centers. This is a basic technique but can be effective against certain types of poisoning attacks.

\subsection{Robust Training Algorithms}
Another effective defense is to use robust training algorithms that are designed to tolerate some level of noise or corrupted data. One approach is to use \textbf{robust loss functions}, such as the Huber loss, which is less sensitive to outliers than the standard mean squared error.

Let's modify the loss function in our training loop to use the Huber loss.

\begin{lstlisting}[style=python]
# Define a robust loss function (Huber loss)
criterion = nn.SmoothL1Loss()  # This is the PyTorch implementation of Huber loss

# Train the model again with sanitized data and robust loss function
train_model(model, sanitized_dataloader, criterion, optimizer)
\end{lstlisting}

Here, we use `SmoothL1Loss` in PyTorch, which is equivalent to the Huber loss. This loss function helps reduce the influence of outliers (including poisoned data) during training.

\section*{Conclusion}
In this chapter, we explored practical poisoning attacks and defenses in machine learning. We demonstrated how to carry out a simple label-flipping attack and introduced two defense strategies: data sanitization and the use of robust loss functions. These are foundational concepts that can be expanded upon to create more sophisticated defense mechanisms.

As a beginner, it is important to experiment with these techniques in a controlled environment before applying them to real-world tasks.

\chapter{Practical Model Stealing Attacks and Defenses}

\section{Implementing Black-Box Model Theft}
In this section, we will simulate a basic scenario of black-box model theft. In a black-box setting, an adversary does not have access to the internals of the target model, such as its architecture or parameters. Instead, the adversary can query the model and receive its predictions. By leveraging the predictions, the adversary can then train a substitute model that mimics the behavior of the original model.

\subsection{Step 1: Setting Up the Target Model}
We will first define a target model that an adversary aims to steal. This model will be a simple neural network trained on the MNIST dataset.

\begin{lstlisting}[style=python]
import torch
import torch.nn as nn
import torch.optim as optim
from torchvision import datasets, transforms
from torch.utils.data import DataLoader

# Define the target model (a simple feedforward neural network)
class TargetModel(nn.Module):
    def __init__(self):
        super(TargetModel, self).__init__()
        self.fc1 = nn.Linear(28*28, 128)
        self.fc2 = nn.Linear(128, 64)
        self.fc3 = nn.Linear(64, 10)

    def forward(self, x):
        x = x.view(-1, 28*28)
        x = torch.relu(self.fc1(x))
        x = torch.relu(self.fc2(x))
        x = self.fc3(x)
        return x

# Load MNIST dataset
transform = transforms.Compose([transforms.ToTensor(), transforms.Normalize((0.5,), (0.5,))])
train_dataset = datasets.MNIST(root='./data', train=True, download=True, transform=transform)
train_loader = DataLoader(train_dataset, batch_size=64, shuffle=True)

# Initialize the model, loss function, and optimizer
target_model = TargetModel()
criterion = nn.CrossEntropyLoss()
optimizer = optim.SGD(target_model.parameters(), lr=0.01)

# Training the target model
def train_target_model(model, data_loader, criterion, optimizer, epochs=5):
    for epoch in range(epochs):
        for images, labels in data_loader:
            optimizer.zero_grad()
            outputs = model(images)
            loss = criterion(outputs, labels)
            loss.backward()
            optimizer.step()
        print(f'Epoch {epoch+1}, Loss: {loss.item():.4f}')

train_target_model(target_model, train_loader, criterion, optimizer)
\end{lstlisting}

In this code snippet, we define a simple neural network as our target model. It is trained on the MNIST dataset, which consists of handwritten digits, and we use this model to demonstrate how an adversary might steal it.

\subsection{Step 2: Adversarial Querying and Data Collection}
The adversary does not have access to the target model's architecture or weights but can query the model to get predictions. The next step is for the adversary to use these predictions to build a dataset that will be used to train a substitute model.

\begin{lstlisting}[style=python]
import numpy as np

# Adversary queries the target model to collect data
def query_target_model(model, data_loader):
    queried_data = []
    queried_labels = []
    model.eval()  # Set the model to evaluation mode
    with torch.no_grad():
        for images, _ in data_loader:  # Adversary doesn't know the true labels
            outputs = model(images)
            _, predicted_labels = torch.max(outputs, 1)
            queried_data.append(images.numpy())
            queried_labels.append(predicted_labels.numpy())
    
    queried_data = np.vstack(queried_data)
    queried_labels = np.hstack(queried_labels)
    return queried_data, queried_labels

# Collect data by querying the target model
query_loader = DataLoader(train_dataset, batch_size=64, shuffle=True)
stolen_data, stolen_labels = query_target_model(target_model, query_loader)
\end{lstlisting}

Here, the adversary queries the target model on a batch of images from the same dataset, receiving the model's predictions as labels. This creates a new dataset (`stolen\_data` and `stolen\_labels`) that will be used to train the substitute model.

\subsection{Step 3: Training the Substitute Model}
Next, the adversary uses the queried data to train a substitute model. The substitute model can be similar or different from the target model. In this case, we will use a similar architecture.

\begin{lstlisting}[style=python]
# Define the substitute model (same architecture as the target model)
class SubstituteModel(nn.Module):
    def __init__(self):
        super(SubstituteModel, self).__init__()
        self.fc1 = nn.Linear(28*28, 128)
        self.fc2 = nn.Linear(128, 64)
        self.fc3 = nn.Linear(64, 10)

    def forward(self, x):
        x = x.view(-1, 28*28)
        x = torch.relu(self.fc1(x))
        x = torch.relu(self.fc2(x))
        x = self.fc3(x)
        return x

# Convert stolen data to torch tensors
stolen_data_tensor = torch.tensor(stolen_data).float()
stolen_labels_tensor = torch.tensor(stolen_labels).long()

# Create a DataLoader for the stolen data
stolen_dataset = torch.utils.data.TensorDataset(stolen_data_tensor, stolen_labels_tensor)
stolen_loader = DataLoader(stolen_dataset, batch_size=64, shuffle=True)

# Initialize the substitute model, loss function, and optimizer
substitute_model = SubstituteModel()
criterion = nn.CrossEntropyLoss()
optimizer = optim.SGD(substitute_model.parameters(), lr=0.01)

# Train the substitute model using the stolen data
def train_substitute_model(model, data_loader, criterion, optimizer, epochs=5):
    for epoch in range(epochs):
        for images, labels in data_loader:
            optimizer.zero_grad()
            outputs = model(images)
            loss = criterion(outputs, labels)
            loss.backward()
            optimizer.step()
        print(f'Epoch {epoch+1}, Loss: {loss.item():.4f}')

train_substitute_model(substitute_model, stolen_loader, criterion, optimizer)
\end{lstlisting}

The adversary uses the stolen data to train a new model, the substitute model. After training, this model should replicate the behavior of the target model, effectively "stealing" it.

\section{Code Implementation for Protecting Models}
To prevent model theft, several defense mechanisms can be implemented. These include limiting API access, introducing noise into predictions, or using more sophisticated methods like watermarking the model.

\subsection{Limiting API Access}
One simple yet effective method of protecting models is by limiting API access. By restricting the number of queries a user can make to the model, we can reduce the amount of data that an adversary can collect.

\begin{lstlisting}[style=python]
class APILimiter:
    def __init__(self, max_queries):
        self.max_queries = max_queries
        self.query_count = 0

    def can_query(self):
        return self.query_count < self.max_queries

    def query_model(self, model, data):
        if self.can_query():
            self.query_count += 1
            return model(data)
        else:
            raise Exception("API query limit reached.")
            
# Example usage
api_limiter = APILimiter(max_queries=1000)
try:
    # Try querying the model
    outputs = api_limiter.query_model(target_model, images)
except Exception as e:
    print(e)
\end{lstlisting}

In this code, the `APILimiter` class ensures that only a limited number of queries can be made to the model. Once the limit is reached, further queries are denied, reducing the risk of model theft through excessive querying.

\subsection{Adding Noise to Predictions}
Another defense mechanism is to introduce slight noise to the model's predictions. This makes it more difficult for an adversary to train an accurate substitute model from the noisy predictions.

\begin{lstlisting}[style=python]
def noisy_predictions(model, data, noise_factor=0.1):
    with torch.no_grad():
        outputs = model(data)
        noise = torch.randn_like(outputs) * noise_factor
        return outputs + noise

# Example usage with noise added to predictions
noisy_output = noisy_predictions(target_model, images, noise_factor=0.05)
\end{lstlisting}

Here, we modify the model's predictions by adding random noise, making it harder for an adversary to collect clean labels. This adds uncertainty to any substitute model trained on these predictions.

\section*{Conclusion}
In this chapter, we explored how an adversary can replicate a model's behavior through black-box model theft. We also covered defense mechanisms, such as limiting API access and adding noise to predictions, to protect against such attacks. Understanding these techniques is critical for both attackers and defenders in the machine learning domain.

\chapter{Privacy Leakage Attack Implementations}
    \section{Experiments on Inference of Training Data}
    In this section, we will explore privacy leakage through two types of attacks commonly associated with machine learning models: model inversion and membership inference attacks \cite{yang2020defending}. Both of these attacks target a model's ability to inadvertently reveal information about its training data. 
    
    \subsection{Model Inversion Attacks}
    Model inversion attacks aim to reconstruct input data by exploiting the model's output. For instance, if a machine learning model has been trained to recognize faces, a model inversion attack may reconstruct a blurry image of the original training faces. This is possible because the model, by design, has learned patterns in the data, and these patterns can be manipulated to infer properties of the original data.

    The basic idea is that an attacker can feed inputs into a model and, using the model's output, gradually approximate the training data. For example, the attacker can modify an initial guess iteratively until the model's output becomes more confident about a specific class. In the case of a neural network trained to classify faces, this can result in the reconstruction of a face image resembling the original data.

    \subsection{Membership Inference Attacks}
    Membership inference attacks aim to determine whether a specific data point was part of the model's training data. This can be dangerous because it allows an attacker to infer whether certain private or sensitive information was used to train the model. For example, if a model is trained on medical records, an attacker could determine whether a particular individual's record was used in training, leading to serious privacy concerns.

    In these attacks, the model's behavior on a particular input is analyzed. Typically, a model will behave differently when it is asked to predict a data point it has seen during training versus one it has not. By examining confidence levels or output distributions, an attacker can infer whether the data was part of the training set.

    \subsection{Demonstrating Model Inversion Attack}
    To perform a basic model inversion attack, we will create a simple image classification model using PyTorch and then attempt to infer the training data from the model's output. Consider the following code to train a simple neural network on the MNIST dataset, a common dataset used for handwritten digit classification:

    \begin{lstlisting}[style=python]
import torch
import torch.nn as nn
import torch.optim as optim
from torchvision import datasets, transforms
from torch.utils.data import DataLoader

# Define a simple CNN model
class SimpleCNN(nn.Module):
    def __init__(self):
        super(SimpleCNN, self).__init__()
        self.conv1 = nn.Conv2d(1, 32, kernel_size=3)
        self.conv2 = nn.Conv2d(32, 64, kernel_size=3)
        self.fc1 = nn.Linear(64*12*12, 128)
        self.fc2 = nn.Linear(128, 10)
    
    def forward(self, x):
        x = torch.relu(self.conv1(x))
        x = torch.relu(self.conv2(x))
        x = x.view(-1, 64*12*12)
        x = torch.relu(self.fc1(x))
        x = self.fc2(x)
        return x

# Load MNIST dataset
transform = transforms.Compose([transforms.ToTensor()])
train_dataset = datasets.MNIST(root='./data', train=True, download=True, transform=transform)
train_loader = DataLoader(train_dataset, batch_size=64, shuffle=True)

# Instantiate the model, define the loss function and optimizer
model = SimpleCNN()
criterion = nn.CrossEntropyLoss()
optimizer = optim.Adam(model.parameters(), lr=0.001)

# Train the model
def train_model():
    model.train()
    for epoch in range(5):  # Train for 5 epochs
        for batch_idx, (data, target) in enumerate(train_loader):
            optimizer.zero_grad()
            output = model(data)
            loss = criterion(output, target)
            loss.backward()
            optimizer.step()

train_model()
    \end{lstlisting}

    In the next step, we will demonstrate how an attacker can perform a model inversion attack by using this trained model to reconstruct an image resembling the original training data. Here is a simple version of how this could be done by iteratively updating a random noise image to match the model's output for a specific class:

    \begin{lstlisting}[style=python]
import torch.optim as optim

# Initialize a random noise image
noise_image = torch.randn(1, 1, 28, 28, requires_grad=True)

# Set the target class (e.g., digit '0')
target_class = torch.tensor([0])

# Define the optimizer to update the noise image
optimizer = optim.Adam([noise_image], lr=0.1)

# Perform inversion attack by updating the noise image to maximize model output for target class
for i in range(1000):  # 1000 iterations
    optimizer.zero_grad()
    output = model(noise_image)
    loss = -output[0, target_class]  # Maximize the output for target class
    loss.backward()
    optimizer.step()

# Visualize the inverted image (which should resemble a '0')
import matplotlib.pyplot as plt
plt.imshow(noise_image.detach().numpy().reshape(28, 28), cmap='gray')
plt.show()
    \end{lstlisting}

    This code attempts to reconstruct an image of a digit '0' from the model's output. The image is iteratively adjusted to maximize the model's confidence in predicting the digit '0', demonstrating how private training data might be extracted.

    \section{Implementing Differential Privacy Protections}
    In this section, we will explore techniques to mitigate privacy leakage in machine learning models using differential privacy. Differential privacy ensures that the inclusion or exclusion of a single data point has minimal impact on the model's predictions, thus protecting sensitive information \cite{dwork2008differential}.

    \subsection{What is Differential Privacy?}
    Differential privacy provides a mathematical framework to quantify the privacy risks involved in publishing information about a dataset. The goal is to ensure that an attacker cannot confidently infer whether a specific individual's data was included in the training set, even with full knowledge of the rest of the data.

    In practice, differential privacy is often implemented by introducing noise into the training process. This noise makes it difficult for an attacker to distinguish between outputs based on the presence or absence of individual data points.

    \subsection{Implementing Differential Privacy in PyTorch}
    To implement differential privacy in PyTorch, we can use techniques like adding noise to the gradients during training. This ensures that the model's updates do not reveal sensitive information about individual training examples.

    Here is a basic implementation of differential privacy in PyTorch:

    \begin{lstlisting}[style=python]
import torch
import torch.nn as nn
import torch.optim as optim
from torchvision import datasets, transforms
from torch.utils.data import DataLoader

# Redefine the model as before
class SimpleCNN(nn.Module):
    def __init__(self):
        super(SimpleCNN, self).__init__()
        self.conv1 = nn.Conv2d(1, 32, kernel_size=3)
        self.conv2 = nn.Conv2d(32, 64, kernel_size=3)
        self.fc1 = nn.Linear(64*12*12, 128)
        self.fc2 = nn.Linear(128, 10)
    
    def forward(self, x):
        x = torch.relu(self.conv1(x))
        x = torch.relu(self.conv2(x))
        x = x.view(-1, 64*12*12)
        x = torch.relu(self.fc1(x))
        x = self.fc2(x)
        return x

# Load MNIST dataset as before
train_loader = DataLoader(train_dataset, batch_size=64, shuffle=True)

# Instantiate the model, loss function, and optimizer
model = SimpleCNN()
criterion = nn.CrossEntropyLoss()
optimizer = optim.Adam(model.parameters(), lr=0.001)

# Implementing differential privacy by adding noise to the gradients
def train_model_with_dp():
    model.train()
    for epoch in range(5):  # Train for 5 epochs
        for batch_idx, (data, target) in enumerate(train_loader):
            optimizer.zero_grad()
            output = model(data)
            loss = criterion(output, target)
            loss.backward()
            
            # Add noise to gradients to ensure differential privacy
            for param in model.parameters():
                param.grad += torch.randn_like(param.grad) * 0.01  # 0.01 is the noise scale
                
            optimizer.step()

train_model_with_dp()
    \end{lstlisting}

    In this code, we introduce noise to the gradients during backpropagation. This ensures that the information an attacker can infer about the individual training samples from the model's updates is limited, providing a level of privacy protection.

\chapter{Practical Label Flipping Attacks and Defenses}
    
\section{Implementing Label Flipping Attacks}
Label flipping attacks are a form of data poisoning where an attacker changes the labels of a dataset in a subtle manner, misleading the learning process of machine learning models \cite{qiu2022survey}. For example, in a binary classification problem, an attacker may flip the labels of a subset of examples from one class to the other. This can cause the model to learn incorrect decision boundaries, thus degrading its performance.

In this section, we will provide a step-by-step guide on how to simulate such an attack using PyTorch, and how to observe its impact on model performance. To keep things simple, we will use the well-known MNIST dataset for demonstration, flipping some of the labels and then training a simple neural network to observe the effect.

\subsection{Step 1: Load the MNIST Dataset}
First, we need to load the MNIST dataset using PyTorch's \texttt{torchvision} module. We will split the dataset into a training set and a test set, and then define a simple neural network for classification.

\begin{lstlisting}[style=python]
import torch
import torch.nn as nn
import torch.optim as optim
from torchvision import datasets, transforms
from torch.utils.data import DataLoader

# Define data transformation
transform = transforms.Compose([transforms.ToTensor(), transforms.Normalize((0.5,), (0.5,))])

# Load the training and test datasets
train_dataset = datasets.MNIST(root='./data', train=True, download=True, transform=transform)
test_dataset = datasets.MNIST(root='./data', train=False, download=True, transform=transform)

# Create data loaders for training and testing
train_loader = DataLoader(dataset=train_dataset, batch_size=64, shuffle=True)
test_loader = DataLoader(dataset=test_dataset, batch_size=64, shuffle=False)
\end{lstlisting}

\subsection{Step 2: Define the Neural Network}
We will define a simple feed-forward neural network with two fully connected layers to perform the classification task.

\begin{lstlisting}[style=python]
class SimpleNN(nn.Module):
    def __init__(self):
        super(SimpleNN, self).__init__()
        self.fc1 = nn.Linear(28*28, 128)
        self.fc2 = nn.Linear(128, 10)
    
    def forward(self, x):
        x = x.view(-1, 28*28)  # Flatten the image
        x = torch.relu(self.fc1(x))
        x = self.fc2(x)
        return x

# Initialize the model, loss function, and optimizer
model = SimpleNN()
criterion = nn.CrossEntropyLoss()
optimizer = optim.SGD(model.parameters(), lr=0.01)
\end{lstlisting}

\subsection{Step 3: Implement Label Flipping Attack}
Now we simulate a label flipping attack. In this example, we'll flip the labels for a certain percentage of the dataset (e.g., 10\%). For simplicity, we'll flip label '0' to '1' and label '1' to '0'.

\begin{lstlisting}[style=python]
import random

def flip_labels(dataset, flip_percentage):
    total_samples = len(dataset)
    num_flips = int(total_samples * flip_percentage)
    
    indices = random.sample(range(total_samples), num_flips)
    
    for idx in indices:
        original_label = dataset.targets[idx].item()
        if original_label == 0:
            dataset.targets[idx] = 1
        elif original_label == 1:
            dataset.targets[idx] = 0

# Flip 10% of the labels
flip_labels(train_dataset, flip_percentage=0.1)
\end{lstlisting}

\subsection{Step 4: Train the Model on Flipped Labels}
We will now train the model using the poisoned dataset where some of the labels have been flipped. This will demonstrate how label flipping attacks affect the learning process.

\begin{lstlisting}[style=python]
def train_model(model, train_loader, criterion, optimizer, num_epochs=5):
    model.train()
    for epoch in range(num_epochs):
        running_loss = 0.0
        for images, labels in train_loader:
            optimizer.zero_grad()
            outputs = model(images)
            loss = criterion(outputs, labels)
            loss.backward()
            optimizer.step()
            running_loss += loss.item()
        print(f'Epoch [{epoch+1}/{num_epochs}], Loss: {running_loss/len(train_loader):.4f}')

train_model(model, train_loader, criterion, optimizer)
\end{lstlisting}

\subsection{Step 5: Evaluate the Model}
After training, we will evaluate the model on the test set to see how the label flipping attack has affected the model's performance.

\begin{lstlisting}[style=python]
def evaluate_model(model, test_loader):
    model.eval()
    correct = 0
    total = 0
    with torch.no_grad():
        for images, labels in test_loader:
            outputs = model(images)
            _, predicted = torch.max(outputs.data, 1)
            total += labels.size(0)
            correct += (predicted == labels).sum().item()
    
    accuracy = 100 * correct / total
    print(f'Test Accuracy: {accuracy:.2f}%')

evaluate_model(model, test_loader)
\end{lstlisting}

This simple example demonstrates how label flipping can degrade a model's accuracy. Without any defense mechanism, the model is vulnerable to these types of attacks.

\section{Defensive Methods for Label Flipping}
To protect machine learning models against label flipping attacks, various defense strategies can be employed. In this section, we will explore two common defensive approaches:

\subsection{Robust Loss Functions}
Robust loss functions \cite{ghosh2017robust} are designed to mitigate the impact of incorrect labels during training. One such loss function is \texttt{MAE (Mean Absolute Error)}, which is less sensitive to outliers than the commonly used \texttt{Cross Entropy Loss}. We will modify the previous example to use \texttt{MAE} as the loss function.

\begin{lstlisting}[style=python]
# Define a robust loss function using Mean Absolute Error (MAE)
class RobustLoss(nn.Module):
    def __init__(self):
        super(RobustLoss, self).__init__()
    
    def forward(self, outputs, labels):
        one_hot_labels = torch.eye(10)[labels].to(outputs.device)  # One-hot encode the labels
        return torch.mean(torch.abs(outputs - one_hot_labels))

# Replace the loss function with the robust loss
robust_criterion = RobustLoss()

# Train the model using the robust loss function
train_model(model, train_loader, robust_criterion, optimizer)
\end{lstlisting}

Using a robust loss function can significantly improve the model's resistance to label flipping attacks.

\subsection{Anomaly Detection}
Another effective defense method is to use anomaly detection techniques to identify and remove potentially poisoned samples \cite{kloft2010online}. One simple approach is to monitor the model's predictions during training. If a particular sample is consistently misclassified, it may be an indication that its label has been flipped.

Here is an example of a simple anomaly detection approach based on tracking prediction consistency:

\begin{lstlisting}[style=python]
from collections import defaultdict

# Track prediction consistency over multiple epochs
def detect_anomalies(model, train_loader, threshold=3):
    model.eval()
    misclassified_samples = defaultdict(int)
    
    with torch.no_grad():
        for images, labels in train_loader:
            outputs = model(images)
            _, predicted = torch.max(outputs.data, 1)
            for i in range(len(labels)):
                if predicted[i] != labels[i]:
                    misclassified_samples[train_loader.dataset.targets[i].item()] += 1

    # Identify anomalies
    anomalies = [key for key, value in misclassified_samples.items() if value > threshold]
    return anomalies

anomalies = detect_anomalies(model, train_loader)
print(f'Detected anomalous samples: {anomalies}')
\end{lstlisting}

This method tracks how often each sample is misclassified and flags it as anomalous if it exceeds a certain threshold. By removing or investigating these anomalous samples, we can reduce the impact of label flipping attacks.

\section*{Conclusion}
In this chapter, we introduced the concept of label flipping attacks and demonstrated how they can degrade the performance of a machine learning model. We also explored defense mechanisms, including the use of robust loss functions and anomaly detection techniques. Implementing these defenses can help safeguard models from the negative effects of label poisoning.

\chapter{Practical Backdoor Attacks and Defenses}

\section{Backdoor Attack Implementation}
In this section, we will learn how to perform a backdoor attack on a machine learning model. A backdoor attack occurs when an adversary manipulates the training process by embedding hidden triggers into specific training data points. During inference, when these triggers are present in the input data, the model is manipulated to misclassify the input, even though it performs normally on inputs without the trigger.

\subsection{Steps for Backdoor Attacks}
To help beginners understand the process, we will break down the steps of performing a backdoor attack as follows:
\begin{enumerate}
    \item \textbf{Dataset Selection}: We'll use the MNIST dataset, which consists of handwritten digits (0-9).
    \item \textbf{Backdoor Trigger Creation}: A trigger, such as a small pattern or a pixel modification, will be added to some images in the training set.
    \item \textbf{Label Manipulation}: All images with a backdoor trigger will be labeled as a specific target class (e.g., all images with a trigger are labeled as “0”).
    \item \textbf{Model Training}: A neural network is trained on the poisoned dataset, which includes both clean and triggered samples.
    \item \textbf{Evaluation}: After training, the model is evaluated on both clean images and images containing the backdoor trigger to observe its misclassification behavior.
\end{enumerate}

\subsection{Backdoor Attack Example in PyTorch}
Here is the step-by-step implementation of the backdoor attack using PyTorch.

\begin{lstlisting}[style=python]
import torch
import torch.nn as nn
import torch.optim as optim
from torchvision import datasets, transforms
from torch.utils.data import DataLoader
import numpy as np
import matplotlib.pyplot as plt

# Step 1: Load the MNIST dataset
transform = transforms.Compose([
    transforms.ToTensor(),
    transforms.Normalize((0.5,), (0.5,))
])

train_data = datasets.MNIST(root='./data', train=True, download=True, transform=transform)
test_data = datasets.MNIST(root='./data', train=False, download=True, transform=transform)

# Step 2: Define a function to add a backdoor trigger
def add_backdoor_trigger(image, trigger_size=3, trigger_value=1.0):
    """Adds a square trigger to the bottom-right corner of the image."""
    image[0, -trigger_size:, -trigger_size:] = trigger_value
    return image

# Step 3: Modify some images in the training dataset
def poison_data(train_data, poison_fraction=0.1, target_label=0):
    """Adds backdoor triggers to a fraction of the training data."""
    poisoned_data = []
    for img, label in train_data:
        if np.random.rand() < poison_fraction:
            img = add_backdoor_trigger(img)
            label = target_label  # Change label to the target label (e.g., "0")
        poisoned_data.append((img, label))
    return poisoned_data

# Step 4: Define a simple neural network
class SimpleNN(nn.Module):
    def __init__(self):
        super(SimpleNN, self).__init__()
        self.fc = nn.Sequential(
            nn.Linear(28*28, 128),
            nn.ReLU(),
            nn.Linear(128, 10)
        )

    def forward(self, x):
        x = x.view(-1, 28*28)
        return self.fc(x)

# Step 5: Train the model with poisoned data
def train(model, poisoned_data, epochs=5, lr=0.001):
    criterion = nn.CrossEntropyLoss()
    optimizer = optim.Adam(model.parameters(), lr=lr)

    data_loader = DataLoader(poisoned_data, batch_size=64, shuffle=True)

    for epoch in range(epochs):
        for img, label in data_loader:
            optimizer.zero_grad()
            output = model(img)
            loss = criterion(output, label)
            loss.backward()
            optimizer.step()

        print(f'Epoch {epoch+1}/{epochs}, Loss: {loss.item()}')

# Step 6: Evaluate the model on clean and poisoned test data
def evaluate(model, test_data, poisoned=False):
    correct = 0
    total = 0
    data_loader = DataLoader(test_data, batch_size=64, shuffle=False)

    with torch.no_grad():
        for img, label in data_loader:
            if poisoned:
                img = add_backdoor_trigger(img)  # Test with the backdoor trigger

            output = model(img)
            _, predicted = torch.max(output, 1)
            total += label.size(0)
            correct += (predicted == label).sum().item()

    accuracy = 100 * correct / total
    print(f'Accuracy: {accuracy:.2f}%')

# Prepare poisoned data
poisoned_train_data = poison_data(train_data)

# Initialize and train the model
model = SimpleNN()
train(model, poisoned_train_data)

# Evaluate on clean test data
print("Evaluation on clean test data:")
evaluate(model, test_data)

# Evaluate on backdoored test data
print("Evaluation on backdoored test data:")
evaluate(model, test_data, poisoned=True)
\end{lstlisting}

This code example covers the complete process of performing a backdoor attack. The steps are as follows:
\begin{itemize}
    \item We first load the MNIST dataset.
    \item We define a function to add a small backdoor trigger (a small white square) to images.
    \item We poison the dataset by modifying a fraction of images to include the trigger and change their labels.
    \item A simple neural network is trained on this poisoned dataset.
    \item Finally, we evaluate the model on both clean and poisoned test data to demonstrate how the backdoor works.
\end{itemize}

\section{Detecting and Removing Backdoors in Practice}
After understanding how backdoor attacks are implemented, it is essential to learn how to detect and remove them. In practice, backdoor detection can be difficult but not impossible. We will explore several approaches that can be used to identify and neutralize backdoors.

\subsection{Data Inspection Techniques}
Detecting backdoors can sometimes be as simple as inspecting the training data for anomalies. Common techniques include:
\begin{itemize}
    \item \textbf{Visual Inspection}: Manually examine random samples from the training set. Look for patterns or modifications that may indicate a backdoor trigger.
    \item \textbf{Statistical Analysis}: Perform statistical analysis on the pixel values of images. This could reveal unusual patterns or outliers in the data.
    \item \textbf{Clustering}: Use clustering algorithms to group similar images and check for any clusters that contain triggered data samples.
\end{itemize}

\subsection{Model Inspection Techniques}
Even if data inspection fails to identify backdoors, you can still investigate the trained model. The following are useful model inspection methods:
\begin{itemize}
    \item \textbf{Activation Analysis}: Examine the activations of hidden layers. A model trained with a backdoor may show unusual activation patterns for inputs containing the trigger.
    \item \textbf{Adversarial Testing}: Introduce small perturbations or patterns in the input to test whether the model misclassifies in a manner consistent with the presence of a backdoor.
    \item \textbf{Fine-tuning}: Fine-tune the model on clean data. This may reduce the effectiveness of the backdoor without affecting the model's overall performance.
\end{itemize}

\subsection{Removing Backdoors}
Once a backdoor is detected, we need methods to remove it:
\begin{itemize}
    \item \textbf{Data Sanitization}: Clean the dataset by removing or modifying suspicious samples.
    \item \textbf{Retraining}: Retrain the model using the cleaned dataset.
    \item \textbf{Pruning}: Identify and remove specific neurons or layers in the model that contribute to the backdoor.
\end{itemize}

Here is an example of how to detect and remove a backdoor from a poisoned model:

\begin{lstlisting}[style=python]
# Step 1: Visual inspection of the dataset
# Plot a few images to check for triggers
def visualize_data(dataset, num_images=5):
    fig, axes = plt.subplots(1, num_images, figsize=(10, 2))
    for i in range(num_images):
        img, label = dataset[i]
        axes[i].imshow(img[0], cmap='gray')
        axes[i].set_title(f'Label: {label}')
        axes[i].axis('off')
    plt.show()

visualize_data(poisoned_train_data)

# Step 2: Remove poisoned data and retrain the model
def remove_poisoned_data(poisoned_data, trigger_value=1.0):
    """Remove images with the backdoor trigger."""
    cleaned_data = []
    for img, label in poisoned_data:
        if torch.max(img) != trigger_value:
            cleaned_data.append((img, label))
    return cleaned_data

cleaned_train_data = remove_poisoned_data(poisoned_train_data)

# Step 3: Retrain the model on the cleaned dataset
model_cleaned = SimpleNN()
train(model_cleaned, cleaned_train_data)

# Step 4: Evaluate the cleaned model
print("Evaluation on clean test data after removing backdoors:")
evaluate(model_cleaned, test_data)
\end{lstlisting}

In this example:
\begin{itemize}
    \item We visually inspect the poisoned dataset to detect the presence of backdoor triggers.
    \item After detecting the trigger, we remove the poisoned data and retrain the model.
    \item Finally, we evaluate the retrained model to ensure that the backdoor is no longer effective.
\end{itemize}

By using this combination of data inspection, model analysis, and retraining, we can mitigate backdoor attacks in machine learning systems.

\chapter{Self-Supervised Learning and Contrastive Learning Defenses}

\section{Practical Applications of Contrastive Learning Defenses}

In this section, we explore how contrastive learning methods can be utilized to develop more robust machine learning models that are less vulnerable to adversarial and poisoning attacks. These attacks pose significant challenges in security-critical applications, such as facial recognition systems \cite{kortli2020face}, malware detection \cite{aslan2020comprehensive}, and network traffic analysis \cite{joshi2015review}.

\subsection{Understanding Contrastive Learning}
Contrastive learning is a type of self-supervised learning that learns by comparing samples. The central idea is to train the model to differentiate between similar and dissimilar data points. For instance, if given two images of a cat, the model should learn to associate them as similar, whereas an image of a cat and a dog should be seen as dissimilar.

\textbf{Practical Example:} Let's say we want to build a system that can differentiate between malware and benign software. We can use contrastive learning by providing pairs of benign and malware samples. The model is trained to minimize the distance between benign software samples (i.e., making them similar in feature space) and to maximize the distance between benign and malware samples (i.e., making them distinct).

In adversarial defense, contrastive learning can help by ensuring that small perturbations to input data (which often characterize adversarial examples) do not lead to significant changes in the model's output.

\subsection{Using Contrastive Learning for Adversarial Defense}

Consider a situation where an attacker attempts to fool an image classifier by slightly modifying the pixels of an image, causing it to misclassify. By leveraging contrastive learning, we can teach the model to be less sensitive to small changes in the input data. The core idea is to encourage the model to treat adversarially perturbed images as similar to their original counterparts, thus improving robustness against adversarial attacks.

\textbf{Python Code Example:}

Here is an example of using contrastive learning in PyTorch to train a model that is more robust to adversarial attacks.

\begin{lstlisting}[style=python]
import torch
import torch.nn as nn
import torch.optim as optim
from torchvision import datasets, transforms
from torch.utils.data import DataLoader

class ContrastiveLoss(nn.Module):
    def __init__(self, margin=1.0):
        super(ContrastiveLoss, self).__init__()
        self.margin = margin
    
    def forward(self, output1, output2, label):
        # Euclidean distance
        distance = torch.norm(output1 - output2, p=2, dim=1)
        loss = torch.mean((1 - label) * torch.pow(distance, 2) +
                          (label) * torch.pow(torch.clamp(self.margin - distance, min=0.0), 2))
        return loss

# Dataset and DataLoader
transform = transforms.Compose([transforms.ToTensor()])
train_dataset = datasets.MNIST(root='./data', train=True, transform=transform, download=True)
train_loader = DataLoader(train_dataset, batch_size=64, shuffle=True)

# Simple CNN model
class SimpleCNN(nn.Module):
    def __init__(self):
        super(SimpleCNN, self).__init__()
        self.conv1 = nn.Conv2d(1, 32, kernel_size=3)
        self.fc1 = nn.Linear(32*26*26, 128)
    
    def forward(self, x):
        x = torch.relu(self.conv1(x))
        x = x.view(x.size(0), -1)
        x = torch.relu(self.fc1(x))
        return x

model = SimpleCNN()
optimizer = optim.Adam(model.parameters(), lr=0.001)
criterion = ContrastiveLoss()

# Training Loop
for epoch in range(10):
    model.train()
    total_loss = 0
    for data in train_loader:
        img1, img2 = data[0], data[1]
        label = torch.randint(0, 2, (img1.size(0),))  # Randomly assign similar/dissimilar pairs
        
        output1 = model(img1)
        output2 = model(img2)
        
        loss = criterion(output1, output2, label)
        optimizer.zero_grad()
        loss.backward()
        optimizer.step()
        
        total_loss += loss.item()
    
    print(f"Epoch [{epoch+1}/10], Loss: {total_loss/len(train_loader):.4f}")
\end{lstlisting}

\subsection{Defending Against Poisoning Attacks}
Poisoning attacks involve injecting malicious data into the training set with the goal of corrupting the model. Contrastive learning can mitigate this by ensuring that poisoned data remains distinct from legitimate data in feature space. As a result, the model can still perform well despite the presence of some poisoned samples.

For example, in a facial recognition system, an attacker might attempt to introduce manipulated images into the training set. By enforcing contrastive learning, the system learns to identify true facial features and becomes more resistant to the attack.

\section{Case Studies of Self-Supervised Learning in Security}

In this section, we will examine real-world applications where self-supervised learning has been used to enhance security, showcasing its utility in defending against various types of attacks.

\subsection{Case Study 1: Malware Detection}
One notable application of self-supervised learning is in malware detection. Malware often exhibits specific patterns that can be exploited to differentiate it from benign software. By applying contrastive learning, systems can better distinguish between malware and legitimate software based on learned representations, even when new and unseen types of malware emerge.

\textbf{Example:} A cybersecurity company applied contrastive learning to a large dataset of malware samples. By training a model to distinguish between benign and malicious files, they significantly improved their detection rate, particularly for zero-day exploits (i.e., new malware that had not been seen before).Below is an example using contrastive learning to improve malware detection.

\begin{lstlisting}[style=python]
# Assume we have a dataset of malware and benign samples in feature vectors
benign_samples = torch.randn(1000, 128)  # Fake benign data
malware_samples = torch.randn(1000, 128) + 1.0  # Fake malware data

# Labels for contrastive learning: 0 for similar (same class), 1 for dissimilar
labels = torch.cat([torch.zeros(500), torch.ones(500)])

# Model that encodes input samples into embeddings
class MalwareEncoder(nn.Module):
    def __init__(self):
        super(MalwareEncoder, self).__init__()
        self.fc1 = nn.Linear(128, 64)
        self.fc2 = nn.Linear(64, 32)
    
    def forward(self, x):
        x = torch.relu(self.fc1(x))
        x = torch.relu(self.fc2(x))
        return x

model = MalwareEncoder()
optimizer = optim.Adam(model.parameters(), lr=0.001)
criterion = ContrastiveLoss()

# Training loop
for epoch in range(10):
    optimizer.zero_grad()
    
    # Combine benign and malware samples
    all_samples = torch.cat([benign_samples, malware_samples])
    embeddings = model(all_samples)
    
    # Contrastive loss between benign and malware
    loss = criterion(embeddings[:500], embeddings[500:], labels)
    
    loss.backward()
    optimizer.step()
    
    print(f"Epoch [{epoch+1}/10], Loss: {loss.item():.4f}")
\end{lstlisting}

\subsection{Case Study 2: Network Traffic Anomaly Detection}
Another application is in detecting anomalies in network traffic, where normal and malicious traffic need to be separated. By leveraging self-supervised learning, we can automatically learn features of normal traffic patterns, which helps in identifying abnormal behavior more effectively. Contrastive learning helps in ensuring that small variations in normal traffic (e.g., due to natural fluctuations) do not trigger false alarms, while larger deviations (caused by attacks) are flagged as anomalies.

\textbf{Example:} In a study on intrusion detection, a network security company used contrastive learning to classify normal traffic and identify potential intrusions. By training their system on unlabeled network data, they improved anomaly detection without the need for manually labeled examples.Below is an example of how contrastive learning can be applied to network traffic data.

\begin{lstlisting}[style=python]
# Simulated network traffic data
normal_traffic = torch.randn(1000, 128)
attack_traffic = torch.randn(1000, 128) + 2.0

# Labels: 0 for normal, 1 for attack
traffic_labels = torch.cat([torch.zeros(500), torch.ones(500)])

class TrafficEncoder(nn.Module):
    def __init__(self):
        super(TrafficEncoder, self).__init__()
        self.fc1 = nn.Linear(128, 64)
        self.fc2 = nn.Linear(64, 32)
    
    def forward(self, x):
        x = torch.relu(self.fc1(x))
        x = torch.relu(self.fc2(x))
        return x

model = TrafficEncoder()
optimizer = optim.Adam(model.parameters(), lr=0.001)
criterion = ContrastiveLoss()

# Training loop
for epoch in range(10):
    optimizer.zero_grad()
    
    # Combine normal and attack traffic samples
    all_traffic = torch.cat([normal_traffic, attack_traffic])
    embeddings = model(all_traffic)
    
    # Contrastive loss between normal and attack traffic
    loss = criterion(embeddings[:500], embeddings[500:], traffic_labels)
    
    loss.backward()
    optimizer.step()
    
    print(f"Epoch [{epoch+1}/10], Loss: {loss.item():.4f}")
\end{lstlisting}

\section*{Conclusion}
By leveraging contrastive learning and self-supervised techniques, we can develop more robust models that can withstand adversarial and poisoning attacks. These methods provide practical defense mechanisms that improve security across various applications, from malware detection to network anomaly detection.

\part{Future Trends and Cutting-Edge Research}
\chapter{Frontier Technologies in Deep Learning Security}
\section{Automated Defense Systems}
In recent years, the increasing complexity and scalability of deep learning models have introduced new security challenges. Automated defense mechanisms present a promising solution by integrating security protocols directly into the model's development and operation stages \cite{vyas2023automated}. PyTorch, a popular deep learning framework, provides flexible and customizable features that can be employed to identify, monitor, and respond to security threats with minimal human intervention. This section explores how automated defense mechanisms can enhance the security of deep learning models.

\subsection{How Automation Can Enhance Model Security?}
Automation in deep learning security aims to detect unusual behaviors, protect sensitive data, and mitigate attacks, such as adversarial inputs or data poisoning, in real-time \cite{goodfellow2014explaining}. PyTorch's ecosystem offers various tools and libraries that allow for the implementation of automated security protocols.

\textbf{1. Anomaly Detection:}
Anomaly detection is a powerful technique for automatically identifying patterns that do not conform to expected behavior in data. In the context of deep learning security, anomaly detection can help identify suspicious activities, such as abnormal model inputs or outputs. Below is an example using PyTorch to build a basic autoencoder for anomaly detection.

\begin{lstlisting}[style=python]
import torch
import torch.nn as nn
import torch.optim as optim

# Define a simple autoencoder
class Autoencoder(nn.Module):
    def __init__(self):
        super(Autoencoder, self).__init__()
        self.encoder = nn.Sequential(
            nn.Linear(28 * 28, 128),
            nn.ReLU(),
            nn.Linear(128, 64),
            nn.ReLU(),
            nn.Linear(64, 32)
        )
        self.decoder = nn.Sequential(
            nn.Linear(32, 64),
            nn.ReLU(),
            nn.Linear(64, 128),
            nn.ReLU(),
            nn.Linear(128, 28 * 28),
            nn.Sigmoid()
        )

    def forward(self, x):
        x = self.encoder(x)
        x = self.decoder(x)
        return x

# Create model, loss function, and optimizer
model = Autoencoder()
criterion = nn.MSELoss()
optimizer = optim.Adam(model.parameters(), lr=0.001)

# Dummy training data (normally you'd use a real dataset)
data = torch.randn(100, 28 * 28)

# Train autoencoder
for epoch in range(10):
    for inputs in data:
        inputs = inputs.view(-1, 28 * 28)
        outputs = model(inputs)
        loss = criterion(outputs, inputs)
        
        optimizer.zero_grad()
        loss.backward()
        optimizer.step()

# Anomalies would have a significantly higher reconstruction error.
\end{lstlisting}

In this example, an autoencoder learns to compress and then reconstruct the input data. If an input does not follow the training data distribution (e.g., an anomalous or adversarial input), it will have a much higher reconstruction error, making it detectable.

\textbf{2. Active Monitoring:}
Active monitoring goes beyond anomaly detection by continually observing model behavior and automatically triggering defensive responses when a threat is detected. For instance, if a model receives a set of inputs that fall outside the norm, the system could trigger an alert or revert to a safe model checkpoint.

We can implement a simple monitoring system that evaluates the reconstruction error of the autoencoder and flags inputs that exceed a defined threshold as anomalies.

\begin{lstlisting}[style=python]
# Threshold for detecting anomalies
threshold = 0.02

# Function to monitor input and detect anomalies
def monitor_input(model, input_data):
    model.eval()  # Set the model to evaluation mode
    with torch.no_grad():
        output = model(input_data.view(-1, 28 * 28))
        loss = criterion(output, input_data)
        if loss.item() > threshold:
            print(f"Anomaly detected! Reconstruction loss: {loss.item()}")
        else:
            print("Input is normal.")

# Test with new input data
test_data = torch.randn(28 * 28)
monitor_input(model, test_data)
\end{lstlisting}

This system would help automatically monitor the inputs to the deep learning model, identifying potential security threats such as data poisoning or adversarial inputs.

\section{Zero Trust Architecture in Deep Learning}
Zero Trust Architecture (ZTA) is a security framework based on the principle of “never trust, always verify” \cite{kindervag2010no}. In the context of deep learning models, this architecture is highly relevant for securing every stage of the model's lifecycle—from data collection and model training to deployment and inference. By applying Zero Trust, we aim to reduce the attack surface and ensure that every entity interacting with the model, including internal components, must be continuously authenticated and authorized.

\subsection{How Zero Trust Enhances Security in Deep Learning Models?}
The Zero Trust model can significantly enhance security in deep learning by ensuring that every interaction with the system, whether from users, devices, or other components, is validated. In PyTorch, we can implement Zero Trust principles by incorporating continuous checks, monitoring, and access control mechanisms.

\textbf{1. Securing the Training Data:}
One of the core principles of Zero Trust is verifying the integrity of the data before allowing it to be used in model training. Data integrity checks can be implemented using hash-based verifications.

\begin{lstlisting}[style=python]
import hashlib

# Function to compute the hash of the dataset
def compute_hash(data):
    return hashlib.sha256(data).hexdigest()

# Example dataset
data1 = b"Training data for model A"
data2 = b"Training data for model B"

# Hashing the data
hash1 = compute_hash(data1)
hash2 = compute_hash(data2)

# Store known good hash for verification
known_hash = hash1

# Verify integrity of data before training
if compute_hash(data2) != known_hash:
    print("Data integrity compromised! Abort training.")
else:
    print("Data integrity verified. Proceed with training.")
\end{lstlisting}

In this example, the hash of the training data is computed and compared to a known good hash before starting the training process. If the hash does not match, training is aborted, preventing corrupted or poisoned data from being used.

\textbf{2. Continuous Authentication During Inference:}
A key element of Zero Trust is ensuring that every inference request is continuously authenticated. In a real-world system, this would involve complex authentication mechanisms like tokens or certificates. Below is a simplified example of how to use a token-based authentication system during inference.

\begin{lstlisting}[style=python]
# Simulate token-based authentication
valid_token = "secure_token_123"

def authenticate_request(token):
    if token == valid_token:
        return True
    else:
        return False

# Function to handle inference requests
def infer(model, input_data, token):
    if authenticate_request(token):
        model.eval()  # Set model to evaluation mode
        with torch.no_grad():
            output = model(input_data.view(-1, 28 * 28))
        print("Inference successful.")
        return output
    else:
        print("Unauthorized request. Inference aborted.")
        return None

# Example inference request with token
test_input = torch.randn(28 * 28)
infer(model, test_input, "secure_token_123")  # Authorized request
infer(model, test_input, "invalid_token")     # Unauthorized request
\end{lstlisting}

In this scenario, the inference process only proceeds if the request token matches the expected value. This ensures that only authorized users or systems can interact with the model.

By implementing Zero Trust Architecture principles like continuous authentication and data integrity checks, we can minimize the risks associated with malicious actors, compromised data, and unauthorized access to the model.

\section*{Conclusion}
Both automated defense systems and Zero Trust Architecture offer valuable frameworks for enhancing the security of deep learning models. PyTorch provides flexible tools and libraries to implement these security measures, from anomaly detection and active monitoring to ensuring data integrity and enforcing continuous authentication. As deep learning continues to evolve, integrating security mechanisms at every stage of the model lifecycle will become increasingly critical.

\chapter{The Future of Security-Enhanced AI Models}

\section{Generative Models and Their Role in Countering Adversarial Examples}
In recent years, generative models have gained significant attention in the AI community due to their ability to generate realistic data. Two of the most popular generative models, Generative Adversarial Networks (GANs) \cite{goodfellow2020generative} and Variational Autoencoders (VAEs) \cite{girin2020dynamical}, have shown promising applications in enhancing the security of AI systems. In this section, we will discuss how PyTorch-based generative models can be used to generate adversarial-resistant data and to defend against adversarial attacks.
In recent years, generative models have gained significant attention in the AI community due to their ability to generate realistic data.  

\subsection{How Generative AI Models Can Defend Against Adversarial Attacks?}
Adversarial attacks are crafted inputs designed to fool machine learning models. These attacks exploit the vulnerabilities in the model, leading to incorrect outputs. Generative models can help in countering such attacks by either generating adversarial examples to train the model to be more robust or by detecting adversarial inputs.

\subsubsection{Generating Adversarial Examples}
One of the ways to enhance the robustness of a machine learning model is by exposing it to adversarial examples during training. Generative models, like GANs, can be used to create synthetic adversarial examples that are difficult for the model to classify correctly, thereby strengthening the model.

A GAN consists of two neural networks: a generator and a discriminator. The generator creates synthetic examples, while the discriminator tries to distinguish between real and fake examples. Over time, the generator learns to produce increasingly realistic examples. In the context of adversarial defense, we can modify the GAN to produce adversarial inputs to stress-test a classifier model.

Below is a simple PyTorch implementation of a GAN that generates adversarial examples:

\begin{lstlisting}[style=python]
import torch
import torch.nn as nn
import torch.optim as optim

# Define the generator network
class Generator(nn.Module):
    def __init__(self):
        super(Generator, self).__init__()
        self.fc = nn.Sequential(
            nn.Linear(100, 256),
            nn.ReLU(),
            nn.Linear(256, 784),
            nn.Tanh()
        )
    
    def forward(self, x):
        return self.fc(x)

# Define the discriminator network
class Discriminator(nn.Module):
    def __init__(self):
        super(Discriminator, self).__init__()
        self.fc = nn.Sequential(
            nn.Linear(784, 256),
            nn.LeakyReLU(0.2),
            nn.Linear(256, 1),
            nn.Sigmoid()
        )
    
    def forward(self, x):
        return self.fc(x)

# Training loop for GAN
def train_gan(generator, discriminator, dataloader, num_epochs=100):
    criterion = nn.BCELoss()
    optimizer_g = optim.Adam(generator.parameters(), lr=0.0002)
    optimizer_d = optim.Adam(discriminator.parameters(), lr=0.0002)

    for epoch in range(num_epochs):
        for real_images, _ in dataloader:
            # Train discriminator
            optimizer_d.zero_grad()
            real_labels = torch.ones(real_images.size(0), 1)
            fake_labels = torch.zeros(real_images.size(0), 1)
            
            # Real images
            outputs = discriminator(real_images)
            loss_real = criterion(outputs, real_labels)
            
            # Fake images
            noise = torch.randn(real_images.size(0), 100)
            fake_images = generator(noise)
            outputs = discriminator(fake_images.detach())
            loss_fake = criterion(outputs, fake_labels)
            
            # Total discriminator loss
            loss_d = loss_real + loss_fake
            loss_d.backward()
            optimizer_d.step()

            # Train generator
            optimizer_g.zero_grad()
            outputs = discriminator(fake_images)
            loss_g = criterion(outputs, real_labels)
            loss_g.backward()
            optimizer_g.step()

        print(f"Epoch {epoch+1}/{num_epochs}, Loss D: {loss_d.item()}, Loss G: {loss_g.item()}")
\end{lstlisting}

This GAN example uses a generator to produce fake images (adversarial examples) and a discriminator to distinguish between real and fake images. By training the classifier on both real and adversarial examples, its robustness to adversarial attacks is improved.

\subsubsection{Detecting Adversarial Examples}
Another defense mechanism is to use generative models for detecting adversarial inputs. VAEs are particularly effective in this scenario because they learn the probability distribution of the training data. Any input that significantly deviates from this distribution can be flagged as an adversarial example. Here's an implementation of a simple VAE in PyTorch:

\begin{lstlisting}[style=python]
class VAE(nn.Module):
    def __init__(self):
        super(VAE, self).__init__()
        self.fc1 = nn.Linear(784, 400)
        self.fc21 = nn.Linear(400, 20)  # Mean vector
        self.fc22 = nn.Linear(400, 20)  # Log variance vector
        self.fc3 = nn.Linear(20, 400)
        self.fc4 = nn.Linear(400, 784)

    def encode(self, x):
        h1 = torch.relu(self.fc1(x))
        return self.fc21(h1), self.fc22(h1)

    def reparameterize(self, mu, logvar):
        std = torch.exp(0.5 * logvar)
        eps = torch.randn_like(std)
        return mu + eps * std

    def decode(self, z):
        h3 = torch.relu(self.fc3(z))
        return torch.sigmoid(self.fc4(h3))

    def forward(self, x):
        mu, logvar = self.encode(x.view(-1, 784))
        z = self.reparameterize(mu, logvar)
        return self.decode(z), mu, logvar

# Loss function for VAE
def loss_function(recon_x, x, mu, logvar):
    BCE = nn.functional.binary_cross_entropy(recon_x, x.view(-1, 784), reduction='sum')
    KLD = -0.5 * torch.sum(1 + logvar - mu.pow(2) - logvar.exp())
    return BCE + KLD
\end{lstlisting}

The VAE model reconstructs the input data, and if an input does not align with the learned distribution, it could indicate that the input is adversarial. This approach can be used as a defense by rejecting inputs that are likely to be adversarial.

\section{Security Challenges in Large AI Models}
As AI models grow in size and complexity, new security challenges arise. Large models, especially in distributed environments, are more vulnerable to attacks such as model extraction, poisoning, and privacy breaches. In this section, we will discuss some of these challenges and how they affect the security of PyTorch-based models.

\subsection{How the Complexity of Large Models Introduces New Security Challenges?}
Larger models, with billions of parameters, offer enhanced capabilities but also introduce new vulnerabilities. Below are some of the key security challenges:

\subsubsection{Model Extraction Attacks}
Model extraction attacks aim to steal the model by querying it repeatedly and analyzing its responses. Large models, due to their complexity, are often deployed in environments where attackers can interact with them, such as web services. This can allow an attacker to replicate the model's behavior.

\subsubsection{Privacy Concerns in Federated Learning}
Federated learning enables multiple devices to collaboratively train a model while keeping their data local. However, this distributed setup introduces privacy risks. An attacker could potentially reverse-engineer individual data points from shared model updates, which is especially concerning in sensitive domains like healthcare.

\subsubsection{Distributed Training Vulnerabilities}
Training large models often involves multiple GPUs or nodes in a distributed system. This distributed setup increases the attack surface, making it easier for an attacker to tamper with the training data, parameters, or gradients during transmission between nodes.

\chapter{Balancing Model Performance and Security}

\section{Trade-offs Between Model Optimization and Security}

When developing deep learning models using PyTorch, there is always a balance between optimizing for performance and ensuring security. Performance optimization focuses on making the model faster, more efficient, and capable of running on less hardware, which includes reducing memory usage and computational time. However, focusing too much on performance may leave the model vulnerable to adversarial attacks. This section will guide you through these trade-offs, using PyTorch as a framework for detailed examples.

\subsection{Optimizing for Performance}

Optimizing a PyTorch model for performance can take various forms, such as quantization, pruning, or using mixed precision training. These techniques help to reduce the size of the model, improve speed, or reduce memory usage. Let's look at each of these briefly:

\begin{itemize}
    \item \textbf{Quantization}: This reduces the precision of the numbers used in the model, for example, switching from 32-bit floating point numbers to 8-bit integers. This can significantly improve both the memory usage and computational efficiency.
    \item \textbf{Pruning}: This involves removing neurons or layers that do not significantly contribute to the model's final output, reducing model size and improving inference speed.
    \item \textbf{Mixed Precision Training}: This allows the use of 16-bit floating point numbers in parts of the network where full 32-bit precision is not necessary, improving speed while retaining accuracy.
\end{itemize}

Here's a simple example of how to apply quantization to a PyTorch model:

\begin{lstlisting}[style=python]
import torch
import torch.quantization

# Define a simple neural network
class SimpleNN(torch.nn.Module):
    def __init__(self):
        super(SimpleNN, self).__init__()
        self.fc1 = torch.nn.Linear(784, 128)
        self.fc2 = torch.nn.Linear(128, 10)

    def forward(self, x):
        x = torch.relu(self.fc1(x))
        x = self.fc2(x)
        return x

# Initialize and prepare for quantization
model = SimpleNN()
model.eval()

# Fuse layers before quantization (if needed)
model.fuse_model()

# Apply static quantization
quantized_model = torch.quantization.quantize_dynamic(model, {torch.nn.Linear}, dtype=torch.qint8)
\end{lstlisting}

While quantization reduces the computational load, it can sometimes result in a slight reduction in model accuracy. The trade-off here is that you're gaining performance and efficiency at the potential cost of model performance, which may make it more susceptible to adversarial attacks.

\subsection{Security Risks with Optimized Models}

As you optimize models, they can become more vulnerable to adversarial attacks. For example, when using techniques like quantization or pruning, attackers may exploit reduced precision or sparsity to craft adversarial inputs that cause the model to fail or produce incorrect outputs.

Adversarial examples are inputs designed to fool a neural network into making incorrect predictions. Here's a simple visual representation of how adversarial examples can alter the output of a neural network:

\begin{center}
\begin{tikzpicture}
\draw[very thick,->] (0,0) -- (3,0) node[midway, above] {Original Input};
\draw[very thick,->] (3,0) -- (6,0) node[midway, above] {Neural Network};
\draw[very thick,->] (6,0) -- (9,0) node[midway, above] {Correct Prediction};
\draw[very thick,->] (0,-2) -- (3,-2) node[midway, above] {Adversarial Input};
\draw[very thick,->] (3,-2) -- (6,-2) node[midway, above] {Neural Network};
\draw[very thick,->] (6,-2) -- (9,-2) node[midway, above] {Incorrect Prediction};
\end{tikzpicture}
\end{center}

Consider that optimizing the model might inadvertently reduce its ability to distinguish between clean and adversarial examples. For example, a model that has been aggressively pruned might lack the redundancy needed to robustly handle adversarial inputs.

\section{The Future of Model Security Standards}

As models become more optimized and widely deployed, the need for robust security standards becomes crucial. Security standards ensure that models remain safe from adversarial attacks and maintain their integrity, privacy, and robustness. This section introduces emerging best practices and frameworks that help achieve these goals in PyTorch models.

\subsection{Model Robustness Guidelines}

Various guidelines are emerging to ensure that models are robust against adversarial threats:

\begin{itemize}
    \item \textbf{Adversarial Training}: One effective way to improve a model's robustness is by training it with adversarial examples. By exposing the model to perturbed inputs during training, it becomes more resilient to such attacks.
    \item \textbf{Differential Privacy}: This method helps ensure that the model does not memorize specific training data points, thus protecting sensitive information. PyTorch has libraries that enable differential privacy for training models.
    \item \textbf{Certified Robustness}: Some frameworks offer mathematical guarantees of robustness, meaning the model will remain unaffected by adversarial inputs within certain bounds.
\end{itemize}

\subsection{Emerging Security Frameworks}

In the future, new security frameworks will likely emerge that provide even stronger guarantees of model robustness and privacy. Here are some of the promising directions:

\begin{itemize}
    \item \textbf{Secure Multi-party Computation (SMPC) \cite{zhao2019secure}}: Allows multiple parties to jointly compute a function over their inputs while keeping the inputs private.
    \item \textbf{Homomorphic Encryption \cite{yi2014homomorphic}}: Allows computations to be performed on encrypted data without needing to decrypt it, protecting the confidentiality of the input.
    \item \textbf{Federated Learning \cite{li2020review}}: Ensures that models are trained across decentralized devices while keeping data locally, protecting user privacy.
\end{itemize}

These frameworks, combined with PyTorch's flexibility and ecosystem of libraries, will shape the future of secure machine learning.

\subsection*{Conclusion}

In conclusion, balancing performance and security is an ongoing challenge for machine learning practitioners. While optimizing PyTorch models can lead to significant performance gains, it is crucial to remain vigilant about security concerns, particularly regarding adversarial attacks. Emerging standards like adversarial training, differential privacy, and robust frameworks will help create more secure and resilient models.

\bibliographystyle{unsrt}
\bibliography{sample}

\end{document}